\documentclass[twocolumn,aps,prd,amsmath,amssymb,preprintnumbers,longbibliography]{revtex4-1}
\usepackage{amsmath} \usepackage{amsfonts} \usepackage{amssymb}
\usepackage{bbm}
\usepackage{epsfig}
\usepackage{graphics}
\usepackage{graphicx}
\usepackage{titlesec}
\usepackage{mathtools}
\usepackage{environ}
\usepackage{dsfont}
\usepackage{version}
\usepackage[colorlinks=true]{hyperref}
\hypersetup{urlcolor=black,linkcolor=black,citecolor=black}
\usepackage{natbib}
\usepackage[toc,page]{appendix}

\usepackage{dsfont}

\makeatletter
\renewcommand*\env@matrix[1][c]{\hskip -\arraycolsep
  \let\@ifnextchar\new@ifnextchar
  \array{*\c@MaxMatrixCols #1}}
\makeatother

\makeatletter										%%
\def\p@subsection{\thesection .\,} %%punkt in References
\makeatother                                       %%

\newcommand{\be}{\begin{equation}}
\newcommand{\ee}{\end{equation}}
\newcommand{\ba}{\begin{align}}
\newcommand{\ea}{\end{align}}
\newcommand{\nn}{\nonumber}

\newcommand{\diag}{\text{ diag }}

\newcommand{\gl}{\big(}
\newcommand{\gr}{\big)}

\newcommand{\K}{{\mathcal K}}

\newcommand{\te}{t+\varepsilon}

\titleformat{\subsection}[block]{\normalfont\bfseries}{\thesubsection.}{1ex}{}
\titlespacing{\subsection}{0pt}{10pt}{1pt}[0pt]
\titleformat*{\section}{\large\bfseries}
\renewcommand{\thesubsection}{\arabic{subsection}}

\pdfoutput=1 %%%Reference ohne spaces
\usepackage{todonotes}
\usepackage{braket}

\newcommand{\tr}{\mathrm{tr}}

\newcommand{\exval}[1]{\langle #1 \rangle}

\newcommand{\cD}{\mathcal{D}}

\newcommand{\bel}[1]{\be\label{#1}}
\newcommand{\psibar}{\overline{\psi}}
\newcommand{\zetabar}{\overline{\zeta}}
\newcommand{\Ktil}{\tilde{\K}}
\newcommand{\epstil}{\tilde{\varepsilon}}
\newcommand{\gbar}{\bar{g}}
\newcommand{\ghat}{\widehat{g}}
\newcommand{\subt}[1]{_{\text{#1}}}
\newcommand{\supt}[1]{^{\text{#1}}}
\newcommand{\inn}{\subt{in}}
\newcommand{\ff}{_{f}}
\newcommand{\atil}{\tilde{a}_{\alpha}}

\newcommand{\qq}[1]{``#1"}

\newcommand{\Shat}{\widehat{S}}
\newcommand{\dt}{\partial_{t}}
\newcommand{\eps}{\varepsilon}
\newcommand{\herm}{^{\dagger}}
\newcommand{\hateq}{\widehat{=}}
\newcommand{\pmat}[4]{\begin{pmatrix} #1 & #2 \\ #3 & #4 \end{pmatrix}}
\newcommand{\pvec}[2]{\begin{pmatrix} #1 \\ #2 \end{pmatrix}}
\newcommand{\Dp}{\widehat{D}_{+}}
\newcommand{\Dm}{\widehat{D}_{-}}
\newcommand{\Cp}{\widehat{C}_{+}}
\newcommand{\Cm}{\widehat{C}_{-}}
\newcommand{\gtil}{\tilde{g}}

\newcommand{\mtil}{\tilde{m}}
\newcommand{\qtil}{\tilde{q}}
\newcommand{\qbar}{\bar{q}}

\newcommand{\p}{^{\dagger}}
\newcommand{\g}{_{\gamma}}

\newcommand{\Stil}{\widetilde{S}}

\newcommand{\2}{^{2}}

\definecolor{refkey}{rgb}{0,0,1}
\definecolor{labelkey}{rgb}{0,1,0}

\begin{document}

%\preprint{}

\title{\LARGE Fermionic quantum field theories as\\probabilistic cellular automata}

\author{C. Wetterich}

\affiliation{Institut  f\"ur Theoretische Physik\\
Universit\"at Heidelberg\\
Philosophenweg 16, D-69120 Heidelberg}

\begin{abstract}
A class of fermionic quantum field theories with interactions is shown to be equivalent to probabilistic cellular automata, namely cellular automata with a probability distribution for the initial states. Probabilistic cellular automata on a one-dimensional lattice are equivalent to two - dimensional quantum field theories for fermions. They can be viewed as generalized Ising models on a square lattice and therefore as classical statistical systems. As quantum field theories they are quantum systems. Thus quantum mechanics emerges from classical statistics.
As an explicit example for an interacting fermionic quantum field theory we describe a type of discretized Thirring model as a cellular automaton.
The updating rule of the automaton is encoded in the step evolution operator that can be expressed in terms of fermionic annihilation and creation operators.
The complex structure of quantum mechanics is associated to particle -- hole transformations. The naive continuum limit exhibits Lorentz symmetry. We exploit the equivalence to quantum field theory in order to show how quantum concepts as wave functions, density matrix, non-commuting operators for observables and similarity transformations are convenient and useful concepts for the description of probabilistic cellular automata. 
\end{abstract}

\maketitle
\bigskip
\noindent

\section{Introduction\label{sec: introduction}}

We show that a class of 1+1-dimensional discretized fermionic quantum field theories can be described as rather simple probabilistic cellular automata. The latter being classical statistical systems, this is an example how quantum mechanics emerges from classical statistics~\cite{Wetterich:2020kqi},~\cite{CWQMCS}. Besides its conceptual relevance, showing for example that no-go theorems based on Bell's inequalities~\cite{BELL},~\cite{CHSH} cannot apply, we hope that the equivalence can help to solve a class of fermionic quantum field theories.
Probabilistic cellular automata can be seen as an example for probabilistic computing. The quantum concepts used to describe probabilistic cellular automata, as wave functions and non-commuting operators for observables, may become useful tools in this context.

The main idea is rather simple. Consider a quantum system with a finite number of states $\tau$. In our fermionic context these will be configurations of occupation numbers taking the values one or zero. For discretized time with time steps $\eps$ the evolution is described by the step evolution operator $\Shat$. In continuum quantum mechanics $\Shat$ corresponds to the evolution operator evaluated for a finite time difference $\eps$. For particular models the step evolution operator $\Shat$ can be a unique jump operator. This means that the matrix elements $\Shat_{\tau\rho}$ are either one or zero, with a single one in each row and column of the matrix. For a unique jump operator every configuration $\rho$ at time $t$ is mapped to a unique configuration $\tau(\rho)$ at $t+\eps$. Taking the occupation numbers as bits, this is precisely the updating step of an automaton or a step in classical computing. With certain locality properties the automaton is a cellular automaton.

Quantum mechanics is a probabilistic theory. The cellular automaton will therefore be a probabilistic cellular automaton. While the evolution is deterministic in this particular case, the probabilistic aspects enter by a probability distribution over initial conditions. The probabilistic information is encoded in a probability distribution $\{p_\tau\}$, or more conveniently in a wave function $\{q_\tau\}$, with $p_\tau=q_\tau^2$. Viewing the wave function as a real vector, the step evolution operator describes the evolution of the wave function by matrix multiplication
\bel{I1}
q_\tau(t+\eps)=\Shat_{\tau\rho}q_\rho(t)\ .
\ee
The continuity of the probabilistic information or wave function reflects the wave aspects of quantum mechanics, while the discrete occupation numbers encode the particle aspects.

We will see how the probabilistic aspects related to the wave function play a crucial role for the understanding of quantum features. In this respect our approach differs from the interesting attempt by t'Hooft to describe quantum mechanics by deterministic cellular automata ~\cite{HOOFT2,GTH,ELZE,HOOFT3,HOOFT4}.

A probabilistic cellular automaton can be described as a generalized Ising model~\cite{CWFGI}. A chain of bits or occupation numbers $n(x)$, with discrete positions $x$, is equivalent to a chain of Ising spins $s(x)$, $s=2n-1$.
A (classical) cellular automaton~\cite{JVN,ULA,ZUS,GAR,LIRO,TOOM,DKT,WOLF,VICH,PREDU,TOMA,FLN,HED,RICH,AMPA,HPP,CREU} updates the bit configuration at every time step to a definite new bit configuration. This can be cast into the form of an Ising model on a two-dimensional lattice with points $(t,x)$. The interactions between Ising spins at neighboring time layers have to be chosen such that the probability is one for the allowed transitions between two neighboring configurations of the cellular automaton, and zero otherwise.
This is easily cast into the form of a partition function or \qq{functional integral} by choosing for the weight factor $e^{-S}$ an action $S$ that diverges to infinity for the forbidden transitions. The probabilistic aspects are implemented by boundary terms at some initial and final time. This type of generalized Ising model is a simple classical statistical system. The equivalence of a quantum model with a probabilistic cellular automaton is therefore the equivalence of a quantum system with a classical probabilistic system.

Ising models~\cite{LENZ,ISING,BINDER} are a central tool of information theory~\cite{SHA}. One can establish a general \qq{bit-fermion map} between generalized Ising models and Grassmann functional integrals~\cite{CWFGI}. For the general case the weight distribution for Ising spins needs not be positive, and the evolution of the fermionic model described by the Grassmann functional integral needs not to be unitary. The equivalence between a fermionic quantum field theory and a probabilistic cellular automaton can be understood as an example for the bit-fermion map for which the fermionic model has a unitary evolution and the weight distribution for Ising spins is a positive probability distribution. The bit-fermion map differs from other fermionic descriptions of two-dimensional Ising models~\cite{PLECH,BER1,BER2,SAM,ITS,PLE1} or other forms of fermion-boson equivalence in two dimensions~\cite{FUR,NAO,COL,DNS}. It is actually valid in arbitrary dimensions. We refer for the formulation of our models as a generalized Ising model to ref.~\cite{CWPCA} and retain here only the property that a probabilistic cellular automaton is a classical statistical system.

It is rather easy to formulate free fermions in two dimensions as probabilistic cellular automata~\cite{CWFCS,Wetterich:2020kqi}. Fermions simply move on the two-dimensional lattice on straight lines as $t$ increases, either to the right or to the left in $x$. A first example of equivalence of a fermionic quantum field theory with interactions with a probabilistic cellular automaton was established in ref.~\cite{CWPCA}. In the present paper we present a generalization of the treatment of interactions which allows for the implementation of a wider class of interactions.
The cellular automaton alternates a propagation step and an interaction step, somewhat similar to the functional integral description of quantum mechanics with alternating factors for the kinetic and potential energies. We develop an expression of the step evolution operator in terms of fermionic annihilation and creation operators which makes the fermionic interpretation of the cellular automaton rather apparent.

The evolution equation~\eqref{I1} involves a real wave function. We propose that the complex structure characteristic of quantum mechanics can be associated to the particle-hole transformation of the cellular automaton or a switch of sign of the Ising spins in the associated generalized Ising model. This allows us to formulate the probabilistic cellular automaton as a quantum system with the usual complex Hilbert space.
The presence of antiparticles characteristic for fermionic quantum field theories arises naturally in this setting.

%Absatz
A continuum limit becomes possible for sufficiently smooth wave functions (not for deterministic cellular automata). A naive continuum limit simplifies the description considerably, leading to the usual Schrödinger equation for complex wave functions, or von Neumann equations for density matrices. These evolution equations are identical for the probabilistic cellular automaton and the fermionic quantum field theory. In the naive continuum limit our fermionic model is Lorentz invariant. If the naive continuum limit already grasps all essential features, or if a renormalization group running introduces interesting new ones, remains to be investigated. A first short account of some of the results can be found in ref.~\cite{CWFCB}.

The starting point for this work is the Grassmann functional integral for the fermionic quantum field theory. In sect.~\ref{sec: 02} we develop the general formalism how to extract the step evolution operator from this functional integral. In sect.~\ref{sec: 03} we construct fermionic models with interactions for which the step evolution operator is a unique jump operator. These are particular discretized Thirring-type models in 1+1-dimensions~\cite{THI,KLA,AAR,FAIV}. We investigate Weyl, Majorana and Dirac fermions in appendix~\ref{ap.C}.
The equivalent cellular automaton is presented in sect.~\ref{sec: 04}. The updating rule is rather simple. In a first step the bits or particles are moving either one step to the right or to the left. Besides the right-movers and left-movers the bits or particles also come in two colors, say red and green. The interaction is implemented in a second step. Whenever a single right-mover and a single left-mover meet at a point $x$, their colors are switched between red and green. We establish the close correspondence of general features of the updating rule and symmetries of the associated Thirring type model in appendix~\ref{ap.D}.

In sect.~\ref{sec: 05} we turn to the probabilistic aspects of the cellular automaton. We introduce the wave function and establish that it is the same for the automaton and the fermion model. We discuss the density matrix, with details given in appendix~\ref{ap.F}. In appendix~\ref{ap.G} we demonstrate the appearance of non-commuting operators for the cellular automaton. These operators are in complete analogy to the fermionic quantum field theory. Sect.~\ref{sec: 06} introduces the complex structure related to the particle-hole transformation. The complex wave function for the cellular automaton establishes the complete applicability of the quantum formulation to this classical statistical system~\cite{CWIT,CWQF}. The corresponding complex hermitian operators are discussed in appendix~\ref{ap.H}.
Sect.~\ref{sec: 07} expresses the step evolution operator in terms of a Hamiltonian. This Hamiltonian is expressed in terms of fermionic annihilation and creation operators, underlining the fermionic interpretation of the cellular automaton. The Hamiltonian can be used for a continuous time evolution that coincides with the automaton at discrete time intervals. The continuum limit further simplifies the description. In sec.~\ref{sec: 08} we discuss different possible vacua and the corresponding one-particle excitations. Sect.~\ref{sec: 09} introduces the momentum and position operators for the one-particle states, and establishes the corresponding uncertainty relation. We discuss our results and possible extensions in sect.~\ref{sec: 10}.

\section{Step evolution operator for models of fermions\label{sec: 02}}

%Absatz
A key quantity for our investigation is the step evolution operator.
For a discrete formulation of functional integrals the step evolution operator $\widehat{S}$ corresponds to the transfer matrix~\cite{BAX,FUC} with a particular normalization. According to this normalization the largest absolute values among the eigenvalues of $\widehat{S}$ are equal to one. The step evolution operator describes the propagation of the local probabilistic information on a \qq{time}-hypersurface to a neighboring hypersurface.
If $\widehat{S}$ is an orthogonal matrix no information is lost. In the presence of a complex structure an orthogonal matrix that is compatible with the complex structure is equivalent to a unitary matrix in the complex picture. Orthogonal step evolution operators generate then a unitary evolution. 

%Absatz
Unique jump matrices have precisely one element equal to one in each row and column. They are orthogonal matrices. If the step evolution operator is a unique jump matrix it describes an automaton. Each local bit-configuration on a hypersurface is mapped to precisely one other local bit-configuration on a neighboring hypersurface. The hypersurfaces can be associated with the time steps of an automaton. Then the step evolution operator $\widehat{S}(t)$ describes how each microscopic state at $t$ is mapped precisely to another microscopic state at $t+\tilde{\varepsilon}$. This extends to probabilistic states as given by a probability distribution over the microscopic states. 

%Absatz
We consider here models with one space-dimension. The fermionic occupation numbers or bits $n_{\gamma}(x)$ are located on the discrete positions $x$ of a chain. For a suitable step evolution operator these positions can be associated with the cells of a cellular automaton. This requires that the updating of the bits in the cell $x$ is only influenced by the configurations of bits in a few neighboring cells. The local fermionic quantum field theories discussed in the present paper realize this cellular automaton property.

%Absatz 
For a given fermionic quantum field theory specified by a Grassmann functional integral we need to extract the associated step evolution operator. Inversely, one may construct for a given step evolution operator the associated quantum field theory. A general formalism for the extraction of the step evolution operator for Grassmann functional integrals for fermionic models has been developed in ref.~\cite{CWPT,CWFCS,CWFGI}. We briefly summarize it here. We specialize to alternating sequences of kinetic operators that describe the change of location of particles, and interaction operators. Both are unique jump operators. In consequence, the unitary evolution is guaranteed and the models correspond to cellular automata. We construct, in particular, the step evolution operator for a particular Thirring-type model. This demonstrates that our setting covers fermionic quantum field theories with non-trivial interactions.

\subsection{Grassmann functional integral\label{subsec: Grassmann functional integral}}
\medskip

Consider a Grassmann functional integral
\begin{equation}\label{01}
Z=\int \mathcal{D}\psi \exp(-S[\psi])=\int\mathcal{D}\psi w[\psi]\;,
\end{equation}
with action
\begin{equation}\label{02}
S=\sum_{t}\mathcal{L}(t)\;.
\end{equation}
For $\mathcal{L}(t)$ involving only even powers of Grassmann variables the weight functional $w[\psi]$ can be written as a product of commuting time local factors $\tilde{\mathcal{K}}(t)$ , 
\be\label{04}
w[\psi]=\exp(-S[\psi])=\!\prod_{t}\tilde{\K}(t)\, , \, \; \tilde{\K}(t)=\exp\lbrace-\mathcal{L}(t)\rbrace\;.
\ee
For our models each local factor depends on two sets of Grassmann variables
$\psi_{\alpha}(t+\tilde{\varepsilon})=\psi_{\gamma}(t+\tilde{\varepsilon}, x)$ and $\psi_{\beta}(t)=\psi_{\delta}(t, y)$ at neighboring $t+\tilde{\varepsilon}$ and $t$. We do not impose space-locality at this stage and leave the range of $x$, $y$ free for the moment.

%Absatz
An element of the local Grassmann algebra at $t$ can be written as a linear combination of Grassmann basis functions, 
\be\label{05}
g(t)=q_{\tau}(t)g_{\tau}(t)\; .
\ee
(We use summation over double indices if not specified otherwise.)
The basis functions $g_{\tau}(t)=g_{\tau}[\psi(t)]$ are products of Grassmann variables $\psi_{\alpha}(t)$
\be\label{06}
g_{\tau}(t)=\tilde{s}_{\tau}\prod_{\alpha=1}^{M}\tilde{a}_{\alpha}\;,
\ee
with $\tilde{a}_{\alpha}=1$ or $\tilde{a}_{\alpha}=\psi_{\alpha}$, and $\tilde{s}_{\tau}=\pm 1$ some conveniently chosen signs. The two possibilities for $\tilde{a}_{\alpha}=\tilde{a}_{\gamma}(x)$ correspond to the two possibilities of a fermion of type $\gamma$ to be present at $x$ or not, where the precise association will be specified later.
For $\alpha=1\dots M$ there a $2^{M}$ basis functions, $\tau=1\dots N , N=2^{M}$. They correspond to the $2^{M}$ microstates for $M$ fermionic degrees of freedom. These microstates can also be interpreted as the configurations for $M$ classical bits or Ising spins, which leads to the equivalence with generalized Ising models~\cite{CWFGI}.

For the convenience of manipulating signs we also define (no sum over $\tau$ here) a second set of basis functions
\be\label{07}
g_{\tau}'=\varepsilon_{\tau}g_{\tau}(t)\,,\quad \varepsilon_{\tau}=(-1)^{\tfrac{m_{\tau}(m_{\tau}-1)}{2}}\;,
\ee
with $m_{\tau}$ the number of $\psi$ - factors in $g_{\tau}$. For arbitrary Grassmann variables $\psi_{\alpha}$ and $\varphi_{\alpha}$ we observe the identity
\be\label{08}
\begin{split}
\exp(\psi_{\alpha}\varphi_{\alpha})&=\prod_{\alpha}(1+\psi_{\alpha}\varphi_{\alpha})= \sum_{\tau}\varepsilon_{\tau}g_{\tau}(\psi)g_{\tau}(\varphi)\\
&=\sum_{\tau}g_{\tau}(\psi)g_{\tau}'(\varphi)\;.
\end{split}
\ee
In this way the signs $\varepsilon_{\tau}$ which arise from anticommutators of Grassmann variables are absorbed by the definition of $g'_{\tau}$. We observe that $g_{\tau}'$ can be obtained from $g_{\tau}$ by a total reordering of all Grassmann variables.

\subsection{Step evolution operator\label{subsec: Step evolution operator}}
\medskip
\noindent

%Absatz
Grassmann functionals have a modular two property~\cite{CWFGI} since only a sequence of two unit step operators reproduces the identical local Grassmann algebra, $g(t+2\tilde{\varepsilon})=g(t)$. This feature is conveniently encoded in the use of different basis functions for even and odd $t$. Here we consider discrete time steps, $t=t_{0}+m\tilde{\varepsilon}$, and denote by odd or even $t$ the integer $\mtil$ being odd or even.
The ``transfer matrix" $\widehat{T}_{\tau\rho}(t)$ is defined for $t$ odd by the double expansion of the local factor~$\tilde{\K}(t)$ in basis functions at $t$ and $t+\tilde{\varepsilon}$,
\be\label{09}
\tilde{\K}(t)=g_{\tau}(t+\tilde{\varepsilon})\widehat{T}_{\tau\rho}(t)g_{\rho}'(t)\;.
\ee
Adding a constant to $\mathcal{L}(t)$ multiplies $\widehat{T}_{\tau\rho}(t)$ by a constant factor. This freedom is used to normalize $\widehat{T}_{\tau\rho}(t)$ such that its largest eigenvalues obey $|\lambda_{i}|=1$.
Here ``largest" means the largest absolute size. For our models there will be more than a single largest eigenvalue. With this normalization the transfer matrix becomes the ``step evolution operator" $\widehat{S}_{\tau\rho}(t)$. We implicitly assume in the following a suitable normalization of $\mathcal{L}(t)$ such that 
\be\label{10}
\tilde{\K}(t)=g_{\tau}(t+\tilde{\varepsilon})\widehat{S}_{\tau\rho}(t)g_{\rho}'(t)\;.
\ee

%Absatz
In view of the modulo two properties of Grassmann functional integrals~\cite{CWFCS,CWFGI,Wetterich:2010ni} we define the step evolution operator for even $t$ by an expansion in conjugate basis functions, 
\be\label{11}
\tilde{\K}(t)=\overline{g}_{\tau}'(t+\tilde{\varepsilon})\widehat{S}_{\tau\rho}(t)\overline{g}_{\rho}(t)\;.
\ee
The conjugate basis functions are defined by the relation
\be\label{12}
\int\mathcal{D}\psi \overline{g}_{\tau}(\psi)g_{\rho}(\psi)=\delta_{\tau\rho}\;.
\ee
Up to signs the map from $g_{\tau}$ to $\overline{g}_{\tau}$ exchanges factors of one and $\psi_{\alpha}$ in eq.~\eqref{06}. 
The association of ``occupied" and ``empty" to $1$ and $\psi_{\alpha}$ therefore switches between even and odd $t$.

We also employ (no sum here)
\be\label{13}
\overline{g}_{\tau}'=\varepsilon_{\tau}'\overline{g}_{\tau}
\ee
obeying
\be\label{14}
\exp(\psi_{\alpha}\varphi_{\alpha})=\sum_{\tau}\overline{g}_{\tau}'(\psi)\overline{g}_{\tau}(\varphi)\;.
\ee
This fixes
\be\label{15}
\varepsilon_{\tau}'=(-1)^{\tfrac{m_{\tau}'(m_{\tau}'-1)}{2}}=(-1)^{m_{\tau}}\varepsilon_{\tau}\eta_{M}(-1)^{Mm_{\tau}}
\;,
\ee
with $m_{\tau}'=M-m_{\tau}$ the number of $\psi$ factors in $\overline{g}_{\tau}$ and $\eta_{M}=1$ for $M=0,1 \bmod 4$ and $\eta_{M}=-1$ for $M=2,3 \mod 4$. For $\overline{g}_{\tau}'$ we observe a relation similar to eq.~\eqref{12}
\be\label{16}
\int\mathcal{D}\psi g_{\tau}'(\psi)\overline{g}_{\rho}'(\psi)=\eta_{M}\delta_{\tau\rho}\;.
\ee

%Absatz
An explicit expression for $M=1$ reads
\begin{align}\label{17}
g_{1}=g_{1}'=\overline{g}_{2}=\overline{g}_{2}'=1 \,, \nonumber \\ 
g_{2}=g_{2}'=\overline{g}_{1}=\overline{g}_{1}'=\psi \,, 
\end{align}
while for $M=2$ one has
\begin{align}\label{18}
&g_{1}=\phantom{-}g_{1}'=\phantom{-}\overline{g}_{4}=\phantom{-}\overline{g}_{4}'=1 \,, \nonumber \\ 
&g_{2}=\phantom{-}g_{2}'=-\overline{g}_{3}=-\overline{g}_{3}'=\psi_{2} \,, \nonumber \\ 
&g_{3}=\phantom{-}g_{3}'=\phantom{-}\overline{g}_{2}=\phantom{-}\overline{g}_{2}'=\psi_{1} \,, \nonumber \\ 
&g_{4}=-g_{4}'=\phantom{-}\overline{g}_{1}=-\overline{g}_{1}'=\psi_{1}\psi_{2} \;.
\end{align}
The general relation between $\overline{g}_{\tau}$ and $g_{\tau}$ is given by
\be\label{19}
\overline{g}_{\tau}(\varphi)=\eta_{M}\int\mathcal{D}\psi\exp(\psi_{\alpha}\varphi_{\alpha})g_{\tau}'(\psi)\;.
\ee
This results in
\begin{align}\label{20}
&\overline{g}_{\tau}'(\varphi)=\int\mathcal{D}\psi\exp(\varphi_{\alpha}\psi_{\alpha})g_{\tau}(\psi) \quad \text{for $M$ even} \nn \\
&\overline{g}_{\tau}'(\varphi)=\int\mathcal{D}\psi\exp(\psi_{\alpha}\varphi_{\alpha})g_{\tau}(\psi)  \quad\text{for $M$ odd}
\end{align}
In the following we will focus on $M=4 \mod 4$ where $\eta_{M}=1$. For an arbitrary number of fermionic species this can be realized by a suitable number of space points.

%Absatz
For the product of two neighboring local factors one has for $t$ odd
\be\label{21}
\tilde{\mathcal{K}}(t+\tilde{\varepsilon})\tilde{\mathcal{K}}(t)=\overline{g}_{\tau}'(t+2\tilde{\varepsilon})\widehat{S}_{\tau\alpha}(t+\tilde{\varepsilon})F_{\alpha\beta}(t+\tilde{\varepsilon})\widehat{S}_{\beta\rho}(t)g_{\rho}'(t)\,,
\ee
with
\be\label{22}
F_{\alpha\beta}(t)=\overline{g}_{\alpha}(t)g_{\beta}(t) \,, \quad \int\mathcal{D}\psi(t)F_{\alpha\beta}(t)=\delta_{\alpha\beta}\;.
\ee
It is the simplicity of the second relation\eqref{22} that justifies the use of the conjugate basis functions.
The integration of the product~\eqref{21} over the common Grassmann variables $\psi(t+\epstil)$ yields a matrix multiplication of the step evolution operator
\be\label{23}
\int\mathcal{D}\psi(t+\tilde{\varepsilon})\tilde{\mathcal{K}}(t+\tilde{\varepsilon})\tilde{\mathcal{K}}(t)=\overline{g}_{\tau}'(t+2\tilde{\varepsilon})\bigl(\widehat{S}(t+\tilde{\varepsilon})\widehat{S}(t)\bigr)_{\tau\rho} g_{\rho}'(t)\;.
\ee
Similarly, one obtains ($t$ odd)
\be\label{24}
\tilde{\mathcal{K}}(t)\tilde{\mathcal{K}}(t-\tilde{\varepsilon})=g_{\tau}(t+\tilde{\varepsilon})\widehat{S}_{\tau\alpha}(t)\tilde{F}_{\alpha\beta}(t)\widehat{S}_{\beta\rho}(t-\tilde{\varepsilon})\overline{g}_{\rho}(t-\tilde{\varepsilon})
\ee
with
\be\label{25}
\tilde{F}_{\alpha\beta}(t)=g_{\alpha}'(t)\overline{g}_{\beta}'(t)\,,\quad \int\mathcal{D}\psi(t)\tilde{F}_{\alpha\beta}(t)=\eta_{M}\delta_{\alpha\beta}\;.
\ee
Again, integrating the intermediate Grassmann variable results in a matrix product
\be\label{26}
\int\mathcal{D}\psi(t)\tilde{\mathcal{K}}(t)\tilde{\mathcal{K}}(t-\tilde{\varepsilon})=\eta_{M}g_{\tau}(t+\tilde{\varepsilon})\bigl(\widehat{S}(t)\widehat{S}(t-\tilde{\varepsilon})\bigr)_{\tau\rho} \overline{g}_{\rho}(t-\tilde{\varepsilon})\,.
\ee

%Absatz
The product structure extends to longer chains of neighboring local factors. 
Employing the relations~\eqref{22},~\eqref{25} the integration over intermediate Grassmann variables results in matrix multiplication of the step evolution operators.
For initial time $t_\textup{in}$ even one can express the partition function by a chain of ordered matrix products of step evolution operators
\be\label{27}
\begin{split}
Z=\int\mathcal{D}\psi(t\ff)\mathcal{D}\psi(t_\textup{in})g_{\tau}(t\ff)\bigl(\widehat{S}(t\ff-\tilde{\varepsilon}) \dots \\
\widehat{S}(t_\textup{in}+1)\widehat{S}(t_\textup{in})\bigr)_{\tau\rho}\overline{g}_{\rho}(t_\textup{in})\;.
\end{split}
\ee
Here we have assumed an odd number of time points $M_{t}$. (Only for both $M=2,3 \bmod 4$ and $M_{t}=3 \bmod 4$ there is an additional factor $\eta_{M}$). In the following we consider $M=4 \bmod 4$, such that $\int\mathcal{D}\psi(t)$ commutes with all Grassmann variables $\psi_{\alpha}(t'\neq t)$ and
\be\label{29a}
\eta_{M}=1 \quad, \quad\quad\varepsilon_{\tau}'=(-1)^{m_{\tau}}\varepsilon_{\tau}\;.
\ee

We can write  eq.~\eqref{27} in the form
\be\label{28}
Z= \tr \bigl{\lbrace} \widehat{S}(t\ff-\tilde{\varepsilon}) \dots \widehat{S}(t_\textup{in})\widehat{\mathcal{B}}\,\bigr{\rbrace}\,,
\ee
where the boundary matrix is given for open boundary conditions by
\be\label{29}
\widehat{\mathcal{B}}_{\rho\tau}=\int\mathcal{D}\psi(t\ff)\mathcal{D}\psi (t_\textup{in})g_{\tau}(t\ff)\overline{g}_{\rho}(t_\textup{in}).
\ee
If one adds in $Z$ boundary factors
\begin{align}
\label{28AA}
g\subt{in}[\psi(t\subt{in})]=&\tilde{q}_{\tau}(t\subt{in})g_{\tau}[\psi(t\subt{in})]\ ,\nn\\
\gbar_f[\psi(t_f)]=&\bar{q}_{\tau}(t_f)\gbar_{\tau}[\psi(t_f)]\ ,
\end{align}
the boundary matrix becomes
\bel{28AB}
\widehat{\mathcal{B}}_{\rho\tau}=\tilde{q}_{\rho}(t\subt{in})\bar{q}_{\tau}(t_f)\ .
\ee
For mixed boundary conditions this matrix can be generalized further.

%Absatz
A this stage we have formally constructed for every sequence of step evolution operators $\widehat{S}(t)$ the associated local factors $\tilde{\mathcal{K}}(t)$, and therefore $\mathcal{L}(t)$ and the functional integral~\eqref{01}~-~\eqref{04}. The opposite direction is formally straightforward.
The step evolution operator obtains from $\tilde{\mathcal{K}}(t)$ for $t$~odd as
\be\label{30}
\widehat{S}_{\tau\rho}(t)=\int\mathcal{D}\psi(t+\tilde{\varepsilon})\mathcal{D}\psi(t)\overline{g}_{\tau}(t+\tilde{\varepsilon})\tilde{\mathcal{K}}(t)\overline{g}_{\rho}'(t)\,,
\ee
while for even $t$ one has
\be\label{31}
\widehat{S}_{\tau\rho}(t)=\int\mathcal{D}\psi(t+\tilde{\varepsilon})\mathcal{D}\psi(t)g_{\tau}'(t+\tilde{\varepsilon})\tilde{\mathcal{K}}(t)g_{\rho}(t)\,.
\ee

%Absatz
In particular, the unit step evolution operator $\widehat{S}_{\tau\rho}(t)=\delta_{\tau\rho}$ obtains by virtue of eqs.~\eqref{08}~\eqref{14} for
\be\label{32}
\tilde{\mathcal{K}}_{id}(t)=\exp\Bigl{\lbrace}\sum_{x}\psi_{\gamma}(t+\tilde{\varepsilon}, x)\psi_\gamma(t, x)\Bigr{\rbrace}\,.
\ee
We emphasize that a static state (unit step evolution operator) does not correspond to a unit local factor. It rather involves an action with a time derivative according to
\begin{align}\label{32a}
S_{0}=&\sum_{t}\mathcal{L}_{0}(t)\;\widehat{=}\;\frac{1}{\tilde{\varepsilon}}\int dt \mathcal{L}_{0}(t)\nn\\
=&-\frac{1}{\tilde{\varepsilon}}\int_{t}\sum_{x}\psi_{\gamma}(t+\tilde{\varepsilon},x)\psi_{\gamma}(t,x)\nn\\
=&-\int_{t}\sum_{x}\frac{1}{\tilde{\varepsilon}}\Big{[}\psi_{\gamma}(t+\tilde{\varepsilon},x)-\psi_{\gamma}(t,x)\Big{]}\psi_{\gamma}(t,x)\nn\\
\widehat{=}&-\int_{t}\sum_{x}\partial_{t}\psi_{\gamma}(t,x)\psi_{\gamma}(t,x)=\int_{t}\sum_{x}\psi_{\gamma}(t,x)\partial_{t}\psi_{\gamma}(t,x)\;.
\end{align}
Such a term is typically part of the action for models of non-relativistic or relativistic fermions.

\subsection{Propagating fermions\label{subsec: Propagating fermions}}
\medskip

%Absatz
For a two-dimensional system we define the "right transport operator" by a local factor
\be\label{33}
\tilde{\mathcal{K}}_{R}(t)=\exp\Bigl{\lbrace}\sum_{x}\psi_{\gamma}(t+\tilde{\varepsilon}, x+\varepsilon)\psi_\gamma(t, x)\Bigr{\rbrace}\,.
\ee
The corresponding step evolution operator is a unique jump operator that maps any state $\rho$ at $t$ to precisely one state $\tau=\overline{\tau}(\rho) $ at $t+\tilde{\varepsilon}$,
\be\label{34}
\widehat{S}_{\tau\rho}=\delta_{\tau,\overline{\tau}(\rho)}\,.
\ee
Comparing with eq.~\eqref{32} a particle or empty place (hole) at $(t,x)$ is now found at $t+\epstil$ at the position $x+\varepsilon$ instead of staying at $x$ for the identity. More in detail, $\overline{\tau}(\rho)$ is obtained from $\rho$ by shifting each occupation number one place in $x$ to the right. Eq.~\eqref{34} is easily established by a change of variables in eq.~\eqref{30}
\be\label{35}
\psi_{\gamma}(t+\tilde{\varepsilon}, x+\varepsilon)=\varphi_{\gamma}(t+\tilde{\varepsilon},x)\,.
\ee
The factor $\overline{g}_{\tau}\bigl(\psi_{\gamma}(t+\tilde{\varepsilon},x)\bigr)$ in eq.~\eqref{30} becomes
\be\label{36}
\overline{g}_{\tau}\bigl(\psi_{\gamma}(t+\tilde{\varepsilon},x)\bigr)=\overline{g}_{\tau}\bigl(\varphi_{\gamma}(t+\tilde{\varepsilon},x-\varepsilon)\bigr)=\overline{g}_{\overline{\rho}(\tau)}\bigl(\varphi_{\gamma}(t+\tilde{\varepsilon},x)\bigr)\;.
\ee
Here $\overline{\rho}(\tau)$ is the inverse of $\overline{\tau}(\rho)$ and shifts all occupation numbers one place to the left. For the Grassmann integral one has $\int\mathcal{D}\psi(t+\tilde{\varepsilon})=\int\mathcal{D}\varphi(t+\tilde{\varepsilon})$ and in terms of the variable $\varphi$ eqs.~\eqref{32},~\eqref{34} become
\be\label{37}
\widehat{S}_{\tau\rho}=\delta_{\overline{\rho}(\tau),\rho}=\delta_{\tau,\overline{\tau}(\rho)}\,.
\ee
The same arguments holds for $g_{\tau}'\bigl(\psi(t+\varepsilon)\bigr)$ in eq.~\eqref{31} , such that eq.~\eqref{34} holds for both $t$ odd or even.

%Absatz
We can define a similar "left-transport operator" by a local factor
\be\label{38}
\tilde{\mathcal{K}}_{L}(t)=\exp\Bigl{\lbrace}\sum_{x}\psi_{\gamma}(t+\tilde{\varepsilon}, x-\varepsilon)\psi_\gamma(t, x)\Bigr{\rbrace}\,.
\ee
It leads to a step evolution operator similar to eq.~\eqref{34}, where $\overline{\tau}(\rho)$ moves now the position of all occupation numbers one place in $x$ to left. Both the left-transport and the right transport operator are unique jump operators and correspond to simple cellular automata. From our construction it is clear how many other cellular automata can be obtained by replacing in eq.~\eqref{33} the variable $\psi_{\gamma}(t+\tilde{\varepsilon}, x+\varepsilon)$ by $\psi_{\beta}(t+\tilde{\varepsilon}, x')$. We will investigate below more general cellular automata with a more complex form of $\tilde{\mathcal{K}}$.

%Absatz
For the particular type of cellular automata corresponding to eq.~\eqref{33} we can decompose the local factor into space-local simple pieces, as seen easily by writing
\begin{align}\label{39}
\tilde{\mathcal{K}}_{R}(t)&=\prod_{x}\prod_{\gamma}\tilde{\mathcal{K}}_{\gamma}(t, x)\,,\nn\\ \tilde{\mathcal{K}}_{\gamma}(t, x)&=\exp\bigl{\lbrace}\psi_{\gamma}(t+\tilde{\varepsilon}, x+\varepsilon)\psi_{\gamma}(t,x)\bigr{\rbrace}\;.
\end{align}
The factors $\tilde{\mathcal{K}}_{\gamma}(t, x)$ for different $\gamma$ have no common Grassmann variable, such that the integrals in eq.~\eqref{30} can be done blockwise. Each block involves only two different Grassmann variables and we can expand (no sum over $\gamma$) 
\be\label{40}
\tilde{\mathcal{K}}_{\gamma}(t, x)=1+\psi_{\gamma}(t+\tilde{\varepsilon}, x+\varepsilon)\psi_{\gamma}(t,x)=1+\varphi\psi \;,
\ee
with $\varphi=\psi_{\gamma}(t+\tilde{\varepsilon}, x+\varepsilon)$ , $\psi=\psi_{\gamma}(t,x)$ . The elements of the Grassmann algebra for $\psi$ (and similar for $\varphi$) are given by eq.~\eqref{17}, and we can establish eq.~\eqref{34} directly by factorizing the Grassmann basis functions appropriately.

%Absatz
A massless Dirac spinor in two dimensions consists of two Weyl spinors, one left moving and the other right moving. We consider here two different Dirac spinors, represented by four different Grassmann variables $\psi_{\gamma}(t,x)$, $\gamma=(\eta, a)$, $\eta=L,R=1, 2$, $a=1,2$. The action for the two free massless Dirac spinors is given by 
\be\label{41}
\begin{split}
\mathcal{L}_\textup{free}(t)=-&\sum_{x}\bigl{\lbrace}\psi_{R, a}(t+\tilde{\varepsilon}, x+\varepsilon)\psi_{R, a}(t,x)\\
&+\psi_{L,a}(t+\tilde{\varepsilon}, x-\varepsilon)\psi_{L, a}(t,x)\bigr{\rbrace}\;.
\end{split}
\ee
This is a simple cellular automaton of the type discussed above.  The two species $a=1,2=r,g$ may be associated with colors, say red for $a=1$ and green for $a=2$.
The complex structure related to Dirac spinors will be discussed in sect.~\ref{sec: 06}.

\section{Interacting fermionic quantum field theories\label{sec: 03}}

Cellular automata for free fermionic quantum field theories in 1+1-dimensions are rather simple \cite{CWQFT,Wetterich:2011dt}. The new feature in refs.~\cite{CWPCA,CWFCB} and the present work is the construction of cellular automata for fermionic models with interactions. We propose here a general strategy of alternating step evolution operators for the propagation and the interaction. This guarantees a unitary evolution by the simple property that each one of the steps is a unique jump operation. The procedure ressembles somewhat the construction of the Feynman path integral by an alternating sequence of momentum and position eigenstates.

%Absatz
Arbitrary fermionic quantum field theories do not lead to a unique jump matrix for the step evolution operator. They can therefore not be associated with an automaton. The task of the present section is therefore the establishment of a family of fermionic quantum field theories which realize a unique jump step evolution operator. This requires that the local factor $\tilde{\mathcal{K}}(t)$ connects a unique Grassmann basis element at $t+\varepsilon$ to each Grassmann basis element at $t$. In other words, each basis element $g'_{\rho}(t)$ should be multiplied by a single element $g_{\tau}(t+\tilde{\varepsilon})$ and not by a sum of such elements. This places restrictions $\mathcal{L}(t)$ that we will discuss in detail in the present section.

%Absatz
We also will introduce later a complex structure. The associated complex picture has the standard properties of quantum field theories in Minkowski space. In particular, we construct a model for which the naive continuum limit is invariant under Lorentz transformations.

\subsection{Fermion interaction and conditional jumps\label{subsec: Fermion interaction and conditional jumps}}

%Absatz
We start with the interaction part of the step evolution operator. For this purpose we choose space-local interactions where at every position $x$ the jump is independent of the configurations of occupation numbers at all other positions $y\neq x$. In this case the local factor $\tilde{\mathcal{K}}(t)$ factorizes into a product of independent factors
\be\label{42}
\tilde{\mathcal{K}}(t)=\sum_{x}\tilde{\mathcal{K}}_i(t,x)\,,
\ee
where $\tilde{\mathcal{K}}_i(t,x)$ involves only the two sets of Grassmann variables $\psi_{\gamma}(t+\tilde{\varepsilon},x)$ and $\psi_{\gamma}(t,x)$ at the given position $x$. Accordingly, the step evolution operator is a direct product 
\be \label{43}
 \widehat{S}_\textup{int}=\widehat{S}(x=1)\otimes\widehat{S}(x=2)\dots\otimes\widehat{S}(x=M_{x})\;.
\ee
Each factor $\widehat{S}(x)$ acts only on the configurations of occupation numbers at $x$. For the two Dirac spinors with $\gamma=1\dots 4$ each factor $\widehat{S}(x)$ is a $16\times 16$ matrix. The matrix $\widehat{S}_\textup{int} $ is therefore a $(16\cdot 2^{M_{x}})\times (16\cdot 2^{M_{x}})$ matrix, with $M_{x}$ the number of $x$- points.

%Absatz
We can discuss each factor $ \tilde{\mathcal{K}} _i(t,x)$ or $\widehat{S}(t,x)$ separately. We label the four internal states $\gamma=1\dots 4$ by $(R1, R2, L1, L2)$ and the 16 states $\tau$ by ordered sequences of occupation numbers. For the example $\tau=(1,0,0,1)$ a particle $R1$ and a particle $L2$ is present, while no particle $R2$ or $L1$ is present. The indices $R$ and $L$ will later be associated to right-movers and left-movers in the propagation step. The colors $1, 2$ may be taken as red and green.

%Absatz
We realize interactions by conditional jumps, as for our first example: Under the condition that precisely two particles are present, namely one left mover and one right mover with different colors, the colors are exchanged. This amounts to a switch of occupation numbers
\be\label{44}
(1,0,0,1)\leftrightarrow(0,1,1,0)\ .
\ee
All other states remain invariant. This process describes the two-particle scatterings 
\begin{align}\label{45}
R1+L2\rightarrow R2+L1 \,, \nn \\
R2+L1\rightarrow R1+L2\;.
\end{align}
If a third or fourth particle is present, no scattering occurs.
We will later add the scattering process for which two green particles transform into two red particles and vice versa. For the moment we discuss only the process~\eqref{45}.
The step evolution operator is a unit matrix except for the sectors of the states $\tau$ with occupation numbers $(1,0,0,1)$ and $(0,1,1,0)$. In this sector the diagonal elements vanish, and one has
\be\label{46}
\widehat{S}_{(1001),(0110)}=\widehat{S}_{(0110),(1001)}=1\;.
\ee
Repeating the switch yields the identity 
\be\label{47}
\widehat{S}^{\,2}(x)=1 \quad,\quad  \quad\widehat{S}_\textup{int}^{\,2}=1\;.
\ee

%Absatz
For the computation of the corresponding local factor $\tilde{\mathcal{K}}_i(t,x)$ we can fix the sign convention for the Grassmann basis functions $g_{\tau}$ by convenience. All the relations discussed above hold independently of the choice of the signs $\tilde{s}_{\tau}$ in eq.~\eqref{06}. We could even use different sign conventions for different $t$. This freedom of the choice of local sign conventions corresponds to a discrete local gauge symmetry of the weight function~\citep{CWIT, CWQF}. We want to keep the relations \eqref{08}~\eqref{14}, and therefore restrict the possibilities to a free global choice of signs which is the same for all $t$.
For the example of the vacuum state for $M=2$ we choose
\begin{align}\label{48}
&g_{(0000)}=g_{(0000)}'=\psi_{1}\psi_{2}\psi_{3}\psi_{4}\,, \nn\\
&\overline{g}_{(0000)}=\overline{g}_{(0000)}'=1\;.
\end{align}
For the Grassmann integral we employ the ordering
\be\label{49}
\int\mathcal{D}\psi=\int d\psi_{4}\int d\psi_{3}\int d\psi_{2}\int d\psi_{1}\;,
\ee
such that $\int\mathcal{D}\psi g_{(0000)}=1$. For the two-particle states relevant for our purpose we chose the basis functions 
\begin{align}\label{50}
&g_{(1001)}=-g_{(1001)}'=\psi_{2}\psi_{3}\,,\nn\\
&g_{(0110)}=-g_{(0110)}'=\psi_{1}\psi_{4}\;.
\end{align}
According to eq.~\eqref{10} the contribution of the two-particle sector to $\tilde{\mathcal{K}}_\textup{int}(x,t)$ reads
\begin{align}\label{51}
\Delta \tilde{\mathcal{K}}_\textup{int}(t,x&)=-\psi_{1}(t+\tilde{\varepsilon},x)\psi_{4}(t+\tilde{\varepsilon},x)\psi_{2}(t,x)\psi_{3}(t,x) \nn\\
&-\psi_{2}(t+\tilde{\varepsilon},x)\psi_{3}(t+\tilde{\varepsilon},x)\psi_{1}(t,x)\psi_{4}(t,x)\;.
\end{align}
Our general conventions for Grassmann basis functions are displayed in appendix~\ref{ap.A}.

%Absatz
We have to combine the contribution~\eqref{51} with the contribution of the unit operator for all other states. For this purpose we first subtract from $\Delta \tilde{\mathcal{K}}_\textup{int}$ the contribution of the unit operator in this particular two-particle sector by defining
\begin{align}\label{52}
\tilde{D}_{\phantom{0}}(t,x)=&\:\tilde{D}_{1}(t,x)-\tilde{D}_{2}(t,x)\,, \nn \\
\tilde{D}_{1}(t,x)=&\:\psi_{1}'\psi_{4}'\psi_{2}\psi_{3}+\psi_{2}'\psi_{3}'\psi_{1}\psi_{4}\,, \nn \\
\tilde{D}_{2}(t,x)=&\:\psi_{1}'\psi_{4}'\psi_{1}\psi_{4}+\psi_{2}'\psi_{3}'\psi_{2}\psi_{3}\,, \nn \\
\tilde{D}{\phantom{0}}(t,x)=&-(\psi_{1}'\psi_{4}'-\psi_{2}'\psi_{3}')(\psi_{1}\psi_{4}-\psi_{2}\psi_{3})\;,
\end{align}
where $\psi_{\gamma}'=\psi_{\gamma}'(t+\tilde{\varepsilon},x), \quad \psi_{\gamma}=\psi_{\gamma}(t,x)$. In terms of $\tilde{D}$ we can write the local factor as 
\be\label{55}
\tilde{\mathcal{K}}_{i}(t,x)=\exp\lbrace\psi_{\gamma}'\psi_{\gamma}\rbrace-\tilde{D}\;.
\ee
The first term produces the unit matrix, while the second term subtracts the unit matrix in the sector of the states $(1,0,0,1)$ and $(0,1,1,0)$ and replaces it by the exchange of colors. 

%Absatz
Next we write the local factor in exponential form in order to have it as a piece of the action. For this purpose we observe the identities 
\begin{align}\label{53}
&\tilde{D}_{1}\tilde{D}_{2}=0, \quad\tilde{D}^{2}=\tilde{D}_{1}^{2}+\tilde{D}_{2}^{2}=2\tilde{D}_{1}^{2} \,, \nn \\ &\tilde{D}^{3}=0, \quad
\tilde{D}^{2}=4\psi_{1}'\psi_{2}'\psi_{3}'\psi_{4}'\psi_{1}\psi_{2}\psi_{3}\psi_{4} \;.
\end{align}
In terms of $\tilde{D}$ we can write the local factor in exponential form
\be\label{54}
\tilde{\mathcal{K}}_i(t,x)=\exp\bigl{\lbrace}\psi_{\gamma}'\psi_{\gamma}-\tilde{D}-\dfrac{1}{2}\tilde{D}^{2}+\psi_{\gamma}'\psi_{\gamma}\tilde{D}-\dfrac{1}{2}(\psi_{\gamma}'\psi_{\gamma})^{2}\tilde{D}\bigr{\rbrace}\;.
\ee
Indeed, the expansion of the exponential yields eq.~\eqref{55}.
With
\be\label{56}
\begin{split}
&\phantom{-}\psi_{\gamma}'\psi_{\gamma}\tilde{D}_{1}=0\,, \\
&-\psi_{\gamma}'\psi_{\gamma}\tilde{D}=\psi_{1}'\psi_{2}'\psi_{3}' \psi_{1}\psi_{2}\psi_{3}+ \psi_{1}'\psi_{2}'\psi_{4}' \psi_{1}\psi_{2}\psi_{4} \\&\quad\quad \quad\quad \quad  + \psi_{1}'\psi_{3}'\psi_{4}' \psi_{1}\psi_{3}\psi_{4}+ \psi_{2}'\psi_{3}'\psi_{4}'\psi_{2}\psi_{3}\psi_{4}\;,
\end{split}
\end{equation}
and
\be\label{57}
(\psi_{\gamma}'\psi_{\gamma})^{2}\tilde{D}=\tilde{D}^{2}
\ee
one obtains $\tilde{\mathcal{K}}_{i}(t,x)=\exp\bigl{\lbrace} -\mathcal{L}_{i}(t,x)\bigr{\rbrace}$ with
\be\label{58}
\mathcal{L}_{i}(t,x)=(-\psi_{\gamma}'\psi_{\gamma}+\tilde{D})(1+\tilde{D})\;.
\ee
This yields for the interaction part of the action
\bel{59A}
\mathcal{L}\subt{int}(t)=\sum_x\mathcal{L}_i(t,x)\ .
\ee

\subsection{Interacting fermionic quantum field theory\label{subsec: Interacting fermionic quantum field theory}}

A quantum field theory for interacting fermions combines the interaction with the propagation of fermions. This can be done by the use of a sequence of alternating local factors. We use the free propagation of Dirac fermions for $t$ even, and the interaction for $t$ odd. A pair of neighboring local factors reads for even $t$
\be\label{59}
\begin{split}
\!\!\!\tilde{\mathcal{K}}(t+\tilde{\varepsilon})\tilde{\mathcal{K}}(t)&=\exp\bigl{\lbrace} -\mathcal{L}_\textup{int}(t+\tilde{\varepsilon})\bigr{\rbrace} \exp\bigl{\lbrace} -\mathcal{L}_\textup{free}(t)\bigr{\rbrace} \\
&=\exp\Bigl{\lbrace} -\sum_{x}\bigl[\mathcal{L}_{i}(t+\tilde{\varepsilon}, x)+\mathcal{L}\ff(t, x)\bigr]\Bigr{\rbrace}\;,
\end{split}
\ee
with $\mathcal{L}_{i}(t+\tilde{\varepsilon}, x)$ given by eq.~\eqref{58} shifted to $t+\tilde{\varepsilon}$, and $\mathcal{L}\ff(t,x)$  extracted from eq.~\eqref{41},
\be\label{60}
\begin{split}
\mathcal{L}\ff(t,x)=-\psi_{R a}(t+\tilde{\varepsilon}, x+\varepsilon)\psi_{R a}(t,x)\\
-\psi_{L a}(t+\tilde{\varepsilon}, x-\varepsilon)\psi_{L a}(t,x)\;.
\end{split}
\ee

%Absatz
We could integrate over the variables $\psi(t+\tilde{\varepsilon})$ and obtain with eq.~\eqref{26}
\be\label{61}
\int\mathcal{D}\psi(t+\tilde{\varepsilon})\tilde{\mathcal{K}}(t+\tilde{\varepsilon})\tilde{\mathcal{K}}(t)=g_{\tau}(t+2\tilde{\varepsilon})(\widehat{S}_\textup{int}\widehat{S}_\textup{free})_{\tau\rho}\overline{g}_{\rho}(t)\;.
\ee
Since both $\widehat{S}_\textup{int}$ and $\widehat{S}_\textup{free}$ are unique jump operators, this also holds for the product. The product matrix
\be\label{62}
\widehat{S}=\widehat{S}_\textup{int}\widehat{S}_\textup{free}
\ee
describes the step evolution operator of a cellular automaton discussed in ref.~\citep{CWPCA, Wetterich:2020kqi}.
Repeating the alternating chain with integration over intermediate Grassmann variables produces matrix chains of $\widehat{S}$ ,
\be\label{63}
\begin{split}
\int\mathcal{D}\psi(t+3\tilde{\varepsilon})&\mathcal{D}\psi(t+2\tilde{\varepsilon})\mathcal{D}\psi(t+\tilde{\varepsilon}) \\
&\times\tilde{\mathcal{K}}(t+3\tilde{\varepsilon})\tilde{\mathcal{K}}(t+2\tilde{\varepsilon})\tilde{\mathcal{K}}(t+\tilde{\varepsilon})
\tilde{\mathcal{K}}(t)\\
=g_{\tau}(t&+4\tilde{\varepsilon})(\widehat{S}^{2})_{\tau\rho}\overline{g}_{\rho}(t)\;.
\end{split}
\ee
The action~\eqref{02}, with $\mathcal{L}(t)$ given by $\overline{\mathcal{L}}_{i}$ or $\overline{\mathcal{L}}\ff$
   for odd and even $t$, respectively, produces the same chain of step evolution operators as a cellular automaton with the corresponding rules of exchange of colors and propagation.
With an implementation of boundary conditions and observables in the fermionic representation to be discussed below, the fermionic model~\eqref{59} is exactly equivalent to a probabilistic cellular automaton.

%Absatz
We choose $\tilde{\varepsilon}=\varepsilon/2$ and rename
\bel{68A}
\psi_{\gamma}(t+\tilde{\varepsilon},x) =\overline{\psi}_{\gamma}(t+\varepsilon , x)\ .
\ee
With these definitions we can employ a coarse grained lattice with $t$ corresponding to even $t$ on the original lattice.
The lattice distance on the coarse grained lattice is the same $\varepsilon
$ in both directions. (We will in the following use integers $m$ for the sites of the coarse grained lattice, corresponding to even $m$ an the original lattice.) We can summarize the action of the discrete fermionic quantum field theory as
\begin{align}\label{69}
\mathcal{L}(t-\varepsilon)=&-\sum_{x}\Big{\lbrace}\overline{\psi}_{Ra}(t,x)\psi_{Ra}(t-\varepsilon,x-\varepsilon)\nn\\
&+\overline{\psi}_{La}(t,x)\psi_{La}(t-\varepsilon,x+\varepsilon)\nn\\
&+\big{[}\psi_{Ra}(t,x)\overline{\psi}_{Ra}(t,x)+\psi_{La}(t,x)\overline{\psi}_{La}(t,x)\nn\\
&-\overline{D}(t,x)\big{]}\big{(}1+\overline{D}(t,x) \big{)}\Big{\rbrace}\;,
\end{align}
with $\overline{D}(t, x)$ given by $\tilde{D}(t,x)$ in eq.~\eqref{52} with the identifications
\begin{align}\label{70}
\psi'_{1}&=\psi_{R1}(t,x)\,,\quad \psi_{1}=\overline{\psi}_{R1}(t,x)\,,\nn\\
\psi'_{2}&=\psi_{R2}(t,x)\,,\quad \psi_{2}=\overline{\psi}_{R2}(t,x)\,,\nn\\
\psi'_{3}&=\psi_{L1}(t,x)\,,\quad \psi_{3}=\overline{\psi}_{L1}(t,x)\,,\nn\\
\psi'_{4}&=\psi_{L2}(t,x)\,,\quad \psi_{4}=\overline{\psi}_{L2}(t,x)\;.
\end{align}
The first two terms are $\mathcal{L}_{\textup{free}}$ from eq.~\eqref{41}, and the remaining part amounts to $\mathcal{L}\subt{int}$ in eq.~\eqref{59A}.
The action~\eqref{69} contains the same information as the action~\eqref{59} since we only have renamed variables. We show in Appendix~\ref{ap.B} that the variables $\overline{\psi}_{\gamma}(t,x)$ have a close connection to the conjugate Grassmann variables in ref.~\cite{CWFCS}. Since the propagation does not mix even and odd sublattices (cf. appendix~\ref{ap.B}) we omit the odd sublattice in the following.

%Absatz
Lattice derivatives are defined by
\begin{align}\label{71}
(\partial_{t}+\partial_{x})\psi(t,x)=\dfrac{1}{\varepsilon}\bigl[\psi(t,x)-\psi(t-\varepsilon,x-\varepsilon)\bigr]\,, \nn\\
(\partial_{t}-\partial_{x})\psi(t,x)=\dfrac{1}{\varepsilon}\bigl[\psi(t,x)-\psi(t-\varepsilon,x+\varepsilon)\bigr]\;.
\end{align}
In terms of these derivatives the discrete action reads 
\begin{align}\label{73}
\mathcal{L}(t-\varepsilon)&=\sum_{x}\varepsilon \overline{\psi}_{Ra}(t,x)(\partial_{t}+\partial_{x}) \psi_{R a}(t,x)\\
+&\varepsilon \overline{\psi}_{L a} (t,x)(\partial_{t}-\partial_{x}) \psi_{L a}(t,x)
+\overline{D}(t,x)+\Delta\mathcal{L}(t-\varepsilon)\nn
\end{align}
with
\begin{align}\label{73a}
\Delta\mathcal{L}(t-\varepsilon)=\sum_{x}&\overline{D}(t,x)\Big{[}\overline{D}(t,x)-\psi_{R a}(t,x)\overline{\psi}_{R a} (t,x)\nn\\
&-\psi_{L a}(t,x)\overline{\psi}_{L a} (t,x)\Big{]}\ .
\end{align}
The sum is over the space points on the even sublattice. Eq.~\eqref{73} is the fermionic representation of a probabilistic cellular automaton. For this discrete formulation no approximations have been made.

\subsection{Continuum formulation\label{subsec: continuum formulation}}

Our fermionic model can be viewed as a particular discretization of a continuum theory. This discretization regularizes the Grassmann functional integral since only a finite number of Grassmann variables appears. In the continuum limit the number of Grassmann variables goes to infinity. This is realized by taking $\varepsilon\rightarrow 0$ at fixed distances in $t$ and $x$. For a given distance in time or space the number of intermediate lattice points goes to infinity. In the naive continuum limit sums are replaced by integrals, 
\be\label{75}
\int dt\int dx=\int_{t,x}=2\varepsilon^{2}\sum_{t,x}
\ee
Here the factor $2\varepsilon^{2}$ accounts for the fact that $\sum_{t,x}$ only sums over the points of the even sublattice. 

%Absatz
For a continuum version of the classical action that is regularized by our discretization we simply omit higher orders in $\eps$.
Lattice derivatives are replaced by partial derivatives, acting on a continuum of Grassmann variables $\psi_{\gamma}(t,x)$,  $\overline{\psi}_{\gamma}(t,x)$. We also choose a different normalization for the Grassmann variables 
\be\label{76}
\psi(t,x)=\sqrt{2\varepsilon}\psi_{N}(t,x)\ .
\ee
In this way we absorb the factor $(2\varepsilon)^{-1}$ arising from $\sum_{t,x}$ and eq.~\eqref{73}. Expressed in terms of $\psi_{N}$ the interaction factor $\tilde{D}(x)$ is proportional $4\varepsilon^{2}\psi_{N}^{4}$. The continuum limit $\eps\to0$ is taken at fixed $\psi_{N}$.

%Absatz
The continuum version simplifies the action considerably. We can omit in eq.~\eqref{73} the piece $\Delta\mathcal{L}(t-\varepsilon)$. Not writing the index $N$ for the renormalized Grassmann variables explicitly the continuum action takes the simple form of a local fermionic quantum field theory,
\be\label{77}
\begin{split}
S=\int_{t,x}&\bigl{\lbrace}\overline{\psi}_{R a}(t,x)(\partial_{t}+\partial_{x})\psi_{R a}(t,x) \\
&+\overline{\psi}_{L a}(t,x)(\partial_{t}-\partial_{x})\psi_{L a}(t,x)
+2\overline{D}(t,x)\bigr{\rbrace}\;.
\end{split}
\ee
For the local interaction term, 
\be\label{78}
\begin{split}
\overline{D}&=-(\overline{\psi}_{R1}\overline{\psi}_{L2}-\overline{\psi}_{R2}\overline{\psi}_{L1})(\psi_{R1}\psi_{L2}-\psi_{R2}\psi_{L1}) \\
&=\overline{\psi}_{R1}\psi_{R1}\overline{\psi}_{L2}\psi_{L2}+\overline{\psi}_{R2}\psi_{R2}\overline{\psi}_{L1}\psi_{L1}\\
&\quad +\overline{\psi}_{R1}\psi_{L1}\overline{\psi}_{L2}\psi_{R2}+\overline{\psi}_{R2}\psi_{L2}\overline{\psi}_{L1}\psi_{R1}\;,
\end{split}
\ee
all variables correspond to $\psi_{N}$ and are taken at $(t,x)$.

%Absatz
The continuum version~\eqref{77} can also be considered as a naive continuum limit of the discrete fermionic model. The time continuum limit of a discretized model is more complex, however. It can be encoded in the quantum effective action which obtains by integrating the fluctuations in the functional integral. This process leads to running couplings and possible modifications of the naive continuum limit.

\subsection{Extended interaction\label{subsec: Extended interaction}}

The action~\eqref{77} defines a type of fermionic quantum field theory. We next extend the interaction  by inclusion of an additional scattering process.
Our construction is not limited to the particular "scattering"~\eqref{45}. We may add to the exchange~\eqref{44} a further exchange
\be\label{79}
(1,0,1,0)\leftrightarrow(0,1,0,1)\;.
\ee
This corresponds to the transition
\begin{align}\label{80}
&R1+L1\rightarrow R2+L2\,, \nn\\
&R2+L2\rightarrow R1+L1\;,
\end{align}
and to a modification of the step evolution in the corresponding sector
\be\label{81}
\widehat{S}_{(1010),(0101)}=\widehat{S}_{(0101),(1010)}=1\;.
\ee
The process of two incoming green particles scattered to two outgoing red particles, and similarly with the colors exchanged, is related to the process~\eqref{45} by a type of crossing symmetry, as characteristic for many relativistic quantum field theories. We will see that the combination of the interactions~\eqref{45} and~\eqref{80} leads in the continuum limit to a type of Thirring model.

%Absatz
The construction of the local factor and $\mathcal{L}_{i}$ proceed in complete analogy to the scattering~\eqref{45}. The relevant basis functions are $(\sigma=\pm 1)$
\begin{align}\label{82}
g_{(1010)}=-g'_{(1010)}=\psi_{2}\psi_{4}\;,\nn\\
g_{(0101)}=-g'_{(0101)}=\sigma\psi_{1}\psi_{3}\;.
\end{align}
The conventions in the appendix~\ref{ap.A} correspond to $\sigma=1$. We have added the free sign in order to investigate how different conventions influence the form of the interaction. The scattering process~\eqref{80} adds to $\tilde{D}$ in eq.~\eqref{52} a term
\be\label{83}
\tilde{C}(t,x)=-(\psi_{1}'\psi_{3}'-\sigma\psi_{2}'\psi_{4}')(\psi_{1}\psi_{3}-\sigma\psi_{2}\psi_{4})\;.
\ee
The expression~\eqref{54} remains valid with $\tilde{D}$ replaced by $\tilde{C}+\tilde{D}$. Also eqs.~\eqref{57}~and~\eqref{58} remain valid with $\tilde{D}\rightarrow\tilde{C}+\tilde{D}$. In the continuum version \eqref{77} one replaces again $\overline{D}$ by $\overline{C}+\overline{D}$, with 
\be\label{83a}
\overline{C}(t,x)=-(\overline{\psi}_{R1}\overline{\psi}_{L1}-\sigma\overline{\psi}_{R2}\overline{\psi}_{L2})(\psi_{R1}\psi_{L1}-\sigma\psi_{R2}\psi_{L2})\;,
\ee
and all Grassmann variables corresponding to $\psi_{N}(t,x)$.
Now the combination $2(\overline{D}+\overline{C})$ specifies the interaction term in the fermionic action.

\subsection{Lorentz symmetry\label{subsec: Lorentz symmetry}}

In the continuum version the action becomes invariant under Lorentz transformations. We introduce for each color two-component vectors of Grassmann variables
\be\label{LA}
\psi_{a}=\begin{pmatrix}
\psi_{Ra}\\ \psi_{La}
\end{pmatrix}\;, \quad \overline{\psi}_{a}=(\overline{\psi}_{La},  \,-\overline{\psi}_{Ra})\;.
\ee
The action takes the familiar form
\be\label{LB}
S=\int_{t,x}\Big{\lbrace} -\overline{\psi}_{a}\gamma^{\mu}\partial_{\mu}\psi_{a}+\mathcal{L}_i\Big{\rbrace}\;,
\ee
with
\be\label{LC}
\mathcal{L}_i=2(\overline{C}+\overline{D})\;.
\ee
Here the Dirac matrices are given by the Pauli matrices
\be\label{90}
\gamma^{0}=-i\tau_{2}\,,\quad \gamma_{1}=\tau_{1}\,,\quad \lbrace\gamma^{\mu},\gamma^{\nu}\rbrace=2\eta^{\mu\nu}\;,
\ee
with Lorentz signature $\eta_{00}=-1$, $\eta_{11}=1$,  $\partial_{0}=\partial_{t}$, $\partial_{1}=\partial_{x}$,
\be\label{91}
\eta_{\mu\nu}=\diag (-1,1)\,,\quad \gamma_{\mu}=\eta_{\mu\nu}\gamma^{\nu}\;.
\ee

%Absatz
The Lorentz transformations act on the coordinates in the usual way. At this point we employ the continuum limit since the lattice coordinates admit only a discrete subgroup. Beyond the transformation of coordinates the fermion doublets transform as spinors
\be\label{92}
\delta\psi=-\eta{\Sigma}^{01}\psi\,,\quad \delta\overline{\psi}=\eta\overline{\psi}{\Sigma}^{01}\;,
\ee
with infinitesimal transformation parameter $\eta$ and generator
\be\label{93}
{\Sigma}^{01}=\dfrac{1}{4}[\gamma^{0},\gamma^{1}]=-\dfrac{1}{2}\tau_{3}.
\ee

%Absatz
We show in appendix~\ref{ap.C} that the interaction part reads  for $\sigma=-1$ 
\be\label{LJ}
\mathcal{L}_{\textup{int}}\!=\!\!\int_x\!\!\Big\{-\tfrac{1}{2}\overline{\psi}_{a}\gamma^{\mu}\psi_{a}\overline{\psi}_{b}\gamma_{\mu}\psi_{b}+\overline{\psi}_{a}\gamma^{\mu}\psi_{b}\varepsilon^{ab}\overline{\psi}_{c}\gamma_{\mu}\psi_{d}\varepsilon^{cd}\Big\}\,\!,\!\!\!
\ee
with antisymmetric tensor $\varepsilon_{12}=-\varepsilon_{21}=1$. The action~\eqref{LJ} is invariant under a global $SO(2)$ - symmetry of continuous color rotations.
An equivalent abelian color symmetry for $\sigma=1$ transforms
\bel{99B}
\pvec{\psi_1}{\psi_2}\to e^{i\beta\tau_1}\pvec{\psi_1}{\psi_2}\ ,\quad \pvec{\psibar_1}{\psibar_2}\to e^{-i\beta\tau_1}\pvec{\psibar_1}{\psibar_2}\ .
\ee
We conclude that the conventions for Grassmann basis elements affect the concrete expression for the action, but do not alter the physical content of the model.

%Absatz
The interaction~\eqref{LJ} defines a type of Thirring model~\cite{THI,KLA,AAR,FAIV} with two colors. By a different ordering of the Grassmann variables we can equivalently write it as a colored Grass-Neveu model~\cite{GN,WWE,RWP,SZKSR}.
For this purpose one expresses $2(\overline{D}+\overline{C})$ in terms of the Lorentz-scalars
\begin{align}
\label{105A}
&\varphi_{a b}=\overline{\psi}_{a}\psi_{b}=(\overline{\psi}_{La}\psi_{Rb}-\overline{\psi}_{Ra}\psi_{Lb})\ , \nn\\
&\tilde{\varphi}_{a b}=\overline{\psi}_{a}\overline{\gamma}\psi_{b}=\overline{\psi}_{La}\psi_{Rb}+\overline{\psi}_{Ra}\psi_{Lb})\ .
\end{align}
An explicit expression can be found in appendix~\ref{ap.C}. There we also discuss the decomposition of Dirac fermions into Weyl and Majorana fermions. 

%Absatz
It is instructive to consider the ``continuum constraint" $\overline{\psi}_{\gamma}=\psi_{\gamma}$, for which the interaction term simplifies
\bel{105B}
\overline{D}=-\sigma \overline{C}=2\psi_{R1}\psi_{L1}\psi_{L2}\psi_{R2}\ .
\ee
Combination into complex Grassmann variables
\bel{105C}
\zeta_{R}=\psi_{R1}+i\psi_{R2}\quad , \quad\quad \zeta_{L}=\psi_{L1}+i\psi_{L2}
\ee
yields for $\sigma=-1$
\bel{105D}
\mathcal{L}_{i}=2(\overline{C}+\overline{D})=-(\zeta_{L}^{*}\zeta_{R}-\zeta_{R}^{*}\zeta_{L})^{2}=-(\overline{\zeta}\zeta)^{2}\ ,
\ee
where the two-component spinors, $\zeta$, $\overline{\zeta}$ are formed in analogy to eq.~\eqref{LA}. This is precisely the interaction of the Gross-Neveu model with a particular value of the coupling.

%Absatz
If we omit the extension of the scattering process~\eqref{80} by setting $\overline{C}=0$, the coupling strength in eq.~\eqref{105D} is reduced by a factor two, and similarly if we only keep the scattering process~\eqref{80} and omit the scattering~\eqref{45} by setting $\overline{D}=0$. These different automata can be considered as different lattice-regularizations of the continuum Gross-Neveu model. This raises the question of the true continuum limit. Do these discrete lattice regularizations of the Gross-Neveu model belong to the same universality class as a continuum regularization which preserves Lorentz symmetry? Is Lorentz symmetry restored in the continuum limit? Is the naive continuum limit a valid approximation to the effective action? Are the particular values of the coupling for which one obtains an automaton singled out for the continuum limit?

%Absatz
We finally observe that the fermionic action appears in the functional integral by the factor $e^{-S}$ and not as $e^{iS}$ as in the usual formulation of quantum field theory with a Minkowski signature. Nevertheless, the model has a unitary evolution. The analytic continuation of the action to euclidean signature differs from the usual setting by an additional overall factor $i$.

\section{Cellular automaton for the fermion model\label{sec: 04}}

We have established that the step evolution operator for a discretization of the particular Thirring type model \eqref{LJ} or equivalent Gross-Neveu model \eqref{105D} is a unique jump matrix. We consider the combination of the evolution steps at $t$ and $t+\tilde{\varepsilon}$ as a single combined evolution step from $t$ to $t+2\tilde{\varepsilon}=t+\varepsilon$ , according to eq.~\eqref{62}. The first operator $\widehat{S}_\textup{free}$ moves right-movers one place in $x$ to the right, and left-movers one place to the left. The second factor $\widehat{S}_\textup{int}$ exchanges at each location $x$ the colors of all particles if precisely one left mover and one right mover is present at $x$. Otherwise the colors are kept. This constitutes a simple rule for a cellular automaton.

%Absatz
With both the Thirring-type model and the associated cellular automaton having the same evolution rule according to identical step evolution operators it only remains to identify the probabilistic information in the wave function of the Thirring model with the one of a probabilistic cellular automaton. The present section will discuss in some more detail the properties of the "updating rule" encoded in the step evolution operator, while we turn to the probabilistic aspects in sect.~\ref{sec: 05}.

\subsection{Different lattice representations\label{subsec: Different lattice representations}}

The square lattice can be decomposed into two sublattices: the even sublattice with $m_{t}+m_{x}$ even, and the odd sublattice with $m_{t}+m_{x}$ odd. Since all particles move on diagonals the dynamics on the even and odd sublattice is completely disconnected and we have defined our model only on the even sublattice. On each lattice point one can have up to four particles, left- or right movers, red or green. Starting from the even sublattice only, we can now redistribute the particles between the sublattices, by putting on the even sublattice the right movers and on the odd lattice sites the left movers. In this picture the whole lattice is used, with each lattice site occupied by up to two particles. On the even sublattice one has green or red right movers, and on the odd sublattice we place the green or red left-movers.

\begin{figure}[h]
%\resizebox{18pc}{!}{%
%\vspace*{-7pt}
\centering\includegraphics[width=3in]{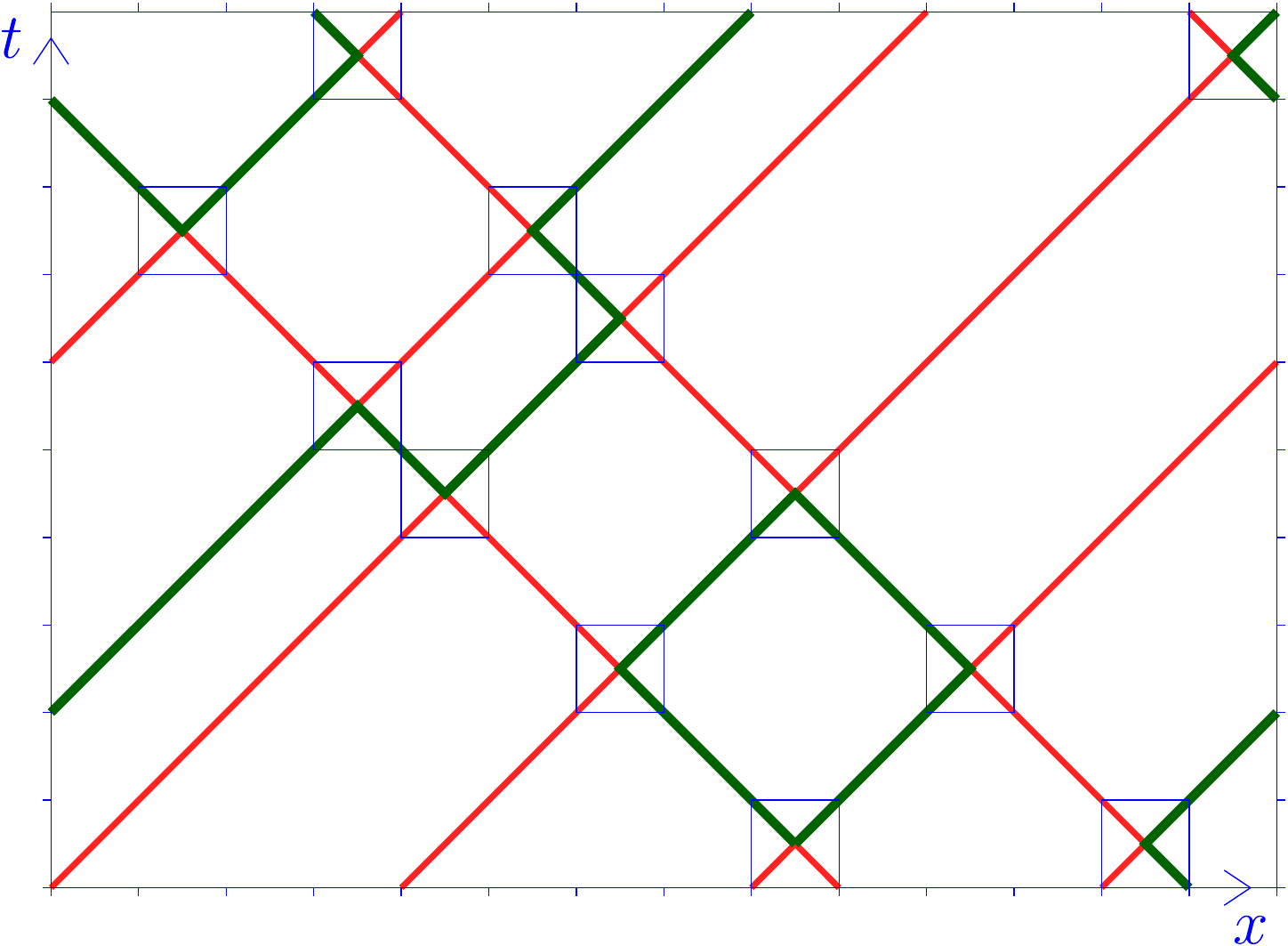}
%%% where xxxxxx name represents "figurename.eps"
\vspace*{-5pt}
%\caption{captionxxx.}\label{Fig.2}
%\end{figure}
%\begin{figure}[h]
%\centering
%\includegraphics[scale=0.10]{fig1}
\caption{Cellular automaton for interacting fermions. Single occupied red or green lines scatter at the squares. We have not indicated empty lines, or left- or right moving lines which are doubly occupied by one red and one green particle. These lines are straight without scattering.}\label{fig: 01}
\end{figure}

%Absatz
This redistribution does not change the dynamics, as illustrated in Fig.~\ref{fig: 01}. In this figure we display the lines of single occupied sites, occupied either by a red or green particle. There are additional lines for empty sites, or sites occupied by both a red and a green particle. These lines do not scatter and are not shown in the figure. The original lattice has points at the centers of the squares surrounding the crosses. Up to four particles can occupy a site, and only the sites of an even sublattice are occupied by the squares. 
\noindent
%\resizebox{18pc}{!}{\begin{figure}[!h]
After the redistribution the lattice sites are at the corners of the squares surrounding the scattering points. The distance between two points is now given by $\varepsilon$. Only up to two particles can occupy each site. On the initial horizontal line at $t=0$ the right movers occupy the sites with even $m_{x}$, and the left movers the sites with odd $m_{x}$. This is interchanged at $t+\varepsilon$. Now the right movers are on the sites with odd $m_{x}$, and the left movers on the sites with even $m_{x}$. 

\subsection{Properties and symmetries of cellular\\ \hspace*{10pt} automaton\label{subsec: Properties and symmetries of cellular automaton}}

From Fig.~\ref{fig: 01} one can easily see a few characteristic features of this cellular automaton.
\begin{enumerate}
\item The total number of right movers and the total number of left movers are conserved separately as time increases. (There is the same number on each hypersurface with given $t$.)
This implies, of course, conserved total particle number,
\bel{117A}
N_p=\sum_x\sum_\gamma n_\gamma(x)\ .
\ee
\label{i01: 01}
\item If we disregard the color, all particles move on straight lines, with velocity $c=1$. They move either to the left or to the right.\label{i01: 02}
\item Doubly occupied lines, with both a red and a green particle moving in the same direction, do not undergo scattering. They move as free "composite states" or "bound states". At most one right-moving and one left-moving composite state can be present at each site. Possible occupation numbers for these composite states are one or zero, as for fermions.
\label{i01: 03}
\item The single occupied lines change color whenever they encounter another single occupied line. The scattering concerns the internal degrees of freedom. The interaction changes the color. In  the four corners surrounding each square for a scattering event one has precisely a total number of four particles, two red and two green, two left movers and two right movers. 
A line with a given color never ends, but it can move backwards in time. Loops or closed lines with a given color are possible.\label{i01: 04}
\item The picture can be  rotated by $\pi/2$ without changing the dynamical rules. The dynamics has a type of "crossing symmetry". If a red and a green particle can scatter into a green and a red particle, there is also a scattering of two green particles into two red particles, and vice versa.\label{i01: 05}
\item The dynamics is invariant under a reflection in $t$ (time reversal symmetry) and in $x$ (parity).\label{i01: 06}
\item The number of red and green particles is not conserved separately. Two red particles can become two green particles. Since changes are always by two particles, and even (odd) number of red particles remains even (odd), and similar for the green particles.
\label{i01: 07}
\item The dynamics is invariant under an exchange of colors $E$. Exchanging the two colors in Fig.~\ref{fig: 01} produces again a diagram allowed by the dynamics. 
\label{i01: 08}
\item A single particle line for occupied red sites can also be seen as a line of single empty green sites or green holes. The symmetry $F$ exchanges a red particle and a green hole as well as a green particle and a red hole. This transformation changes double occupied lines into empty lines, and vice versa. Since these lines do not under go scattering, the dynamics is invariant under the symmetry $F$. Combining the symmetry $F$ with the color exchange symmetry $E$ one obtains particle-hole symmetry $K$. Under this symmetry each particle $\gamma$ is mapped to a hole $\gamma$ and vice versa, e.g. red right moving particles transform to red right moving holes etc. The dynamics is invariant under particle-hole symmetry. 
\label{i01: 09}
\end{enumerate}
These properties have their correspondence in the fermionic quantum field theory. In the appendix~\ref{ap.D} we list the symmetries of the Thirring-type model. This establishes direct relations for all the nine points. General properties as conserved quantum numbers for the automaton find a direct root in continuous or discrete symmetries in the fermionic language.

\subsection{Simple evolution of deterministic automaton}
\label{subsec: Simple evolution of deterministic automaton}

Many of the features discussed so far are rather easily extended to more complex automata. The particular automaton discussed here has a rather simple structure which makes it suitable for a discussion of general concepts, since the latter often find simple concrete realizations. For a deterministic cellular automaton with a sharp initial state $\tau_0$ at $t\inn$ we can compute the configuration $\tau_n$ for $t=t\inn+n\eps$ in a straightforward way.

%Absatz
If at $t\inn$ we have at position $x$ both a right-moving red and a right-moving green particle, we will find at $t$ at the position $x+n\eps$ both the red and green right-moving particles. This follows from the observation that doubly occupied lines do not scatter. The same holds if at $t\inn$ neither a right-moving red particle not a right-moving green particle is present at $x$. Since empty lines do not scatter, we infer for $t$ that at $x+n\eps$ no right-movers are present. The argument extends similarly to left-movers, with $x+n\eps$ replaced by $x-n\eps$.

%Absatz
What remains are positions $x$ at $t\inn$ for which only a single right-moving particle and/or a single left-moving particle is present. Single particles follow straight lines, only the color can change due to scattering. For every single right-mover at $(t\inn,x)$ we find a single right-mover at $(t\inn+n\eps, x+n\eps)$, and for every single left-mover at $(t\inn,x)$ one has a single left-mover at $(t\inn+n\eps, x-n\eps)$. The numbers of right-movers and left-movers at $(t,x)$ are easily determined in this way for any initial configuration $\tau_0$. What remains is the determination of the color of the single right-movers and single left-movers at $(t,x)$.

%Absatz
For this purpose we observe that a single right-moving particle line changes color whenever it crosses a single left-moving particle line, and conserves color for every crossing of a doubly occupied or empty left-moving line. This can be seen directly from Fig.~\ref{fig: 01}. Indeed, according to our updating rule, a change of color occurs if the line crosses a left-moving line of either a single red or a single green particle. The color change occurs independently of the color of the encountered single left-moving particle. The same rule holds for the color of single left-moving particles. The color is switched whenever a single right-moving particle of arbitrary color is crossed.

%Absatz
The color of a single right-mover at $(t,x)$ is determined by the color of the right-mover at $(t\inn, x-n\eps)$ and the number of color switches. We have the same color at $(t,x)$ and $(t\inn, x-n\eps)$ if the number of crossed single left-movers is even, while a color switch occurs for an odd number of crossed single left-moving lines. For the counting of the number of switches we define the \qq{backwards light cone} of a single right-mover at $(t,x)$ by the interval $[x-n\eps, x+n\eps]$.

\begin{figure}[h]
\centering\includegraphics[width=.5\textwidth]{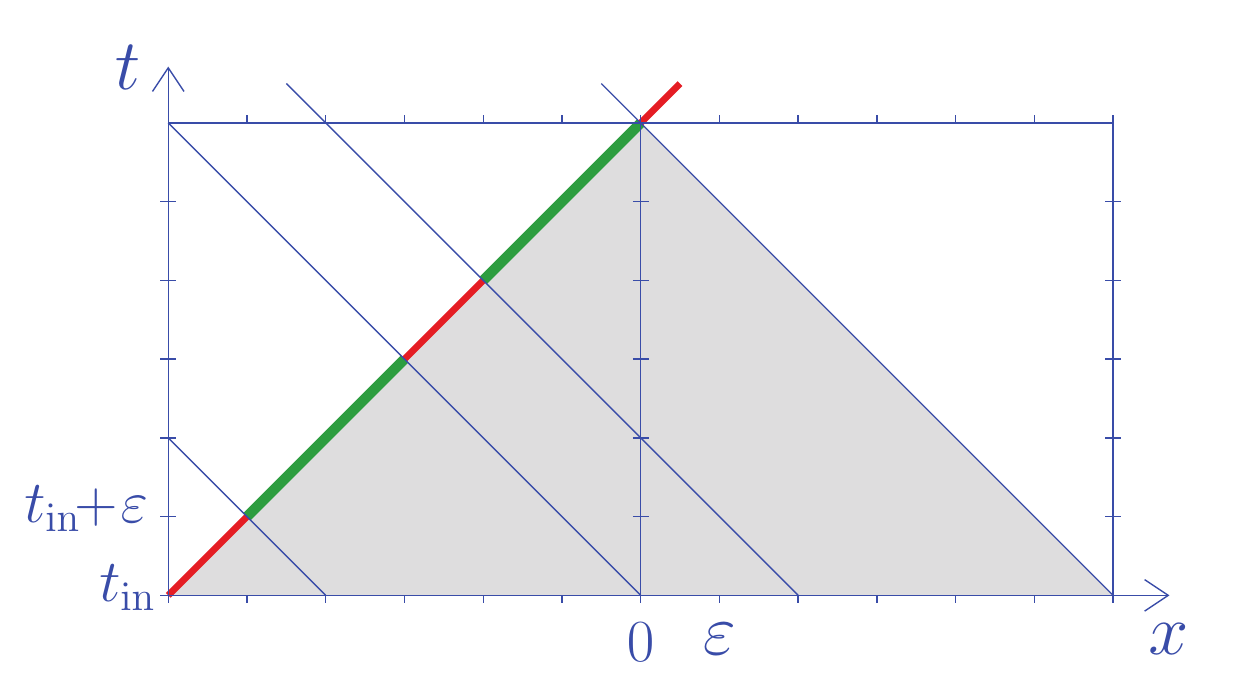}
\caption{Color switches for a single right-mover. The past light cone of the particle at ($t, x=0$) is shaded.}\label{fig: YA}
\end{figure}

One of the boundaries of the light cone is the past trajectory of the single right-mover, whereas the other corresponds to the past trajectory of a left-mover. Every single left-mover at $t-n\eps$ in the interval $[x-(n-2)\eps, x+n\eps]$ will cross the right-moving single particle line in the time interval $[t-(n-1)\eps,t]=[t\inn+\eps, t]$.
The number of color switches is therefore given by the number of positions in the interval $[x-(n-2)\eps, x+n\eps]$ for which a single left-mover is present at $t\inn$. This is easily visualized in Fig.~\ref{fig: YA}. The analogous rule holds for the color switches for a single left-mover at $(t,x)$. The number of switches corresponds to the number of single right-movers in the interval $[x-n\eps, x+(n-2)\eps]$ at $t\inn$.

%Absatz
We can use time-reversal invariance in order to construct for any given configuration $\tau$ at $t$ the corresponding configuration for $t'<t$. With a single combinatorial algorithm for determining for every configuration $\tau$ at $t$ the configuration $\tau_0$ at $t\inn$ from which it originates, we can focus on the probabilistic aspects of the cellular automaton. In principle, the problem is simple since $p_\tau(t)=p_{\tau_0}(t\inn)$. For a very large number of time steps and positions $x$ the relevant light cones become large. One would like to find some type of continuum formulation. We will see that the concepts of quantum mechanics as wave functions and a density matrix are rather useful in this context.

\subsection{Automaton with shifted blocks\label{subsec: Automaton with shifted blocks}}
There is an alternative view on the automaton of our model. We can start at even $t_\textup{in}$ in the picture where left and right movers are situated on different sublattices. For even $t$ we define blocks $B(t,x)$ consisting of four sites $(t,x), (t+\varepsilon,x),(t,x+\varepsilon) $ and $(t+\varepsilon,x+\varepsilon)$.
The evolution from $t$ to $t+\varepsilon$ can then be described separately in each block $B(t,x)$. Each block defines a small automaton with four variables (occupation numbers) at both $t$ and $t+\varepsilon$, namely the red and green particles at $x$ and $x+\varepsilon$. The rules for the automaton have to specify how each one of the sixteen configurations is mapped from $t$ to $t+\varepsilon$. The rule is that all configurations at $(t,x)$ are transported to $(t+\varepsilon,x+\varepsilon)$, and all configurations at $(t, x+\varepsilon)$ are transported to $(t+\varepsilon,x)$, with one exception: Configurations with one particle at $x$ and one particle at $x+\varepsilon$ change color when they are transported to the diagonally opposite sites at $t+\varepsilon$. In other words, all particles are transported on the crossed diagonals in the block. The color of the particles remains the same except for the interchange in case of single occupancy at $x$ and $x+\varepsilon$. The squares on the bottom line in Fig.~\ref{fig: 01} correspond to blocks for which a color exchange occurs. Drawing a square in the lower left corner would feature a red particle moving on the diagonal without color change. 

%Absatz
For odd $t+\varepsilon$ we again define blocks $B(t+\varepsilon,x)$, consisting now of the sites $(t,x), (t+\varepsilon,x), (t,x-\varepsilon), (t+\varepsilon,x-\varepsilon)$. As compared to even $t$, the blocks are shifted one place to the left. Otherwise the same rules for the automaton of each block apply. Alternating the positions of the blocks between even and odd $t$ we reconstruct all rules of the full cellular automaton. The advantage of this formulation is that at each time step it is sufficient to solve the translation between $16\times 16$ unique jump step evolution operator $\widehat{S}(t,x)$ and a representation in terms of Grassmann variables. The overall step evolution operator obtains as a direct product over all blocks,
\be\label{T14}
\widehat{S}(t)=\prod_{x}\widehat{S}(t,x)\;,
\ee
similar to eq.~\eqref{43}. The difference is that we have now only half the number of blocks, while the kinetic and interaction part are treated in common within each block. This setting is described in ref.\cite{CWPCA}.
The cellular automaton is precisely the same as the one corresponding to the particular Thirring model discussed in the present paper. Also the associated fermionic model is the same.

\section{Probabilistic cellular automata and evolution of the wave function\label{sec: 05}}

For probabilistic cellular automata the initial conditions are given by a probability distribution of initial conditions. This can be described by a wave function which plays the same role as in quantum mechanics. The evolution of the wave function follows a discrete Schrödinger equation.
We will first discuss the concept of a wave function for probabilistic cellular automata. Subsequently, we show that this wave function is the same for the fermionic model. This demonstrates that an interacting quantum theory, more specifically a fermionic quantum field theory with interactions, is precisely equivalent to a probabilistic cellular automaton.

\subsection{Initial conditions\label{subsec: Initial conditions}}

For a deterministic cellular automaton the initial state at some initial time $t_{\textup{in}}$ is given by precisely one specific configuration $\overline{\rho}$. This configuration is propagated by the rules of the automaton to any later time $t$, such that the configuration $\tau$ at $t$ is uniquely determined. A convenient description uses an $N$-component real vector $q_\textup{in}$ with components $(q_\textup{in})_{\rho}$. The initial state is specified by $q_{\rho}(t_\textup{in})=\delta_{\rho,\overline{\rho}}$, such that only the $\overline{\rho}\:\!$-component of $q(t_\textup{in})$ differs from zero. 
The initial microscopic state $\overline{\rho}$ is transformed at each time step by the rules of the cellular automaton. Each step corresponds to a (matrix - ) multiplication of the vector $q$ by the step evolution operator $\widehat{S}$. After a certain number of steps one arrives at $t\ff$ at a vector $q(t\ff)$. Only one component of this vector differs from zero. This indicates the microscopic state which is reached by the action of the automaton.

%Absatz
For a probabilistic cellular automaton the initial condition specifies a probability $p_{\rho}(t_\textup{in})$ for every possible initial configuration $\rho$. It obeys the standard laws of probability theory, 
$p_{\rho}(t_\textup{in})\geq 0, \sum_{\rho}p_{\rho}(t_\textup{in})=1$. Each configuration $\rho$ propagates by the deterministic rules of the automaton to a specific configuration $\tau(t,\overline{\rho})$ at later $t$. The probability to find the configuration $\tau$ at $t$,  $p_{\tau}(t)$, is precisely the probability $p_{\overline{\rho}}(t_\textup{in})$ of the initial configuration from which it originated,
\be\label{PC01}
p_{\tau}(t)=p_{\overline{\rho}(\tau)}(t_\textup{in})\;.
\ee
This transformation of the probability distribution defines the probabilistic cellular automaton. The probability distribution at any $t$ is, in principle, calculable from the initial probability distribution. We may again use the vector $q_{\textup{in}}$ for the specification of the initial condition. It is defined by the relation $p_{\rho}(t_{\textup{in}})=(q_{\rho}(t_{\textup{in}}))^{2}$.
In contrast to the deterministic cellular automaton more than one component of $q(t_{\textup{in}})$ can differ from zero. We will see that the evolution rule of multiplication with the step evolution operator is the same for probabilistic and deterministic cellular automata.

\subsection{Wave function for cellular automaton\label{subsec: Wave Function for cellular automaton}}

The specification of the probability distribution by a wave function $q(t)$ can be used for every time $t$,
\be\label{PC02}
p_{\tau}(t)=(q_{\tau}(t))^{2}\;.
\ee
The positivity of the probabilities is guaranteed, and the normalization requires that $q(t)$ is a unit vector 
\be\label{PC03}
q_{\tau}(t)q_{\tau}(t)=1\;.
\ee
The use of the ``classical wave function" $q(t)$~\cite{CWQPCS} instead of the probability distribution~$p(t)$ offers both technical and conceptual advantages~\cite{Wetterich:2020kqi}.

%Absatz
The evolution law for the wave function can be written in terms of the step evolution operator $\widehat{S}(t)$ by matrix multiplication
\be\label{PC04}
q(t+\varepsilon)=\widehat{S}(t)q(t)\,,\quad q_{\tau}(t+\varepsilon)=\widehat{S}_{\tau\rho}(t)q_{\rho}(t)\;.
\ee
Indeed, with
\be\label{PC05}
\widehat{S}_{\tau\rho}(t)=\delta_{\tau, \overline{\tau}(\rho)}=\delta_{\overline{\rho}(\tau),\rho}\;,
\ee
the step evolution operator differs from zero only if the configuration $\tau$ at $t+\varepsilon$ equals the configuration $\overline{\tau}(\rho)$ associated to the configuration $\rho$ at $t$ by the rule of the automaton. Eq.~\eqref{PC04} implies
\begin{align}\label{PC06}
q_{\tau}(t+\varepsilon)&=q_{\overline{\rho}(\tau)}(t)\,,\nn\\
p_{\tau}(t+\varepsilon)&=p_{\overline{\rho}(\tau)}(t)\;,
\end{align}
thus producing the rule for the probabilistic cellular automaton. Following the evolution for a sequence  of time step yields eq.~\eqref{PC01}.

%Absatz
The vector $q(t)$ ressembles the wave function of quantum mechanics in a real representation. Any complex wave function $\psi_{Q}$ has an associated real representation with twice the number of components. With
\be\label{PC07}
\psi_Q(t)=q_{R}(t)+iq_{I}(t)\;,
\ee
the real representation reads
\be\label{PC08}
q(t)=\begin{pmatrix}
q_{R}(t)\\q_{I}(t)
\end{pmatrix}\;.
\ee
Whenever the evolution of $q(t)$ is compatible with the complex structure~\eqref{PC07} the normalization of the wave function, $\psi_{Q}^{*}\psi_{Q}=1$,  is guaranteed by the relation~\eqref{PC03}. It is conserved by the evolution ~\eqref{PC04} since $\widehat{S}$ is an orthogonal matrix, such that the length or norm of $q(t+\varepsilon)$ is the same as the one of $q(t)$. 
The evolution is therefore unitary. Also the relation between the wave function or probability amplitude $q_{\tau}(t)$ and the probabilities $p_{\tau}(t)$ in eq.~\eqref{PC02} is the same as for quantum mechanics.

\subsection{Wave function for fermionic quantum field\\ \hspace*{10pt} theory}
\label{subsec: Wave function for fermionic quantum field theory}

%Absatz
The Grassmann wave function for the fermionic description is an element of the real Grassmann algebra constructed over the variables $\psi_{\alpha}(t)$ at a given $t$, $g(t)=g[\psi(t)]$. We can expand it in the basis of the fermionic quantum field theory with the functions $g_{\tau}(t)=g_{\tau}[\psi(t)]$,
\be\label{G1}
g(t)=q_{\tau}(t)g_{\tau}(t)=q_{\tau}(t)g_{\tau}[\psi(t)]\ .
\ee
We will see that the coefficients $q_{\tau}(t)$ are the components of the quantum wave function if $g(t)$ is properly normalized. They will be in one-to-one correspondence with the classical wave function for the probabilistic cellular automaton. This allows the identification of the fermionic quantum field theory with the cellular automaton. The wave function $q(t)$ is a real unit vector with $N=2^{M}$ components, as appropriate for a quantum field theory of Dirac fermions, with $M=4M_{x}$ and $M_{x}$ the number of space points. We will later decompose $q(t)$ into sectors with a fixed particle number. The wave function for the one-particle sector $q_{\gamma}^{(1)}(t,x)$ will be a real four-component function of $t$ and $x$.

%Absatz
The time evolution of the Grassmann wave function $g(t)$ obeys
\begin{align}\label{G2}
g(&\te)=\int\cD\overline{\psi}(\te)\cD\psi(t)\Ktil_{\text{int}}\gl t+\frac{\varepsilon}{2}\gr\Ktil_{\text{free}}(t)g(t)\nn\\
=&g(t+2\epstil)=\int\cD\psi(t+\epstil)\cD\psi(t)\Ktil(t+\epstil)\Ktil(t)g(t)\ ,
\end{align}
where the second line refers to the original formulation before coarse graining. Insertion of
\begin{align}
\label{G3}
\Ktil(t)=&\gbar_{\tau}'(t+\epstil)\widehat{S}_{\tau\rho}^{\text{free}}\gbar_{\rho}(t)\ ,\nn\\
\Ktil(t+\epstil)=&g_{\tau}(t+2\epstil)\widehat{S}_{\tau\rho}^{\text{int}}g_{\rho}'(t+\epstil)\ ,
\end{align}
yields, in close analogy to sect~\ref{sec: 02},
\be\label{G4}
g(t+2\epstil)=q_{\tau}(t+2\epstil)g_{\tau}(t+2\epstil)\ ,
\ee
with evolution of the wave function according to
\be\label{G5}
q_{\tau}(t+2\epstil)=S_{\tau\rho}\supt{int}S_{\rho\sigma}\supt{free}q_{\sigma}(t)\ .
\ee
In the coarse grained language this is identical to eq.~\eqref{PC04}, such that the wave function of the fermionic system follows the same evolution as the one for the cellular automaton.

%Absatz
The evolution law~\eqref{G2} follows directly from the Grassmann functional integral by a partial integration over the variables at $t'<t$,
\begin{align}
\label{G6}
g(t)=\int&\cD\psi(t\subt{in}\leq t'<t)\cD\psibar(t\subt{in}<t'\leq t)\nn\\
&\times\Ktil(t-\epstil)\dots\Ktil(t\subt{in})g(t\subt{in})\ .
\end{align}
The step to $g(t+2\epstil)$ involves two additional $\Ktil$-factors and two additional integrations, proving eq.~\eqref{G2}. 

%Absatz
We have implemented initial conditions at $t\subt{in}$ by an additional factor $g(t\subt{in})$, which encodes the wave function $g_{\tau}(t\subt{in})$ according to
\bel{G6A}
g(t\subt{in})=g\inn[\psi(t\inn)]=q_{\tau}(t\subt{in})g_{\tau}[\psi(t\subt{in})]\ .
\ee
This factor could be seen as an additional part of $\Ktil(t\inn)$. We prefer to have it separated in the notation. In this case the integrand of the partition function $Z$ in eq.~\eqref{01}, and accordingly the weight distribution $w[\psi]$ in eq.~\eqref{04}. are multiplied by an additional factor $g(t\inn)$. We can implement boundary conditions at $t\ff$ by a further boundary factor $\widehat{g}(t\ff)$. 
We will typically choose these ``final boundary conditions" such that the conjugate wave function agrees with the wave function, but more general choices are possible as well.

%Absatz
The conjugate Grassmann wave function is introduced in complete analogy to the Grassmann wave function, evolving no backwards from the final boundary condition. We present details in appendix~\ref{ap.E}. Quite generally, the pair of Grassmann wave function and conjugate Grassmann wave function permit the evaluation of expectation values for time-local observables in the fermionic quantum field theory. This can be connected directly to the Grassmann functional expression for observables in terms of Grassmann operators. 

\subsection{Density matrix for cellular automaton.}

%Absatz
For a pure state we define a symmetric real density matrix for the cellular automaton as a bilinear in the wave function,
\bel{DM1a}
\widehat{\rho}_{\alpha\beta}(t)=q_{\alpha}(t)q_{\beta}(t)\ .
\ee
Its relation to the density matrix in the fermionic model represented by a Grassmann functional is established in the appendix~\ref{ap.F}. The density matrix can be extended to more general boundary conditions for mixed states. Its evolution obeys a discretized Newmann equation
\bel{DM2a}
\widehat{\rho}(t+\varepsilon)=\widehat{S}(t)\widehat{\rho}(t)\widehat{S}^{T}(t)\ .
\ee
The diagonal elements are the time-local probabilities $p_{\alpha}(t)=\widehat{\rho}_{\alpha\alpha}(t)$.

%Absatz
The density matrix $\widehat{\rho}(t)$ is the central object which specifies the time-local probabilistic information for the cellular automaton. Once known at a given time $t$ all expectation values of time-local observables can be computed from it. No additional information on the past properties of the automaton for $t'<t$ are needed. With the evolution equation~\eqref{DM2a} the density matrix can be computed for $t'>t$, allowing for predictions in terms of the state of $t$. The density matrix contains probabilistic information beyond the time-local probabilities $p_{\alpha}(t)$.  This is stored in the off-diagonal elements of $\widehat{\rho}(t)$. This additional information allows the computation of expectation values of observables beyond those that are functions of occupation numbers at $t$.

\subsection{Operator for observables}

%Absatz
In quantum mechanics one associates to some observable $A$ a hermitian operator $\widehat{A}$ such that its expectation value is given for all $t$ by the quantum rule
\bel{DM3a}
\langle A(t)\rangle =\tr\big{\lbrace}\widehat{A}\rho(t)\big{\rbrace}\ .
\ee
Here $\rho(t)$ is the quantum density matrix. It is a hermitian complex matrix which is normalized, $\tr \rho =1$, and positive in the sense that all its eigenvalues are positive semidefinite. Expressing the complex quantities in terms of real quantities the density becomes a real symmetric matrix $\widehat{\rho}(t)$, and similarly the operators are real symmetric matrices. These structures are found in a completely analogous way for probabilistic automata. 

%Absatz
We discuss in the appendix~\ref{ap.G} how observables for the automaton are mapped to operators. This includes observables involving occupation numbers at different times. The quantum law~\eqref{DM3a} for expectation values follows directly from the general classical statistical rule for expectation values in probabilistic systems. Also may powerful methods of quantum physics, as a change of basis, can be directly implemented for the probabilistic automaton. Quantum mechanics is characterized by non-commuting operators for observables. We know that such observables, as the momentum observable not commuting with the position observable for a particle, play an important role in quantum mechanics. The momentum is a key quantity to characterize the single-particle state in a fermionic quantum field theory, with extensions to many-particle states. It can be expected to be also a useful quantity for the associated probabilistic automaton. This will be discussed in sect.~\ref{sec: 09}. The momentum observable is represented by an operator that does not commute with operators for the occupation numbers.

%Absatz
Finally, one would like to make the step from a real formulation to a complex formulation and see how density matrix and operators are mapped to hermitian complex matrices. This requires the introduction of a suitable complex structure in the next section.

\section{Particle-hole symmetry and\\complex structure}
\label{sec: 06}

%Absatz
Particle-hole symmetry is a key ingredient for fermionic quantum field theories. The complex structure of quantum mechanics can be based on this structure. Central symmetries as charge conjugation $C$, time reversal $T$, and $CPT$ are directly connected to particle-hole symmetry. For fermionic quantum field theories the presence of antiparticles emerges naturally in this context. In turn, particle-hole symmetry reflects the modulo-two property of the Grassmann functional integral. The particle-hole transformation can be formulated on the level of the wave function. It therefore applies directly to the cellular automaton. The propagation and scattering of our cellular automaton is invariant under the exchange of particles and holes and therefore realizes particle-hole symmetry. The complex structure based on the particle-hole transformation can be extended to include additional discrete transformations acting on internal indices.

\subsection{Particle-hole transformation}
\label{subsec: Particle-hole transformation}

%Absatz
On the level of the Grassmann wave function we define the particle-hole transformation as
\bel{PH2}
g(t)=q_{\tau}(t)g_{\tau}(t)\to g^c(t)=q_{\tau}(t)g_{\tau}^c(t)\ .
\ee
Here $g_\tau^c$ is related to $\gbar_\tau$ or $\gbar_\tau'$ by a sign 
\bel{DS1}
g_\tau^c=\varepsilon_\tau^c\gbar_\tau\ ,\quad \eps_\tau^c=\pm1\ .
\ee
We choose the sign such that $g_\tau^c$ is one of the Grassmann basis elements $g_{\tau^{c}}$, without an additional minus sign.
Every particle (factor $\atil=1$ in $g_{\tau}$) is mapped to a hole ($\atil=\psi_{\alpha}$ in $g_{\tau}^c$), and vice versa. 

Instead of changing the Grassmann basis elements at fixed $q(t)$ we can realize the particle-hole transformation $g(t)\to g^{c}(t)$ also by a map for the wave function at fixed basis elements,
\bel{PH1A}
K(q)=q^{c}\ ,\quad K_{\tau\rho}q_\rho=q_\tau^{c}\ .
\ee
The matrix $K$ is defined by
\bel{PH2A}
g^c=K_{\tau\rho}q_\rho g_\tau=q_\tau g_\tau^c=q_\tau^c g_\tau\ .
\ee
This implies indeed
\bel{PH3A}
K(g)=\gl K(q)\gr_\tau g_\tau=g^c\ .
\ee
With
\bel{PH4A}
K\gl K(g)\gr=g\ ,
\ee
the matrix $K$ describes an involution
\bel{PH5A}
K_{\tau\sigma}K_{\sigma\rho}=\delta_{\tau\rho}\ ,\quad K^2=1\ .
\ee

%Absatz
For a concrete form of the matrix $K$ we need to identify pairs of basis elements $\gl g_\tau,g_\tau^c\gr$ which are mapped into each other by $K$. For this purpose we divide the set of configurations $\{\tau\}$ into two subsets $\{\tau'\}$ and $\{\tau^c\}$, where the elements of $\{\tau^c\}$ obtain from elements of $\{\tau'\}$ by a particle-hole transformation. Correspondingly we group the components of the wave function into two sets. The associated pairs $q_{\tau'}$ and $q_{\tau^c}$ are grouped into two-component vectors, whose components are mapped into each other by $K$,
\bel{PH6A}
\chi_\tau=\pvec{q_{\tau'}}{q_{\tau^c}}=\pvec{q'_\tau}{q_\tau^c}\ ,\quad K\chi_\tau=\pvec{q_{\tau^c}}{q'_\tau}=\pvec{q_\tau^c}{q'_\tau}\ ,
\ee
such that in this subspace one has
\bel{PH7A}
K=\pmat{0}{1}{1}{0}=\tau_1\ .
\ee
The number of independent components $\chi_\tau$ is only half the number of components of $q_\tau$. We may choose for $\{\tau'\}$ all configurations with total particle number $N_p<2M_x$, for which the complement configurations $\tau^c$ obey $N_p>2M_x$. For the remaining \qq{half-filled configurations} with $N_p=2M_x$ we include one half in $\{\tau'\}$ and the other half in $\{\tau^c\}$.

%Absatz
For a given choice of basis the matrix $K$ is uniquely fixed. Particle-hole transformations are therefore realized on the level of wave functions. This formulation directly applies to the probabilistic cellular automaton since it shares the same wave function with the associated fermionic quantum field theory. The particle-hole transformation and the associated complex structure are useful concepts for the classical statistical system of the probabilistic cellular automaton. We note that the choice of signs $\varepsilon_\tau^c$ in eq.~\eqref{DS1} is not unique. Different choices lead to different matrices $K$. We only require the involution property $K^{2}=1$. The grouping into pairs ($q_{\tau}', q_{\tau}^{c}$) remains the same, but for the action of $K$ on a subspace with given $\tau$ one may have to replace  $\tau_{1}$ by $-\tau_{1}$. For the sake of simplicity we will stick here to the definition~\eqref{PH7A}.

%Absatz
A unit step evolution operator reproduces the same Grassmann wave function only after two evolution steps $\epstil$.
After a single evolution step it changes the Grassmann wave function $g[\psi]$ to $\widehat{g}[\,\psibar\,]$ according to
\begin{align}\label{PH1}
\int\cD\psi(t)&\exp\{\psibar_{\alpha}(\te)\psi_{\alpha}(t)\}q_{\tau}(t)g_{\tau}[\psi(t)]\nn\\
=&q_{\tau}(t)\gbar_{\tau}'[\,\psibar(\te)]=\widehat{g}(\te)\ .
\end{align}
For every factor $\tilde{a}_{\alpha}=\psi_{\alpha}$ in $g[\psi]$ one has a factor $\bar{a}_{\alpha}=1$ in $\widehat{g}[\,\psibar\,]$, while a factor $\tilde{a}_{\alpha}=1$ in $g[\psi]$ results in a factor $\psibar_{\alpha}$ for $\widehat{g}[\,\psibar\,]$. If we identify a factor $\atil=1$ in $\ghat[\psi]$ with a present particle, and a factor $\atil=\psi_{\alpha}$ with an absent particle or \qq{hole}, the role of particles and holes is interchanged for a single evolution step $\epstil$. Up to relative minus-signs the unit step evolution operator realizes in a single step $\epstil$ the particle-hole conjugation $K$. This is the reason for our use of coarse graining that groups together two evolution steps $\epstil$ to a combined step $\varepsilon=2\epstil$.

\subsection{Complex Structure}
\label{subsec: Complex Structure}

A general complex structure is defined by a pair of discrete transformations $\gl K, I\gr$ which obey
\bel{CS1AA}
K^2=1\ ,\quad I^2=-1\ ,\quad \big\{K,I\big\}=0\ .
\ee
The involution $K$ realizes the operation of complex conjugation, while $I$ implements the multiplication by $i$. For $K$ we choose the particle-hole transformation~\eqref{PH1A}. Quantities that are even with respect to $K$ are considered as real, while odd quantities become imaginary.
Correspondingly, we define
\bel{CS2A}
\varphi_{R\tau}=\frac{1}{\sqrt{2}}\gl q'_\tau+q_\tau^c\gr\ ,\quad \varphi_{I\tau}=\frac{1}{\sqrt{2}}\gl q'_\tau-q_\tau^{c}\gr\ .
\ee

%Absatz
The map from real to complex wave functions, $q(t)\to\varphi(t)$, is realized by defining the components $\varphi_\tau$ of the complex wave function by
\bel{CS3A}
\varphi_\tau=\varphi_{R\tau}+i\varphi_{I\tau}=\frac{1+i}{\sqrt{2}}q'_\tau+\frac{1-i}{\sqrt{2}}q_\tau^c =e^{\frac{i\pi}{4}}\gl q'_\tau-iq_\tau^c\gr\ .
\ee
For each $\tau$ this is a map $\chi_\tau\to\varphi_\tau$, and we recall that the number of complex components is now $N/2$, $\tau=1\dots N/2$.
Keeping in mind the different ranges for the sums over $\tau$ we observe
\begin{align}\label{CS4A}
\varphi_\tau^*\varphi_\tau=\sum_{\tau=1}^{N/2}\varphi_\tau^*\varphi_\tau&=\sum_{\tau=1}^{N/2}\gl (q'_\tau)^2+(q_\tau^c)^2\gr \nn\\&=\sum_{\tau=1}^{N}q_\tau^2=q_\tau q_\tau=1\ ,
\end{align}
which amounts to a standard normalization of a complex wave function in quantum mechanics
\bel{CS5A}
\varphi\herm(t)\varphi(t)=1\ .
\ee

%Absatz
The map~\eqref{CS3A} also specifies the transformation $I$ as
\bel{CS6A}
I=\pmat{0}{1}{-1}{0}=i\tau_{2}\ .
\ee
With this definition one has
\bel{CS7A}
\varphi(Kq)=\varphi^*\ ,\quad \varphi(Iq)=i\varphi\ ,
\ee
such that the multiplication of $\varphi$ with a complex number can be realized by an appropriate linear transformation of $q$.

%Absatz
This generalizes to the multiplication of $\varphi$ by a complex $N/2\times N/2$-matrix $A$, which is implemented by a multiplication of $q$ by a real $N\times N$-matrix $\widehat{A}$,
\begin{align}\label{CS8A}
&\varphi_A=A\varphi\ ,\quad A=A_R+iA_I\ \leftrightarrow\ q_A=\widehat{A}q\ ,\nn\\ &\widehat{A}=A_R\mathds{1}+A_II\ .
\end{align}
Any real matrix of the form $A_R\mathds{1}+A_II$ is called compatible with the complex structure and associated in the complex picture to the complex matrix $A$. For matrices $\widehat{A}$, $\widehat{B}$ that are compatible with the complex structure the multiplication of $q$ by $\widehat{A}$ in the real basis is mapped to the multiplication of $\varphi$ by $A$ in the complex basis. Also the real matrix product $\widehat{A}\widehat{B}$ is mapped to the complex matrix product $AB$. For symmetric matrices $\widehat{A}^T=\widehat{A}$ the compatibility condition~\eqref{CS8A} implies $A_R^T=A_R$, $A_I^T=-A_I$. The associated complex matrix is therefore hermitian, $A\herm=A$.

%Absatz
For symmetric operators $\widehat{A}$ which are compatible with the complex structure the quantum rule~\eqref{QA4} for expectation values takes in the complex formulation the usual form
\bel{221A}
\langle  A\rangle =\langle \varphi| A|\varphi\rangle=\varphi_{\tau}^{*}A_{\tau\rho}\ \!\varphi_{\rho}\  .
\ee
We could choose a different basis
\bel{DB1}
\chi'=\pvec{\varphi_R}{\varphi_I}\ ,
\ee
which is related to $\chi$ by the similarity transformation~\eqref{CS2A}. In this basis one has
\bel{DB2}
K=\tau_3\ ,\quad I=-i\tau_2\ .
\ee

%Absatz
There are many possibilities to introduce a complex structure~\eqref{CS1AA} by a suitable choice of the discrete transformations $K$ and $I$. In general, the particle-hole transformation and the involution defining the complex conjugation may be different transformations $K'$ and $K$. In particular, we may multiply the particle-hole transformation $K'$ by a change of sign of all Grassmann variables with the color two. Accompanied by a corresponding modification of $I$ this makes the setting compatible with the definition of a complex Dirac spinor in terms of two real Majorana spinors.

%Absatz
A useful complex structure should be compatible with the time evolution in the sense that the step evolution operator is a matrix compatible with the complex structure obeying eq.~\eqref{CS8A}. This requirement restricts the possible complex structures, but is not sufficient to single out a unique one. We may further require that the vacuum state is invariant under the complex conjugation $K$. Thus the useful complex structures may depend on the vacuum state. We discuss this briefly in sec.~\ref{sec: 08}, focusing in the following on the identification of $K$ with the particle-hole transformation.

\subsection{Complex density matrix}
\label{subsec: Complex density matrix}

For a pure state we define the complex density matrix by
\bel{CS9A}
\rho_{\tau\rho}=\varphi_\tau\varphi_\rho^*\ .
\ee
In terms of the real wave function this reads
\bel{CS10A}
\rho_{\tau\rho}=q'_\tau q'_\rho+q_\tau^c q_\rho^c+i\gl q'_\tau q_\rho^c-q'_\rho q_\tau^c\gr\ .
\ee
The symmetric part of $\rho$ is real and the antisymmetric part imaginary, such that $\rho$ is hermitian
\bel{CS11A}
\rho\herm=\rho\ .
\ee
The r.h.s. of eq.~\eqref{CS10A} is a linear combination of matrix elements of the real density matrix $\widehat\rho$.

%Absatz
For a generalization beyond pure states we choose a basis for which the $N/2$ components corresponding to the states $\tau'$ of the complex formulation form the first set of components, and the one for $\tau^c$ the second set, such that the pure state wave function is ordered as
\bel{CS12A}
q=\pvec{q'}{q^c}\ ,
\ee
with $q_\tau'=q_\tau$ for the first $N/2$ components. In this basis we define for general symmetric $\widehat{\rho}$
\bel{CS13A}
\widehat\rho=\pmat{\rho'}{\tilde\rho}{\tilde\rho^T}{\rho^c}\ .
\ee
For the special case of pure states this yields
\bel{CS14A}
\rho_{\tau\rho}'=q_\tau'q_\rho'\ ,\quad \rho_{\tau\rho}^{c}=q_\tau^{c}q_\rho^c\ ,\quad \tilde\rho_{\tau\rho}=q_\tau'q_\rho^c\ ,
\ee
and therefore eq.~\eqref{CS10A} reads
\bel{CS15A}
\rho_{\tau\rho}=\rho_{\tau\rho}'+\rho_{\tau\rho}^c+i\gl\tilde\rho_{\tau\rho}-\tilde\rho_{\tau\rho}^T\gr\ .
\ee

%Absatz
The matrix~\eqref{CS13A} is the most general form of a symmetric real density matrix $\widehat\rho$. We can employ eq.~\eqref{CS15A} for the definition of the complex density matrix $\rho$ for mixed states. For general symmetric mixed state density matrices $\widehat\rho$ the map $\widehat\rho\to\rho$ is not invertible. A general real symmetric $N\times N$-matrix has $N(N+1)/2$ independent real entries, while the number of independent real elements for a general hermitian $N/2\times N/2$-matrix is only $N^2/4$. For a pure state density matrix the relation $\widehat{\rho}\2=\widehat{\rho}$ is mapped to $\rho\2=\rho$, such that $\widehat{\rho}\2$ is mapped to $\rho\2$. For a general mixed state density matrix~\eqref{CS13A} the real square $\widehat{\rho}^{2}$ is no longer mapped to the complex square $\rho\2$.
We can generalize the map~\eqref{CS15A} to non-symmetric real $\widehat\rho$ by replacing $\tilde\rho^T$ in eqs.~\eqref{CS13A}~\eqref{CS15A} by an independent matrix $\tilde\rho'$. If $\widehat\rho$ is not symmetric, $\rho$ is not hermitian.

\subsection{Unitary evolution}
\label{subsec: Unitary evolution}

Let us next discuss the compatibility of the step evolution operator with the complex structure. Since the particle-hole conjugation is a map on configurations or wave functions at a point $t$, it remains on the same sublattice for $x$, e.g. even $x$ for even $t$. After the action of the step evolution operator $\Shat(t)$ the configuration remains on the even sublattice, involving now odd $x$ for odd $t+\eps$. Only after two steps the configurations are again on the same even sublattice for $x$, such that the evolution can be compared with the action of the particle-hole transformation. For this reason we define here, with a slight abuse of notation and for even $t$,
\bel{CS15AA}%Double A
\Shat=\Shat(t+\eps)\Shat(t)\ .
\ee

A general step evolution operator $\Shat$ reads in the basis~\eqref{CS12A}
\bel{CS16A}
\Shat=\pmat{\Shat'}{\Stil}{\Stil'}{\Shat^c}\ ,
\ee
where the orthogonality $\Shat^T\Shat=1$ restricts the $N/2\times N/2$-matrices $\Shat'$, $\Shat^c$, $\Stil$ and $\Stil'$. The evolution law~\eqref{DM21}, $\widehat\rho(t+2\eps)=\Shat\widehat\rho(t)\Shat^T$, is compatible with the complex structure provided that
\bel{CS17A}
\Shat^c=\Shat'\ ,\quad \Stil'=-\Stil\ .
\ee
In this case one has
\bel{CS18A}
\Shat=\Shat'\mathds{1}+\Stil I\ ,
\ee
and we can map $\Shat$ to a complex $N/2\times N/2$-matrix $U$ defined by
\bel{CS19A}
U=\Shat'+i\Stil\ .
\ee
According to eq.~\eqref{CS8A} the time evolution of the complex wave function $\varphi$ reads
\bel{CS20A}
\varphi(t+2\eps)=U(t)\varphi(t)\ .
\ee
This translates to the evolution law for the complex density matrix
\bel{CS4}
\rho(t+2\eps)=U\rho(t)U^{\dagger}\ .
\ee

%Absatz
With the condition~\eqref{CS17A} the orthogonality of $\Shat$ translates to
\begin{align}
\label{CS21A}
&\Shat'^T\Shat'+\Stil^T\Stil=1\ ,\nn\\
&\Shat'^T\Stil=\Stil^T\Shat'=\gl\Shat'^T\Stil\gr^T\ .
\end{align}
These relations imply that $U$ is a unitary matrix,
\bel{CS22A}
U\herm U=1\ .
\ee
This is easy to understand from eq.~\eqref{CS4A}: Real orthogonal transformations preserve the norm of $q$, and unitary transformations preserve the norm of $\varphi$.

%Absatz
For a suitable choice of bit configurations $\tau'$ and $\tau^{c}$ that are mapped into each other by the particle-hole transformation the step evolution operator $\Shat$ of our model obeys the properties~\eqref{CS21A}. Such a choice is necessary in order to define the action of $\Shat$ in the basis~\eqref{CS12A}. For an appropriate choice we obtain $\Shat^c=\Shat'$ and $\Stil=\Stil'=0$, such that the evolution is indeed unitary for the corresponding complex structure.

%Absatz
Consider first a submanifold of states $\tau'$ that are mapped by $\Shat$ to other states within the same submanifold, and assume that the particle-hole transform $K(\tau)$ of every state in the submanifold yields a state $\tau^c$ that does not belong to the submanifold. This defines the submanifold of states $\{\tau^c\}$ that are in the complement of $\{\tau'\}$ with respect to $K$.
Particle-hole symmetry of the dynamics implies $\Shat^c=\Shat'$ for all states in these submanifolds. For a suitable assignment of states to $\tau'$ and $\tau^c$ this division into submanifolds is complete in the sense that every configuration $\tau$ belongs either to $\{\tau'\}$ or to $\{\tau^c\}$. This is precisely the case if the particle-hole transformation never mixes configurations of the two submanifolds, i.e. for $\Stil=\Stil'=0$. We next establish this property for a suitable assignment of configurations or states.

%Absatz
The step evolution operator for our cellular automaton or the associated fermionic quantum field theory preserves the total particle number $N_p$~\eqref{117A}. We assign states with $N_p>2M_x$ to $\{\tau'\}$. Particle-hole conjugation maps these states to new states with particle number $N_p'=4M_x-N_p$, for which $N_p'<2M_x$. These states belong to the complement $\{\tau^c\}$. The particle number conserving evolution cannot mix states with $N_p<2M_x$ with states for which $N_p'>2M_x$. What remains to be determined is a suitable distribution of the half-filled states to $\{\tau'\}$ and $\{\tau^c\}$ such that the evolution does not mix the associated wave functions $q'$ and $q^c$,
\bel{CS23A}
\Stil=\Stil'=0\ .
\ee
The particle-hole transformation maps particles to holes for each species separately. For a given configuration $\tau$ we can define particle numbers for each species $(N_{R1})_\tau$, $(N_{R2})_\tau$, $(N_{L1})_\tau$ and $(N_{L2})_\tau$. For a state $\tau'$ with given particle numbers $(N_\gamma)_{\tau'}$ the particle-hole conjugate or complementary state $\tau^c$ has particle numbers 
\bel{CS24A}
(N_\gamma)_{\tau^c}=M_x-(N_\gamma)_{\tau'}\ .
\ee
We need to distribute the half-filled states with $N_{R1}+N_{R2}+N_{L1}+N_{L2}=2M_x$ to $\{\tau'\}$ and $\{\tau^c\}$. We associate the states with $N_{R1}+N_{R2}>M_x$ to $\{\tau'\}$, and the states with $N_{R1}+N_{R2}<M_x$ to $\{\tau^c\}$. Since $\Shat$ preserves the total number of right movers $N_{R1}+N_{R2}$, one infers that particle-hole conjugation does not mix these states, in accordance with eq.~\eqref{CS23A}. What remains to be distributed at this stage are only the configurations with an equal number of right movers and left movers, $N_{R1}+N_{R2}=N_{L1}+N_{L2}=M_x$. 

%Absatz
As familiar in quantum mechanics one can make a change of basis by a complex similarity transformation. This does not change the evolution law~\eqref{CS4}.
%Ab hier C53
The step evolution operator $U(t)$ remains unitary in the new basis, but in general no longer real and orthogonal.

\subsection{Complex operators and general boundary conditions}
\label{subsec: complex operators and}

%Absatz
The complex structure extends to the operators associated to observables. We discuss this issue in the appendix~\ref{ap.H}. Not every observable is compatible with the complex structure. For those observables that are compatible with the complex structure the real symmetric operators are mapped to complex hermitian operators. The expectation value of the observable is then given by the usual quantum rule~\eqref{DM3a} in the complex formulation.

%Absatz
We can finally formulate the general restriction for the boundary conditions. Pure state boundary conditions are specified by the choice of initial and final wave functions $\tilde{q}(t_{\textup{in}})$, $\overline{q}(t_{\textup{f}})$. As in quantum mechanics, mixed boundary conditions can be obtained by appropriate weighted sums. The general boundary conditions should be chosen such that $\rho$ is a positive hermitian normalized matrix, i.e. all eigenvalues $\lambda_{\beta}$ of $\rho$ should be positive semidefinite, $\lambda_{\beta}\geq0$. This property, as well as the relations $\tr\rho=1$ and $\rho^{\dagger}=\rho$, are preserved by the unitary evolution~\eqref{CS4}. It is therefore sufficient that they hold at some given $t$, say at $t\inn$ or $t\ff$. For a positive matrix all diagonal elements are positive, $\rho_{\tau\tau}\geq0$ (no sum here), such that the conditions for a probabilistic setting are obeyed. In particular, for $\tilde{q}(t)=\bar{q}(t)=q(t)$ the density matrix is a real pure state density matrix with one eigenvalue one and all other eigenvalues zero. This particular boundary condition corresponds to $\ghat(t\ff)=q_{\tau}(t\ff)g_{\tau}'[\,\psibar(t\ff)]$.

\section{Continuous evolution and Hamilton operator}
\label{sec: 07}

The step evolution operator describes the unitary evolution of quantum mechanics in discrete time steps. One can construct an associated continuous time evolution, given by a Schrödinger or von-Neumann equation, which reproduces the discrete time evolution for all discrete times $t=t\inn+m\varepsilon$. In the continuum limit the discreteness of the time evolution plays no longer a role. As an input for this section we will only use the step evolution operator $\Shat$ such that all results apply equally to the cellular automaton and the fermionic quantum field theory.

%Absatz
We will express the step evolution operator and the associated Hamiltonian in terms of fermionic annihilation and creation operators. This makes the fermionic content of our probabilistic automaton directly visible, without the need of an explicite use of the bit-fermion map to a Grassmann functional integral. The derivation of the corresponding expression is, however, rather complex for the propagation part. The use of the relation between annihilation and creation operators on one side, and Grassmann variables on the other side, is a useful tool in this context.

\subsection{Hamilton operator}
\label{subsec: Hamilton operator}

The Hamilton operator $H$ is related to the step evolution operator $\Shat$ by
\bel{CE1}
\Shat=\exp\gl-i\varepsilon H\gr\ .
\ee
Since $\Shat$ is a unitary (in our case orthogonal) matrix, $H=H^{\dagger}$ is hermitian.
The von-Neumann equation for the evolution of the density matrix,
\bel{CE2}
i\dt\rho=[H,\rho]\ ,
\ee
has for a time-independent hermitian Hamilton operator $H$ the solution
\bel{CE3}
\rho(t)=U(t,t_{1})\rho(t_{1})U^{\dagger}(t,t_{1})\ ,
\ee
with
\bel{CE4}
U(t,t_{1})=\exp\gl-i(t-t_{1})H\gr\ .
\ee
For $t-t_{1}=\varepsilon$ one has $U(t,t_{1})=\widehat{S}(t_{1})$ according to eq.~\eqref{CE1}. The von Neumann equation~\eqref{CE3} reproduces the discrete evolution equation~\eqref{CS4} if we take $H$ piecewise constant in the intervals between the discrete time points $t=t\subt{in}+m_{t}\varepsilon$.

%Absatz
For $H$ we can make the ansatz
\bel{CE5}
H=H\subt{free}+H\subt{int}+\Delta H\ ,
\ee
with
\bel{CE6}
\Shat\subt{free}=\exp\gl-i\eps H\subt{free}\gr\ ,\quad \Shat\subt{int}=\exp\gl-i\eps H\subt{int}\gr\ .
\ee
This implies for $\Delta H$ the relation
\begin{align}
\label{CE7}
\exp\Big{\{}-&i\eps\gl H\subt{int}+H\subt{free}+\Delta H\gr\Big{\}}\nn\\
&=\exp\gl-i\eps H\subt{int}\gr\exp\gl-i\eps H\subt{free}\gr\ ,
\end{align}
and therefore
\bel{CE8}
\Delta H=\mathcal{O}\gl\eps[H\subt{int},H\subt{free}]\gr\ .
\ee
This may suggest that in the continuum limit $\eps\to0$ the commutator term can be neglected, and $\Delta H$ can be omitted.

%Absatz
The issue is not as straightforward as in the usual functional integral formulation for quantum field theories. The reason is that the step evolution operator for an automaton is not a small deviation from unity of the order $\varepsilon$ since we deal with discrete jumps. For example, we will see that the interaction part $H\subt{int}$ involves a factor $\varepsilon^{-1}$. The neglection of the commutator term $\Delta H$ may be justified for a sufficiently smooth wave function. In this case we have the possibility that the wave function changes only by a small amount $\sim\varepsilon$ for one step of the evolution from $t$ to $t+\varepsilon$. If the part $\sim \Delta H$ only induces changes $\sim \varepsilon\2$ it can indeed be neglected for $\varepsilon\rightarrow 0$.  The continuous character of the wave function, and therefore the probabilistic character of the automaton, are crucial in this respect.

\subsection{Annihilation and creation operators}
\label{subsec: Annihilation and creation operators}

We can express $\Shat\subt{free}$ and $\Shat\subt{int}$ in terms of fermionic annihilation operators $a_{\alpha}$ and creation operators $a_{\alpha}^{\dagger}$, which obey the usual anti-commutation relations
\bel{CE9}
\{a_{\alpha}\herm,a_{\beta}\}=\delta_{\alpha,\beta}\ ,\quad \{a_{\alpha},a_{\beta}\}=\{a_{\alpha}\herm,a_{\beta}\herm\}=0\ .
\ee
Their action on the Grassmann wave function can be represented as
\bel{CE10}
a_{\alpha}\hateq\psi_{\alpha}\ ,\quad a_{\alpha}\herm\hateq\frac{\partial}{\partial\psi_{\alpha}}\ .
\ee

%Absatz
We want to implement here the action on the wave function and density matrix and therefore need a suitable representation in the chosen basis. For a single two-state system we employ the real $2\times 2$-matrices
\bel{CE11}
a=\pmat{0}{0}{1}{0}\ ,\quad a\herm=\pmat{0}{1}{0}{0}\ ,\quad a\herm a=\pmat{1}{0}{0}{0}\ .
\ee
For the sixteen local states for the fermions $\psi_{Ra}$, $\psi_{La}$ we use
\begin{align}
\label{CE12}
a\subt{R1}=&a\otimes1\otimes1\otimes1\ ,\quad a\subt{R2}=\tau_{3}\otimes a\otimes1\otimes1\ ,\nn\\
a\subt{L1}=&\tau_{3}\otimes\tau_{3}\otimes a\otimes1\ ,\quad a\subt{L2}=\tau_{3}\otimes\tau_{3}\otimes\tau_{3}\otimes a\ .
\end{align}
Finally, the anti-commutation relations for annihilation operators at different $x$ are implemented by introducing the $16\times16$-matrix $T_{3}$,
\bel{CE13}
T_{3}=\tau_{3}\otimes\tau_{3}\otimes\tau_{3}\otimes\tau_{3}\ ,\quad \{T_{3},a_{\gamma}\}=0\ ,
\ee
and taking direct products of the $16\times16$-matrices $a_{\gamma}$, $T_{3}$,  and $1$,
\begin{align}
\label{CE14}
a_{\gamma}(x\inn)=&a_{\gamma}\otimes1\otimes1\otimes1\dots\ ,\nn\\
a_{\gamma}(x\inn+\eps)=&T_{3}\otimes a_{\gamma}\otimes1\otimes1\dots\ ,\nn\\
a_{\gamma}(x\inn+2\eps)=&T_{3}\otimes T_{3}\otimes a_{\gamma}\otimes1\dots\ .
\end{align}
The creation operators $a_{\gamma}\herm(x)$ are the hermitian conjugated (transposed in our case) of the annihilation operators $a_{\gamma}(x)$. Further details can be found in appendix~\ref{ap.A}.

%Absatz
We will exploit the close connection between Grassmann variables and annihilation/creation operators by choosing sign conventions such that
\begin{align}
\label{CE14A}
g_{\rho}[\psi](a_{\alpha})_{\rho\tau}=&\psi_{\alpha}g_{\tau}[\psi]\ ,\nn\\
g_{\rho}[\psi](a_{\alpha}\herm)_{\rho\tau}=&\frac{\partial}{\partial\psi_{\alpha}}g_{\tau}[\psi]\ .
\end{align}
Indeed, the product $\psi_{\alpha}g_{\tau}$ vanishes if $g_{\tau}$ contains already a factor $\psi_{\alpha}$. In this case the basis function $g_{\tau}$ corresponds to a state for which no particle $\alpha$ is present. The annihilation operator $a_{\alpha}$ yields zero, as it should be. If $g_{\tau}$ contains no factor $\psi_{\alpha}$ it describes a state with a particle $\alpha$ present. After multiplication with $\psi_{\alpha}$ the product $\psi_{\alpha}g_{\tau}$ is a new basis function $g_{\tau'}$ up to a possible minus sign. For this basis function $g_{\tau'}$ no particle $\alpha$ is present. Thus multiplication with $\psi_{\alpha}$ annihilates a particle $\alpha$, leading to a state with no particle $\alpha$ present. This is precisely the action of the annihilation operator $a_{\alpha}$ which therefore transforms $g_{\rho}(a_{\alpha})_{\rho\tau}=\eta g_{\tau'}$, $\eta=\pm1$. 
We conclude the general relation (no sum over $\tau$, $\alpha$)
\bel{CE14B}
g_{\rho}(a_{\alpha})_{\rho\tau}=\eta_{\tau\alpha}\psi_{\alpha}g_{\tau}\ ,\quad \eta_{\tau\alpha}^{2}=1\ .
\ee

%Absatz
The argument for the creation operator is similar, with $\partial/\partial\psi_{\alpha}g_{\tau}$ vanishing if a particle $\alpha$ is present (no factor of $\psi_{\alpha}$ in $g_{\tau}$), and creating a particle if no particle is present in $g_{\tau}$ (eliminating the factor $\psi_{\alpha}$). One concludes
\bel{CE14C}
g_{\rho}(a_{\alpha}\herm)_{\rho\tau}=\eta_{\tau\alpha}\frac{\partial}{\partial\psi_{\alpha}}g_{\tau}\ .
\ee
The occurrence of the same factor $\eta_{\tau\alpha}$ as in eq.~\eqref{CE14B} follows from (no sum over $\alpha$)
\bel{CE14D}
g_{\rho}(a_{\alpha}\herm a_{\alpha})_{\rho\tau}=g_{\rho}(\widehat{n}_{\alpha})_{\rho\tau}=\frac{\partial}{\partial\psi_{\alpha}}\psi_{\alpha}g_{\tau}\ ,
\ee
with $(n_{\alpha})_{\tau\rho}=(1,0)$ the particle occupation number associated to $g_{\tau}$. The anti-commutation relations~\eqref{CE9} remain unaffected if we multiply both $a_{\alpha}$ and $a_{\alpha}\herm$ by $(-1)$. This freedom of choice of a sign for $a_{\alpha}$, together with the freedom of choice for the sign $\tilde{s}_{\tau}$ for $g_{\tau}$ in eq.~\eqref{06}, permits us to choose conventions for which $\eta_{\tau\alpha}=1$.

%Absatz
We can define the sign $\tilde{s}_{\tau}$ by taking a plus sign if the factors $\psi_{\beta}$ in $g_{\tau}$ are ordered with the lowest $\beta$ to the left, as in eq.~\eqref{18}. All states can be obtained from the completely filled state $g_{1}=1$ by consecutive application of the annihilation operator $\psi_{\beta}$, starting with the largest $\beta$ and continuing with decreasing $\beta$. The fact that $a_{\alpha}$, $a_{\beta}\herm$ obey the same anti-commutation relations as $\psi_{\alpha}$, $\partial/\partial\psi_{\beta}$ guarantees the consistency of the sign convention. More details on sign conventions can be found in the appendix~\ref{ap.A}, where we show that the above convention is compatible with the prescription~\eqref{CE11}-\eqref{CE14}.

\subsection{Interaction Hamiltonian}
\label{subsec: Interaction Hamiltonian}

We next express the interaction part of the step evolution operator and Hamiltonian in terms of annihilation and creation operators. For this purpose we first consider the $16\times 16$-matrix
\bel{CE15}
\Dp=a_{R2}\herm a_{L1}\herm a_{L2}a_{R1}=a\otimes a\herm\otimes a\herm\otimes a\ .
\ee
Its action on a local state produces zero except for the state $(1\ 0\ 0\ 1)$ for which
\bel{CE16}
\Dp(1\ 0\ 0\ 1)=(0\ 1\ 1\ 0)\ .
\ee
Similarly,
\bel{CE16A}
\Dm=\Dp\herm=a_{R1}\herm a_{L2}\herm a_{L1}a_{R2}=a\herm\otimes a\otimes a\otimes a\herm
\ee
acts as
\bel{CE17}
\Dm(0\ 1\ 1\ 0)=(1\ 0\ 0\ 1)\ .
\ee
The sum $\widehat{D}=\Dp+\Dm$ interchanges the corresponding two-particle states
\bel{CE18}
\widehat{D}:\ (1\ 0\ 0\ 1)\leftrightarrow(0\ 1\ 1\ 0)\ ,\quad \widehat{D}^{2}=1\ ,
\ee
while it yields zero for all other local states. As a result, the operator
\bel{CE19}
\Shat_{D}(t)=\exp\gl-i\alpha\widehat{D}t\gr
\ee
equals unity for all states except $\varphi_{1}=(1\ 0\ 0\ 1)$ and $\varphi_{2}=(0\ 1\ 1\ 0)$, while
\bel{CE20}
\Shat_{D}(t)\pvec{\varphi_{1}}{\varphi_{2}}=\cos(\alpha t)\pvec{\varphi_{1}}{\varphi_{2}}-i\sin(\alpha t)\pvec{\varphi_{2}}{\varphi_{1}}\ .
\ee
In particular, for $\alpha=\pi/(2\eps)$ one has
\bel{CE21}
\Shat_{D}(\eps)\pvec{\varphi_{1}}{\varphi_{2}}=-i\pvec{\varphi_{2}}{\varphi_{1}}\ .
\ee
Up to the factor $-i$ this describes the color exchange process~\eqref{44} for an incoming red right mover and an incoming green left mover, together with the process for interchanged colors.

The color exchange process for an incoming right mover and incoming left mover of the same color can be described by defining the matrices
\begin{align}
\label{CE22}
\Cp=&a_{R2}\herm a_{L2}\herm a_{R1}a_{L1}=-a\otimes a\herm\otimes a\otimes a\herm\nn\\
\Cm=&\Cp\herm=a_{L1}\herm a_{R1}\herm a_{L2}a_{R2}=-a\herm\otimes a\otimes a\herm\otimes a\ .
\end{align}
The matrix 
\bel{CE23}
\widehat{C}=\Cp+\Cm
\ee
annihilates all states except
\bel{CE24}
\varphi_{3}=(1\ 0\ 1\ 0)\ ,\quad \varphi_{4}=(0\ 1\ 0\ 1)\ ,
\ee
which are interchanged
\bel{CE25}
\widehat{C}\pvec{\varphi_{3}}{\varphi_{4}}=-\pvec{\varphi_{4}}{\varphi_{3}}\ .
\ee
With
\bel{CE26}
\Shat_{C}(t)=\exp\gl-\frac{i\pi t}{2\eps}\widehat{C}\gr
\ee
one finds
\bel{CE27}
\Shat_{C}(t)\pvec{\varphi_{3}}{\varphi_{4}}=\cos\Big{(}\frac{\pi t}{2\eps}\Big{)}\pvec{\varphi_{3}}{\varphi_{4}}+i\sin\Big{(}\frac{\pi t}{2\eps}\Big{)}\pvec{\varphi_{4}}{\varphi_{3}}\ ,
\ee
while for all other states except $\varphi_{3}$, $\varphi_{4}$ this operator acts as unity. In particular, for $t=\eps$ one realizes the exchange~\eqref{79},
\bel{CE28}
\Shat_{C}(\eps)\pvec{\varphi_{3}}{\varphi_{4}}=i\pvec{\varphi_{4}}{\varphi_{3}}\ .
\ee

%Absatz
Since $\widehat{D}$ and $\widehat{C}$ act on different sets of states those operators commute,
\bel{CE29}
[\widehat{D},\widehat{C}]=0\ ,\quad \widehat{D}_{\pm}\widehat{C}_{\pm}=\widehat{C}_{\pm}\widehat{D}_{\pm}=0\ ,
\ee
resulting in
\bel{CE30}
\Shat_{i}(t)=\exp\Big{\{}-\frac{i\pi t}{2\eps}\gl\widehat{D}+\widehat{C}\gr\Big{\}}=\Shat_{D}(t)\Shat_{C}(t)\ .
\ee
For $t=\eps$ we define
\bel{CE31}
\Shat_{i}=\Shat_{i}(\eps)
\ee
which is unity for all states except $\varphi_{1}$, $\varphi_{2}$, $\varphi_{3}$, $\varphi_{4}$ for which it acts as
\bel{CE32}
\Shat_{i}
\begin{pmatrix}
\varphi_{1} \\ \varphi_{2} \\ \varphi_{3} \\ \varphi_{4}
\end{pmatrix}
=
\begin{pmatrix}
\varphi_{2} \\ \varphi_{1} \\ \varphi_{4} \\ \varphi_{3}\ .
\end{pmatrix}
\ee
This is precisely the local color exchange of the cellular automaton. Furthermore, both $\widehat{D}$ and $\widehat{C}$ involve even numbers of creation and annihilation operators, such that the even products of $T_{3}$ in eq.~\eqref{CE14} are unity. We conclude that $\Shat\subt{int}$ is a direct product of local factors
\bel{CE33}
\Shat\subt{int}=\Shat_{i}(x\inn)\otimes\Shat_{i}(x\inn+\eps)\otimes\Shat_{i}(x\inn+2\eps)\dots\ .
\ee

%Absatz
We can identify the interaction Hamiltonian in eq.~\eqref{CE6} as
\begin{align}
\label{CE34}
H\subt{int}=\sum_{x}\frac{\pi}{2\eps}\gl&\widehat{D}(x)+\widehat{C}(x)\gr\nn\\
=\frac{\pi}{2\eps}\sum_{x}\Big{\{}&a_{R2}\herm(x) a_{L1}\herm(x) a_{L2}(x)a_{R1}(x)\nn\\
+&a_{R1}\herm(x) a_{L2}\herm(x) a_{L1}(x)a_{R2}(x)\nn\\
+&a_{R2}\herm(x) a_{L2}\herm(x) a_{R1}(x)a_{L1}(x)\nn\\
+&a_{L1}\herm(x) a_{R1}\herm(x) a_{L2}(x)a_{R2}(x)\Big{\}}\ .
\end{align}
The interaction Hamiltonian is hermitian. In the form
\begin{align}
\label{CE35}
H\subt{int}=-\frac{\pi}{2\eps}\sum_{x}&\big{[}a_{R1}\herm(x)a_{R2}(x)-a_{R2}\herm(x)a_{R1}(x)\big{]}\nn\\
\times&\big{[}a_{L1}\herm(x)a_{L2}(x)-a_{L2}\herm(x)a_{L1}(x)\big{]}
\end{align}
the symmetries $L\leftrightarrow R$, as well as color exchange $1\leftrightarrow2$, are directly visible. We observe the prefactor $1/\varepsilon$. In distinction to the Grassmann variables we cannot absorb this factor by an arbitrary multiplicative renormalization of the annihilation and creation operators. Their normalization is fixed by the inhomogeneous anticommutation rule.
We will encounter later a different continuum normalization for which the anticommutator is $\sim\delta(x-y)$.

%Absatz
The interaction Hamiltonian is not uniquely fixed by eq.~\eqref{CE6}. We can add a piece
\bel{290A}
H\subt{int}'=\frac{2\pi}{\varepsilon}\sum_{x} f(x)\ ,
\ee
with $f(x)$ an integer function of 
\begin{align}\label{290B}
\widehat{n}_{R}(x)&=a_{R1}^{\dagger}(x)a_{R1}(x)+a_{R2}^{\dagger}(x)a_{R2}(x)\ , \nn\\
\widehat{n}_{L}(x)&=a_{L1}^{\dagger}(x)a_{L1}(x)+a_{L2}^{\dagger}(x)a_{L2}(x)\ .
\end{align}
Indeed, $\widehat{n}_{R}(x)$ and $ \widehat{n}_{L}(x)$ commute with $ H\subt{int}$, such that
\bel{290D}
\exp\big{[}-i\varepsilon(H\subt{int}+H\subt{int}')\big{]}=\widehat{S}\subt{int}\prod_{x}\exp\big{(}-2\pi i f(x)\big{)}=\widehat{S}\subt{int}\ .
\ee

\subsection{Right and left transport}
\label{subsec: Right and left transport}

We next turn to the free or kinetic step evolution operator $\Shat\subt{free}$. Despite the very simple structure of the right- or left-transport operators the expression of the Hamiltonian in terms of annihilation and creation operators needs some care. Otherwise the unique jump property is lost.

%Absatz
Since $\Shat\subt{free}$ is a product of independent factors for the left and right movers, and for the two colors, we can write
\begin{align}
\label{CE36}
\Shat\subt{free}=&\Shat_{1}^{(R)}\otimes\Shat_{2}^{(R)}\otimes\Shat_{1}^{(L)}\otimes\Shat_{2}^{(L)}\nn\\
H\subt{free}=&\sum_{a=1,2}\gl H_{a}^{(R)}+H_{a}^{(L)}\gr\ .
\end{align}
The two colors will be distinguished only by the label $a$ of the operators $a_{La}(x)$ etc. We will in the following not write the color labels explicitly.

%Absatz
In the appendix~\ref{ap.I} we proof the useful relations
\begin{align}
\label{CE37}
\Shat^{(R)}=&N\Big{[}\exp\Big{\{}\sum_{x}a_R\herm(x+\eps)\big{[}a_R(x)-a_R(x+\eps)\big{]}\Big{\}}\Big{]}\ ,\nn\\
\Shat^{(L)}=&N\Big{[}\exp\Big{\{}\sum_{x}a_L\herm(x-\eps)\big{[}a_L(x)-a_L(x-\eps)\big{]}\Big{\}}\Big{]}\ .
\end{align}
Here $N$ is an ordering operation for operators which is specified in app.~\ref{ap.I}. For this proof we employ the relations between functions of Grassmann variables and Grassmann derivatives on one side, and functions of annihilation and creation operators on the other side. They are explained in detail in the appendix~\ref{ap.A}.
We stress that eq.~\eqref{CE37} is an identity for matrices. It holds independently of the fermionic language used for the proof, such that it applies immediately to the cellular automaton.

\subsection{Partial continuum limit and loss of unique jump property}
\label{subsec: Partial continuum limit}

For the partial continuum limit the ordering operation for the exponential function can be omitted, while the orthogonality of $\Shat\subt{R,L}$ is maintained. In this limit one finds
\begin{align}
\label{CE63}
\Shat_R=&\exp\Big\{-\frac{1}{2}\sum_xa_R\herm(x)\big[a_R(x+\eps)-a_R(x-\eps)\big]\Big\}\ ,\nn\\
\Shat_L=&\exp\Big\{\frac{1}{2}\sum_xa_L\herm(x)\big[a_L(x+\eps)-a_L(x-\eps)\big]\Big\}\ .
\end{align}
The omission of the ordering operation $N$, which will be motivated below, has an important conceptual consequence. The matrices $\widehat{S}_{R,L}$ are no longer unique jump matrices. A given wave function $q_{\rho}(t)=\delta_{\rho,\overline{\rho}}$ corresponding to a sharp (deterministic) configuration $\overline{\rho}$ is mapped to a sum of nonzero entries for different configurations according to $q_{\tau}(t+\tilde{\varepsilon})=\widehat{S}_{\tau\rho}q_{\rho}(t)=\widehat{S}_{\tau\overline{\rho}}$.  Replacing eq.~\eqref{CE37} by eq.~\eqref{CE63} the evolution is no longer given by a simple updating rule for an automaton. We now encounter a feature characteristic for most quantum systems: a given state undergoes a probabilistic evolution to several different quantum states. In a double slit experiment an incoming particle with a given momentum can pass either in one or the other slit, or even combine both possibilities and produce the characteristic interference. This change of character of the evolution also occurs if we omit the commutator term $\Delta H$ in eq.~\eqref{CE7}.

%Absatz
Because of its conceptual relevance we discuss the partial continuum limit in some detail. Omitting only the ordering operation $N$ one has (not indicating the index $R$ for the fermionic operators)
\begin{align}
\label{CE64}
\Shat_R'=&\exp\Big\{-\sum_xa\herm(x)\big[a(x)-a(x-\eps)\big]\Big\}\ ,\nn\\
\gl\Shat_R^{\prime\text{T}}\gr^{-1}=&\exp\Big\{\sum_xa\herm(x)\big[a(x)-a(x+\eps)\big]\Big\}\ .
\end{align}
The operator $\Shat_R'$ is no longer orthogonal, since
\begin{align}
\label{CE65}
\sum_x\gl a\herm(x)a(x)\gr\supt{T}=&\sum_xa\herm(x)a(x)\ ,\nn\\
\sum_x\gl a\herm(x)a(x-\eps)\gr\supt{T}=&\sum_xa\herm(x-\eps)a(x)\nn\\
=&-\sum_xa\herm(x)a(x+\eps)\ .
\end{align}
The two expressions in eq.~\eqref{CE63} remain identical, however, if we assume coincidence of the two versions of lattice derivatives
\begin{align}
\label{CE66}
\partial_x^{(-)}a(x)=&\frac{1}{\eps}\big[a(x)-a(x-\eps)\big]\ ,\nn\\
\partial_x^{(+)}a(x)=&\frac{1}{\eps}\big[a(x+\eps)-a(x)\big]\ ,
\end{align}
which will be the case effectively if $\Shat_R$ acts on sufficiently smooth wave functions. We can take a partial continuum limit even for discrete $x$ if we take for the exponent the mean of the two expressions in eq.~\eqref{CE64}. This leads to $\Shat_R^{-1}=\Shat_R\supt{T}$ in eq.~\eqref{CE63}, and similar for $\Shat_L$.

%Absatz
The reason for omitting the ordering operation $N$ in the partial continuum limit is the observation that the difference between the ordered exponential in $\Shat_R$ and the standard exponential $\Shat_R'$ only arises from the anticommutators of $\big\{a\herm(x),a(y)\big\}$ at identical points $x=y$. For $x\neq y$ the anticommutator vanishes, similar to $\big\{\psibar(x),\psi'(y)\big\}=0$. The ordering operation has therefore no effect if $x$ and $y$ are sufficiently distant from each other. Consider now a smooth wave function $\tilde{q}$ that varies only over distances $L\gg\eps$. Only this variation will lead in eq.~\eqref{CE37} to a difference of $\Shat_R$ from the unit operator. The sums $\sum_{xy}$ for $T_2$ in eq.~\eqref{CE60} will extend effectively over large ranges of $x-y$, typically $|x-y|>L$. The contribution of strictly identical or neighboring points $y=x$ or $y=x-\eps$, that give rise to $T_{2,\text{c}}$ in eq.~\eqref{CE62}, is suppressed as compared to the total sum by a factor $\eps/L$. It vanishes in the continuum limit $\eps/L\to0$. This argument extends in a similar way to higher orders in the expansion of the exponential. In summary, for sufficiently smooth wave functions the step evolution operator does not retain the detailed information on space differences of the order $\eps$. The smoothness of the probabilistic information, as encoded in a smooth wave function $\tilde{q}$, plays a central role here. As our discussion above shows, the approximation~\eqref{CE63} fails if we consider very sharp wave functions, as the one for a single particle located at $t$ precisely at the position $z$. A systematic discussion of the continuum limit can be found in the appendix~\ref{ap.J}.

%Absatz
As mentioned before, the transition to the partial continuum limit has an important conceptual implication. While $\Shat\subt{R,L}$ remain orthogonal operators, they are no longer unique jump operators. The action of $\Shat\subt{R,L}$ in eq.~\eqref{CE63} does not map a given bit configuration at $t$ uniquely to a new bit configuration at $t+\epstil$. The partial continuum limit of a probabilistic cellular automaton has no longer a deterministic time evolution. The step evolution operator $\Shat$ becomes genuinely probabilistic in the partial continuum limit, mapping a given bit configuration only with certain probabilities to new bit configurations. The different bit configurations to which $\Shat$ maps differ only on distances of the order $\eps$. For a smooth wave function they are sufficiently close to each other such that their difference in a suitable continuum limit plays no role. In a certain sense, the continuum limit is a coarse graining for which the resolution on distances $\varepsilon$ is lost. Nevertheless, following the evolution for many time steps the loss of determinism on the coarse grained level can have important consequences.

\subsection{Kinetic Hamiltonian and continuum limit}
\label{subsec: Kinetic Hamiltonian and continuum limit}

Combining eqs.~\eqref{CE63} with eq.~\eqref{CE6} we can directly extract the kinetic Hamiltonian $H\subt{free}$,
\begin{align}
\label{CE67}
H\supt{(R)}=&-\frac{i}{2\eps}\sum_{x}a_R\herm(x)\big[a_R(x+\eps)-a_R(x-\eps)\big]\ ,\nn\\
H\supt{(L)}=&\frac{i}{2\eps}\sum_{x}a_L\herm(x)\big[a_L(x+\eps)-a_L(x-\eps)\big]\ .
\end{align}
As it should be, the Hamiltonian is hermitian.
In terms of the lattice derivative
\bel{CE68}
\partial_xa(x)=\frac{1}{2\eps}\big[a(x+\eps)-a(x-\eps)\big]
\ee
one obtains
\bel{CE69}
H\supt{(R,L)}=\mp i\sum_xa\subt{R,L}\herm(x)\partial_xa\subt{R,L}(x)\ .
\ee

%Absatz
We recognize a type of lattice momentum operator $-i\partial_x$. Indeed, we can make a lattice Fourier transform
\bel{CE70}
a(x)=M_x^{-\tfrac{1}{2}}\sum_pe^{ipx}a(p)\ ,
\ee
with $M_x$ the number of $x$-points and discrete periodic momenta (period $2\pi/\eps$) which we take in the range ($m$ integer)
\bel{CE71}
p=\frac{2\pi m}{\eps M_x}\ ,\quad |p|\leq\frac{\pi}{\eps}\ .
\ee
With
\bel{CE72}
\sum_xe^{ix(p-p')}=M_x\delta_{p,p'}\ ,\quad \sum_pe^{-ip(x-y)}=M_x\delta_{x,y}\ ,
\ee
the inverse reads
\bel{CE73}
a(q)=M_x^{-\tfrac{1}{2}}\sum_xe^{-ipx}a(x)\ .
\ee

%Absatz
The annihilation and creation operators in the momentum basis obey the usual anti-commutation relations
\bel{CE74}
\big\{a\herm(p),a(q)\big\}=\delta_{p,q}\ ,\quad \big\{a(p),a(q)\big\}=\big\{a\herm(p),a(q)\herm\big\}=0\ .
\ee
In the momentum basis one obtains the simple relation
\bel{CE75}
H\supt{(R)}=\sum_pp\;a_R\herm(p)a_R(p)\ ,\quad H\supt{(L)}=-\sum_pp\;a_L\herm(p)a_L(p)\ ,
\ee
where we employ here the continuum approximation for the definition of the derivative $\partial_x$,
\bel{CE76}
-i\partial_xa(x)=M_x^{-\tfrac{1}{2}}\sum_pp\;e^{ipx}a(p)\ .
\ee
For free right movers one finds the relativistic energy-momentum relation (dispersion relation) $E=P$, while free left movers obey $E=-P$. Momentum eigenstates with positive energy $E>0$ obey $P=p>0$ for right movers and $P=p<0$ for left movers. Correspondingly, for a single right handed or left handed particle the plane wave solutions are
\bel{CE77}
\varphi_R\supt{(1)}\sim e^{ip(x-t)}\ ,\ p>0\ ,\ \ \varphi_L\supt{(1)}\sim e^{ip(x+t)}\ ,\ p<0\ .
\ee

%Absatz
As mentioned above, we can neglect $\Delta H$ for the continuum limit in the time direction. The density matrix obeys then the continuous von-Neumann equation with Hamiltonian
\begin{align}
\label{CE78}
H=&-\frac{i}{2\eps}\sum_x\Big\{a_{Ra}\herm(x)\big[a_{Ra}(x+\eps)-a_{Ra}(x-\eps)\big]\nn\\
&\quad\quad\quad-a_{La}\herm(x)\big[a_{La}(x+\eps)-a_{La}(x-\eps)\big]\Big\}\nn\\
&-\frac{\pi}{2\eps}\sum_x\big[a_{R1}\herm(x)a_{R2}(x)-a_{R2}\herm(x)a_{R1}(x)\big]\nn\\
&\quad\quad\quad\times\big[a_{L1}\herm(x)a_{L2}(x)-a_{L2}\herm(x)a_{L1}(x)\big]\ .
\end{align}

%Absatz
We can further take the continuum limit in the $x$-direction, with $\sum_x=\eps^{-1}\int\text{d}x$, replacing the lattice derivative by a partial derivative,
\begin{align}
\label{CE79}
H=\int\text{d}x&\Big\{-\frac{i}{\eps}\big[a_{Ra}\herm(x)\partial_xa_{Ra}(x)-a_{La}\herm(x)\partial_xa_{La}(x)\big]\nn\\
-\frac{\pi}{2\eps^{2}}&\big[a_{R1}\herm(x)a_{R2}(x)-a_{R2}\herm(x)a_{R1}(x)\big]\nn\\
\times&\big[a_{L1}\herm(x)a_{L2}(x)-a_{L2}\herm(x)a_{L1}(x)\big]\Big\}\ .
\end{align}
The factors of $\eps$ can be absorbed by a renormalization
\bel{CE80}
a\subt{ren}(x)=\frac{1}{\sqrt{\eps}}a(x)\ ,
\ee
which corresponds to the continuum normalization of the anticommutator relations
\bel{CE81}
\big\{a\subt{ren}\herm(x),a\subt{ren}(y)\big\}=\frac{1}{\eps}\delta_{x,y}\sim\delta(x-y)\ .
\ee

%Absatz
The evolution of the complex density matrix for the probabilistic cellular automaton follows the von-Neumann equation for a fermionic quantum field theory. In case of a pure state density matrix $\rho^{2}=\rho$ we can express it in terms of a complex wave function $\varphi$
\bel{CE82}
\rho_{\alpha\beta}=\varphi_\alpha\varphi_{\beta}^*\ ,
\ee
which evolves according to the Schrödinger equation
\bel{CE83}
i\partial_t\varphi=H\varphi\ .
\ee
We can express the continuous evolution in the real formulation as well. For this purpose we replace $i$ by the matrix~$I$ which maps $q'$ to $q^{c}$, resulting in 
\bel{327A}
\partial_{t}q=-I\widehat{H}q=Wq=(W\subt{kin}+W\subt{int})q\ ,
\ee
where $\widehat{H}$ obtains from $H$ in eq.~\eqref{CE79} by the replacement $i\rightarrow I$. The kinetic part is block-diagonal in $q'$ and $q^{c}$
\bel{327B}
W\subt{kin}=-\frac{1}{\varepsilon}\int_{x}\Big{\lbrace}a_{Ra}^{\dagger}(x)\partial_{x}a_{Ra}(x)-a_{La}^{\dagger}(x)\partial_{x}a_{La}(x)\Big{\rbrace}\ ,
\ee
while the interaction part $W\subt{int}$ involves a factor $I$ and therefore rotates between $q'$ and $q^{c}$.

%Absatz
For understanding the approximations involved in the continuum limit it is instructive to integrate the Schrödinger equation~\eqref{327A} from $t$ to $t+\varepsilon$ under the assumption of slowly varying $q(t)$,
\bel{327C}
q(t+\varepsilon)=q(t)+\delta q(t)=\exp (\varepsilon W) q(t)=(1+\varepsilon W)q(t)+O(\varepsilon^{2})\ .
\ee
We focus first on a right-mover of a free theory where in a discrete setting for $x$ one has
\bel{327D}
\varepsilon W=\frac{1}{2}\sum_{x}\Big{(}a^{\dagger}(x+\varepsilon)a(x)+a(x+\varepsilon)a^{\dagger}(x)\Big{)} \ .
\ee
For each term in the sum the wave function changes only for those configurations for which the occupation numbers at $x+\varepsilon$ and $x$ are different. If the components of the wave functions for these pairs of configurations differ in relative size only by $\varepsilon$ the change $\delta q / q$ is of the order $\varepsilon$. In other words, the wave function should be smooth enough such that the derivative operator, 
\begin{align}
\label{327E}
W&=-\frac{1}{2\varepsilon}\sum_{x}a^{\dagger}(x)\Big{[}a(x+\varepsilon)-a(x-\varepsilon)\Big{]}\nn\\
=-&\sum_{x}a^{\dagger}(x)\partial_{x}a(x)=\frac{1}{2\varepsilon}\sum_{x}\Big{[}a^{\dagger}(x+\varepsilon)-a^{\dagger}(x-\varepsilon)\Big{]}a(x)\nn\\
&=\sum_{x}\partial_{x}a^{\dagger}(x)a(x)=-W^{T}\ ,
\end{align}
can be regarded as a quantity of the order one, not diverging for $\varepsilon\rightarrow 0$. In the continuum limit we can then neglect the term $\sim O(\varepsilon^{2})$ in eq.~\eqref{327C}. Since $W$ is an antisymmetric matrix, $\widehat{S}_{R,c}=\exp(\varepsilon W)$ is an orthogonal matrix. In the appendix~\ref{ap.J} we present the operator $W(t)$ and its properties in more detail.

%Absatz
The continuous step evolution operator, 
\bel{327F}
\widehat{S}_{R,c}=1+\varepsilon W \ ,
\ee
differs in structure from the original right-transport operator $\widehat{S}_{R}$. It is now given by unity with an additional contribution $\varepsilon W$ that is considered to be small in the sense that its action on smooth wave functions produces only a small effect. It is no longer a unique jump operator since it has both diagonal entries from the term one, and off-diagonal entries from the term $\varepsilon W$. Nevertheless, for sufficiently smooth wave functions the difference between the action of $\widehat{S}_{R}$ and $\widehat{S}_{R, c}$ on $q$ is of the order $\varepsilon^{2}$ and can be neglected in the continuum limit.

%Absatz
A second condition on wave functions for which a continuum limit is valid requires that the commutator $[ H\subt{int},\, H\subt{free}]$ can be considered as a quantity that remains finite for $\varepsilon\rightarrow 0$ (or diverges less fast than $\varepsilon^{-1}$). In this case the neglection of $\Delta H$ in eq.~\eqref{CE7} is justified. It can happen that $H\subt{int}$ diverges for $\varepsilon\rightarrow 0$, partially depending on the choice of $H\subt{int}'$  in eq.\eqref{290A}. As long as the divergent part commutes with $H\subt{free}$ the continuum limit remains valid. The divergent part of $H\subt{int}$ may lead to overall phase rotations of wave functions on a time scale given by $\varepsilon$ which can be factored out. Typically, the integrated continuous evolution operator
\bel{327G}
\widehat{S}^{c}=\exp\big{[}-i\varepsilon(H\subt{int}+H\subt{free})\big{]}\ ,
\ee
is no longer a unique jump matrix, in contrast to $\widehat{S}$. A given configuration can now either propagate or scatter, reproducing the same result as the automaton only for sufficiently smooth wave functions. 

%Absatz
We conclude that the continuum limit for sufficiently smooth wave functions corresponds to some type of coarse graining. It looses the distinction between two neighboring configurations which may evolve differently by applying the updating rule of the automaton. The two configurations are treated as identical, with certain probabilities to follow either the updating of the first or the second configuration. The continuum limit is no longer an automaton.

\section{Ground states and one particle wave functions}
\label{sec: 08}

Many concepts of quantum field theories apply directly to probabilistic cellular automata. These include the notions of ground state and one-particle excitations or conserved quantities as momentum and charge. On the level of the density matrix for the cellular automaton these features can be discussed in complete analogy to other quantum systems. In the formulation as cellular automata many perhaps somewhat abstract concepts in fermionic quantum field theories find a rather concrete intuitive realization.

\subsection{Ground states}
\label{subsec: Ground states}

Possible ground states correspond to a stationary density matrix. In case of a classical pure state they correspond to time-translation invariant classical wave functions $\qtil(t)$, $\qbar(t)$. We will define here translation invariance by shifts $2\eps$. Thus the criterion for a ground state reads
\begin{align}
\label{GS1}
\widehat{\rho}(t+2\eps)=&\widehat{\rho}(t)\ ,\quad \rho(t+2\eps)=\rho(t)\ ,\nn\\
\qtil(t+2\eps)=&\qtil(t)\ ,\quad \qbar(t+2\eps)=\qbar(t)\ .
\end{align}
First obvious candidates for ground states are the completely empty state, $n_\gamma(t,x)=0$, or the completely filled state $n_\gamma(t,x)=1$.

\begin{figure}[h]
%\resizebox{18pc}{!}{%
%\vspace*{-7pt}
\centering\includegraphics[width=3in]{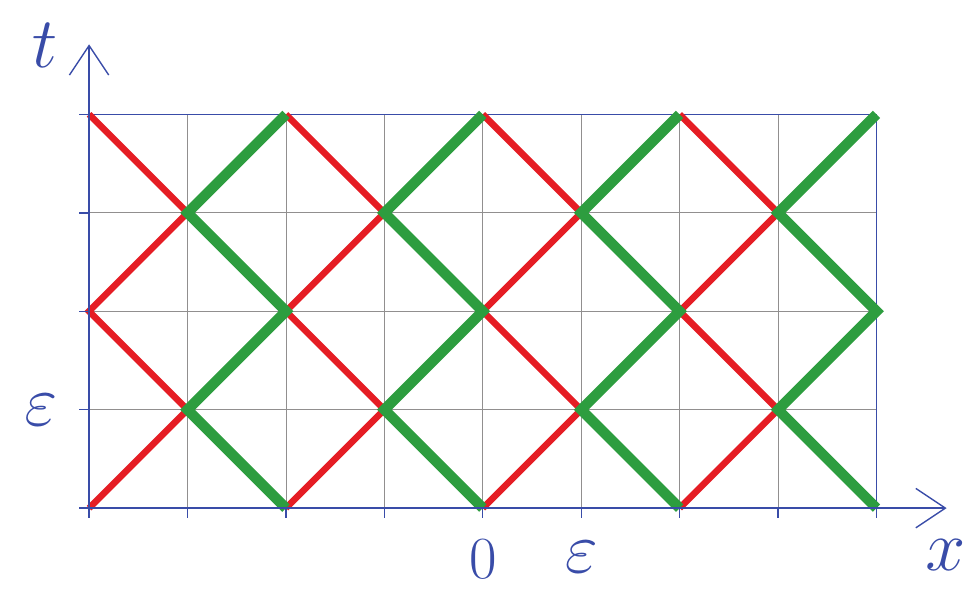}
%%% where xxxxxx name represents "figurename.eps"
\vspace*{-5pt}
%\caption{captionxxx.}\label{Fig.2}
%\end{figure}
%\begin{figure}[h]
%\centering
%\includegraphics[scale=0.10]{fig1}
\caption{Configuration for vacuum $(A)$.}\label{Fig.G1}
\end{figure}

Other possible ground states are half-filled states. For even $t$ we can have at each position $x$ one green left mover and one red right mover. At $t+\eps$ all particles switch color, and at $t+2\eps$ the configuration turns back to the configuration at $t$. Such a state is depicted in Fig.~\ref{Fig.G1}.
%Fig2 HERE (Fig. 4 from 10.1016/j.nuclphysb.2020.115296
%\resizebox{18pc}{!}{\begin{figure}[!h]

%Absatz
There are only one-particle lines, no doubly occupied or empty lines.
We may denote the wave function for this state as $\qtil^{(A)}$, and for a similar state with exchanged color by $\qtil^{(B)}$. We will take for simplicity positive $\qtil^{(A)}(t\inn)$, which equals one for the half-filled configuration $\tau_{A}$ described above, and vanishes for all other $\tau\neq\tau_{A}$, $\qtil_{\tau}^{(A)}(t\inn)=\delta_{\tau,\tau_{A}}$.
We infer
\bel{GS2}
\qtil(t\inn+2m\eps)=\qtil^{(A)}\ ,\quad \qtil(t\inn+(2m+1)\eps)=\qtil^{(B)}\ .
\ee
We note the existence of a similar, but not identical, ground state with $\qtil(t\inn)=\qtil^{(B)}$. The color-exchange symmetry maps one ground state $(A)$ to the other ground state $(B)$. For any given ground state $(A)$ or $(B)$ it is spontaneously broken.

%Absatz
A further half-filled ground state $(C)$ has at $t\inn$ all sites occupied by one red left mover and one red right mover, with associated wave function $\qtil(t\inn)=\qtil^{(C)}$.
Again, by a switch of color of all particles we define the state $\qtil^{(D)}$, with
\bel{GS3}
\qtil(t\inn+2m\eps)=\qtil^{(C)}\ ,\quad \qtil(t\inn+(2m+1)\eps)=\qtil^{(D)}\ .
\ee
The ground state with $\qtil(t\inn)=\qtil^{(D)}$ is closely related, but not identical. Again, any given half-filled ground state $(C)$ or $(D)$ breaks the color exchange symmetry spontaneously.

%Absatz
For the ground states we take the conjugate wave function equal to the classical wave function, $\qbar^{(0)}(t)=\qtil^{(0)}(t)$. This results in a diagonal real density matrix with only one non-vanishing element,
\bel{GS4}
\widehat{\rho}_{\tau\rho}(t)=\rho_{\tau\rho}(t)=\delta_{\tau\rho}\delta_{\tau,\bar{\tau}(t)}\ ,
\ee
with $\bar{\tau}(t)=\tau_{A}$ for $t$ even and $\bar{\tau}(t)=\tau_{B}$ for $t$ odd, and similarly for the other ground states.
For the ground state $\bar\tau(t\inn)=\tau_A$ the ground state wave function obeys
\begin{align}
\label{GS5}
\qtil_\tau^{(0)}(t)=&\qbar_\tau^{(0)}(t)=\delta_{\tau,\tau_A}\ \text{for}\ t=t\inn+2m\eps\nn\\
\qtil_\tau^{(0)}(t)=&\qbar_\tau^{(0)}(t)=\delta_{\tau,\tau_B}\ \text{for}\ t=t\inn+(2m+1)\eps\ .
\end{align}

%Absatz
None of these ground states is invariant under the particle-hole transformation. The completely empty and completely filled states are mapped onto each other, and for the half-filled ground states the particle hole transformation maps $(A)\leftrightarrow (B)$ and $(C)\leftrightarrow (D)$. We can define the wave function of a particle-hole symmetric ground state by the superposition
\bel{321A}
q_{0}(t)=\frac{1}{\sqrt{2}}\bigl{(}q^{(A)}+q^{(B)}\bigr{)}\ ,
\ee
where we identify $\tilde{q}=\bar{q}=q$. For this ground state the probability for the configurations~$\tau_{A}$ and $\tau_{B}$ are both one half. This ground state is invariant under color exchange and obeys $q_{0}(t+\varepsilon)=q_{0}(t)$. If we want to define complex conjugation by the particle-hole transformation, the ground state has a real wave function in the complex language. Similar particle-hole invariant ground states can be obtained by combining $(C)$ and $(D)$ or the totally empty and filled states. One may also consider the half-filled ground state
\bel{321B}
q_{0}(t)=\frac{1}{2}\bigl{(} q^{(A)}+q^{(B)}+q^{(C)}+q^{(D)}\bigr{)}\ ,
\ee
with probability~$1/4$ for the configurations~$\tau_{A}$, $\tau_{B}$, $\tau_{C}$ and $\tau_{D}$.

%Absatz
Finally, another interesting ground state is the half-filled equipartition state. Let us denote by~$\tau{}_{\ i}^{(E)}$ the configurations which have precisely one right-mover and one left-mover on each site $x$.
There are $4^{M_{x}}$ such configurations since at each site there are four possibilities: The right mover can be red or green, and the same holds for the left-mover. With wave functions $q_{i}$ defined by $(q_{i})_{\tau}=\delta_{\tau , \tau{}_{\  i}^{(E)}}$, the equipartition wave function obeys 
\bel{321C}
q^{(E)}(t)=2^{-M_{x}}\sum_{i}q_{i} \ .
\ee
The probabilities to find a given half-filled configuration $\tau{}_{\ i}^{(E)}$ are all equal and given by $4^{-M_{x}}$. The half-filled equipartition state is invariant under the particle-hole transformation. The latter maps each configuration $\tau{}_{\ i}^{(E)}$ to another configuration $\tau{}_{\ j}^{(E)}$ which obtains by switching the color of all particles.

%Absatz
We can start at $t\subt{in}$ with the equipartition~\eqref{321C}. The time evolution of any given configuration $\tau{}_{\ i}^{(E)}$ is rather simple. At every point of the $(t,x)$-lattice a left moving single-particle line crosses a right-moving single particle line. For each given left-moving or right-moving single-particle line the color changes therefore at every time step. In particular, at $t+2\varepsilon$ every color of a right mover at $(t, x)$ is displaced to $(t+2\varepsilon , x+2\varepsilon)$, while the color of a left-mover at $(t,x)$ is found at $(t+2\varepsilon , x- 2\varepsilon)$. This is again one of the configurations $\tau{}_{\ k}^{(E)}$, such that the equipartition wave function is stationary. Actually the equipartition wave function keeps the same value~\eqref{321C} for all $t$ since the additional switch of color from $t$ to $t+\varepsilon$ still remains within the space of configurations $\tau{}_{\ i}^{(E)}$.

%Absatz
The continuum Hamiltonian~\eqref{CE78} constitutes a map within the space of wave functions for configurations with one right-mover and one left-mover at each $x$. The eigenvalues of the Hamiltonian $H_{0}$ restricted to this space correspond to possible vacuum energies. A vacuum wave function that is an eigenstate of $H_{0}$ undergoes a phase rotation with the corresponding eigenvalue. This overall phase rotation can be factored out. If the continuum limit is exact for a given vacuum state the relative phase between $t$ and $t+\varepsilon$ equals zero, according to $q_{0}(t+\varepsilon)=q_{0}(t)$.

\subsection{One particle wave function}
\label{subsec: One particle wave function}

%Absatz
Single particle states are, in general, a complex issue. In a quantum field theory the notion of a single particle corresponds to some local excitation of a vacuum or ground state. Its properties depend on the particular vacuum. For our very simple automaton the task of defining single-particle states is facilitated by the property that each particle line continues as either a right-mover or a left-mover. Only its color can change by scattering. For a vacuum with a fixed particle number we can define a single-particle configuration at $t\subt{in}$ by adding one occupied bit to the vacuum. This must be a right-mover or a left-mover at some position $x$. For vacua with precisely one right-mover and one left-mover at every position $x$ we can follow the trajectory of the ``surplus" bit or particle by spotting at each $t$ the location where three bits are occupied. The corresponding trajectories have to be on the diagonal corresponding to a right-mover or a left-mover. The particle defined in this way undergoes no change of direction. (This can differ if one chooses another form of the interaction.) Since the particle moves with light velocity, no mass term is induced by some form of spontaneous symmetry breaking. The same argument holds for single-hole configurations. 

%Absatz
Despite its simplicity, several interesting features of one-particle states are visible in our model. 
One possibility to define a real one-particle wave function $q^{(1)}(t)$ is by the action of a single creation operator on the ground state
\bel{GS6}
q_\tau^{(1)}(t)=\mathcal{N}\sum_{x,\gamma}q\g^{(1)}(t,x)\gl a\g\herm(x)\gr_{\tau\rho}q_\rho^{(0)}\ .
\ee
The coefficients $q\g^{(1)}(t,x)$ are the one-particle wave function in the occupation number basis. They can be seen as a $4M_x$-component real vector, or as a four-component field which depends on a single discrete space coordinate $x$. 
The normalization factor~$\mathcal{N}$ is introduced to ensure the normalization
\bel{322A}
\sum_{x,\gamma}\bigl{(}q_{\gamma}^{(1)}(t,x)\bigr{)}^{2}=1 \ .
\ee

%Absatz
The definition~\eqref{GS6} is appropriate if the evolution according to eq.~\eqref{PC04} does not change this form of the wave function.
The step evolution operator $\Shat$ of our model does not change the particle number. For ground states with one right-mover and one left-mover on each site the vectors $\gl a\g\herm(x)\gr_{\tau\rho}q_\rho^{(0)}$ form a complete basis of all one-particle states. Then we can write $q^{(1)}(t)$ in the form~\eqref{GS6} for arbitrary $t$. The projection $\Shat_{\gamma\delta}(x,y)$ of the step evolution operator $\Shat$ on the one-particle states follows directly from applying the updating rules on the corresponding one-particle bit configurations.
We can similarly define the single-hole wave function $q^{(c,1)}(t)$ by replacing in eq.~\eqref{GS6} the creation operator $a_{\gamma}^{\dagger}(x)$ by the annihilation operator $a_{\gamma}(x)$, and the coefficients $q_{\gamma}^{(1)}(t,x)$ by the one-hole wave function $q_{\gamma}^{(c,1)}(t,x)$. We may define a generalized one-particle configuration by either an additional bit of type $\gamma$ present or absent at $x$. The corresponding one-particle wave function is the pair $\big{(}q_{\gamma}^{(1)}(t,x), q_{\gamma}^{(c,1)}(t,x)\big{)}$. In this case we adapt the normalization~\eqref{322A} correspondingly
\bel{348A}
\sum_{x,\gamma}\Big{[}\big{(}q_{\gamma}^{(1)}(t,x)\big{)}^{2}+\big{(}q_{\gamma}^{(c,1)}(t,x)\big{)}\Big{]}=1\ .
\ee
The continuum evolution connects $q^{(1)}$ and $q^{(c,1)}$ due to the factor $I$ in the interaction part $W\subt{int}$ in eq.~\eqref{327A}.

%Absatz
For the empty ground state a single particle does not undergo scatterings. The evolution is therefore simple: a single right-moving particle evolves on a straight trajectory to increasing $x$, without changing its color. Similarly, a single left-mover follows a straight trajectory with decreasing $x$.
For half-filled ground states the situation could, in principle, be more complex since scattering could induce a change of color of the generalized particle. For our simple automaton this issue is solved easily for ground states with one right-mover and one left-mover on each site. The single particle encounters a one-particle line at each time step, such that the additional bit or hole changes color at each time step.

%Absatz
The one-particle bit configurations depend on the ground state. For the empty ground state one has a single occupied bit moving on a straight line without scattering, either as a right mover or a left mover, and either red or green.
Adding a particle to the half-filled vacuum $(A)$ one produces a double occupied line. The particles that can be added for the initial state are not arbitrary. At a given position we can only add a green right mover or a red left mover, since for the ground state $(A)$ the green left mover and the red right mover is already occupied. The only non-zero components of the one-particle wave function~\eqref{GS6} for $q^{(0)}=q^{(A)}$ occur for the corresponding values of $\gamma$. The one-particle states involve only half the numbers of species as compared to the empty ground state.

\begin{figure}[h]
%\resizebox{18pc}{!}{%
%\vspace*{-7pt}
\centering\includegraphics[width=3in]{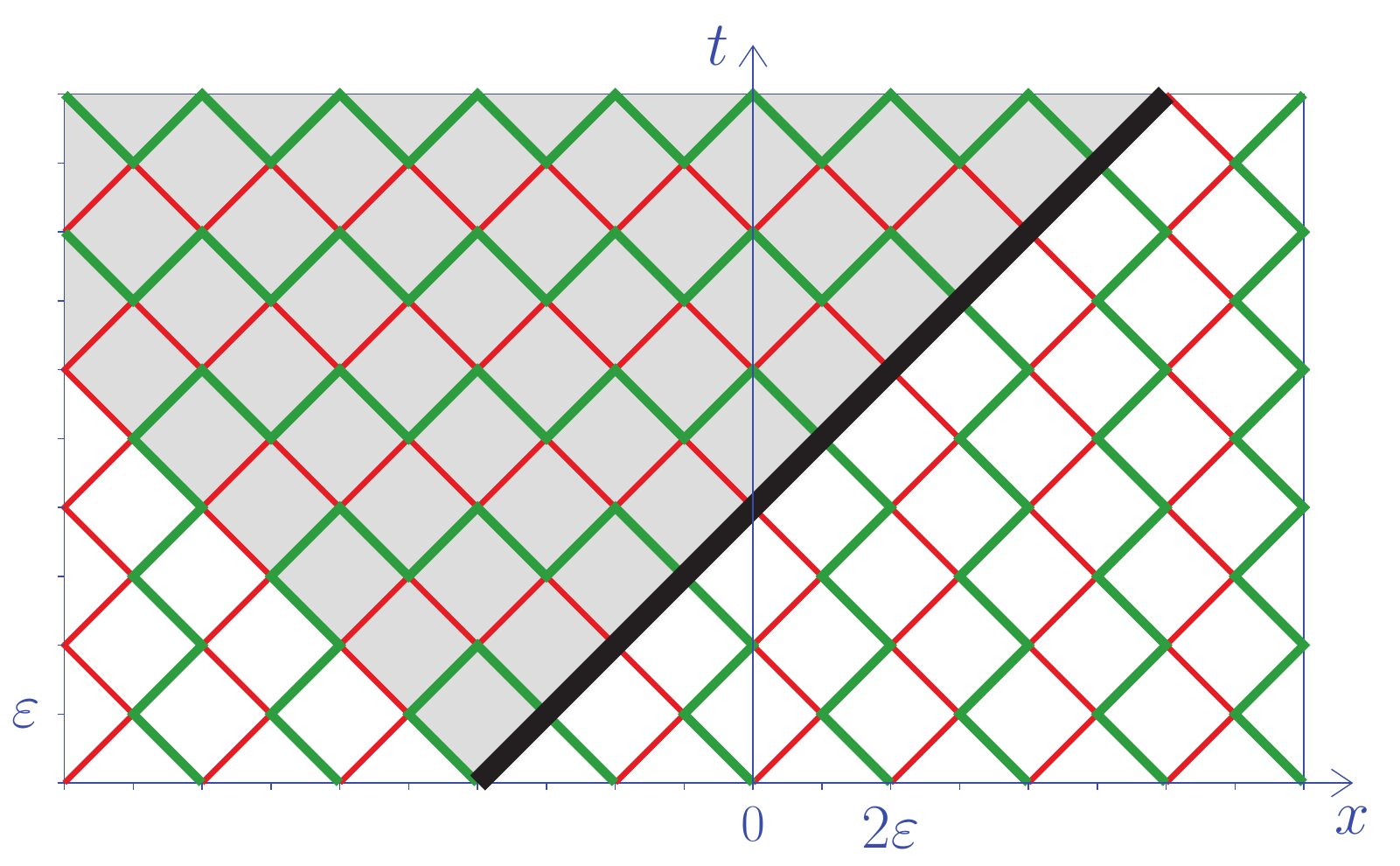}
%%% where xxxxxx name represents "figurename.eps"
\vspace*{-5pt}
%\caption{captionxxx.}\label{Fig.2}
%\end{figure}
%\begin{figure}[h]
%\centering
%\includegraphics[scale=0.10]{fig1}
\caption{Soliton excitation. The black line is doubly occupied. To the left of the doubly occupied line, within the shaded ``light cone", we observe the vacuum $(C)$. Outside this light cone one observes the vacuum (A). The doubly occupied line can be considered as a soliton separating different vacuum states.}\label{fig: G2}
\end{figure}

%Absatz
Adding at $t=0$ to the ground state $(A)$ a green right mover produces the initial point for a double occupied line with one red and one green right mover. This line propagates to the right without scattering, similar to the one-particle lines for the empty ground state. We have depicted the evolution of the corresponding spin configuration in Fig.~\ref{fig: G2}. Similarly, adding a red left mover produces a doubly occupied left moving line without further scattering.
%\resizebox{18pc}{!}{\begin{figure}[!h]

%Absatz
With increasing $t$ the form ~\eqref{GS6} of the one-particle wave function is not preserved. We observe to the left of the double-occupied line the appearance of a new vacuum structure, namely vacuum~$(C)$. The vacuum~$(C)$ is found within a light-cone spreading out from its origin at t=0. 
Outside the light cone one finds the vacuum $(A)$. The double-line of a particle created at $t=0$ is therefore accompanied by a string of alternating red or green particles extending to the other border of the light cone. The wave function for this configuration for $t>0$ can no longer be described by the action of creation operators at various positions $x$, with invariant vacuum otherwise.
%Absatz(?)
The half-filled vacua $(A)$, $(B)$, $(C)$, $(D)$ can actually be present in different regions of space, with appropriate double-occupied or empty lines at the boundaries. In this context double-occupied or empty lines can be interpreted as solitons. With the understanding that to the right and left of the double-occupied or empty lines there are different vacua we can still describe the propagation of the soliton by a suitable one-particle wave function $q_{L, R}^{(1)}(t,x)$.

%Absatz
Finally, for the half-filled equipartition vacuum $(E)$ a single particle constructed according to eq.~\eqref{GS6} corresponds to a double-occupied line with the same vacuum on both sides. Whatever is the distribution of colors for the right-and left-movers for all points $x$ away from the double-occupied line at $t$, there is a corresponding color distribution at $t+\varepsilon$. Since all color distributions have the same weight for the vacuum wave function $q^{(E)}$ in eq.~\eqref{321C}, the wave function differs from the vacuum only at the position $x$ of the double-occupied line, and by a possible normalization factor $\mathcal{N}$.

%Absatz
There are only two different one-particle states of this type - one right-moving and the other left-moving. The double-occupied line involves both a red and a green particle and is therefore "color blind". More in detail, the creation operator~ $a_{R1}\herm(x)$ in eq.~\eqref{321C} creates for all $\tau_{\ i}^{(E)}$ with a green right-mover at $x$ a site with both a red and a green right mover, while it yields zero if $\tau_{\ i}^{(E)}$ involves a red-right mover at $x$. It yields a non-zero result only for half of the $q_{i}^{(E)}$. The creation operator $a_{R2}\herm(x)$ yields a non-zero result for the other half of the $q_{i}^{(E)}$, producing the same double-occupied site at $x$.

%Absatz
We can also construct single-particle states for composite particles. As an example, a composite single-particle excitation in a given vacuum, say vacuum $(A)$, can be constructed by a double-occupied line with a neighboring parallel empty line. This replaces in eq.~\eqref{GS6} the operator~$a\g\herm(x)$ by $a\g\herm(x) a_{\delta}(x+2\varepsilon)$, and correspondingly $q\g ^{(1)}(t,x)$ by $q_{\gamma\delta} ^{(1)}(t,x) $. Outside these two lines the vacuum is not modified. The possible values for $(\gamma , \delta )$ are restricted. For a right-moving particle in vacuum~$(A)$ one has $\gamma=(R,2)$ and $\delta=(R,1)$, while the left-moving particle has $\gamma=(L,1)$, $\delta=(L, 2)$. Other particles or holes are two parallel neighboring double-occupied lines, or two parallel empty lines. Further more extended objects can separate the two parallel lines by a certain distance, with a string of a different vacuum inbetween the two lines.

\subsection{Charges}
\label{subsec: Charges}

For the half filled configuration $\tau^{(A)}$ we may take away a single red right-moving particle. This produces a red hole. The corresponding empty line replaces the double occupied line in Fig.~\ref{fig: G2}. Otherwise Fig.~\ref{fig: G2} remains unchanged. A right-moving red hole propagates in the same way as a green right-moving particle. Similarly, a green left-moving hole behaves as a red left-moving particle.
This generalizes to the other half-filled configurations $\tau_{\ i}^{(E)}$. Right- and left-moving holes can be described similarly to eq.\eqref{GS6} with $a\g\herm (x)$ replaced by the annihilation operator $a\g (x)$.

%Absatz
One may treat both the one-particle state and the one-hole state as a generalized particle, and assign charge $Q=1$ to particles and $Q=-1$ to holes. With total particle number
\bel{GS7}
N_p=\sum\g\sum_x n\g(x)
\ee
taking in the half-filled vacuum the value $N_0=2M_x$ one has
\bel{GS8}
Q=N_p-N_0=N_p-2M_x\ .
\ee
Since $N_p$ is conserved, also the charge $Q$ is conserved.

%Absatz
Instead of left and right moving particles and holes we may speak about left and right moving particles with opposite charge $Q$. For the half-filled vacuum all these generalized one-particle states move on straight lines without scattering. They are rather similar to the one-particle states for the empty vacuum.

%Absatz
We can introduce separate charges $Q_R$ and $Q_L$ for the right movers and left movers
\bel{Q1}
Q_R=N_{R1}+N_{R2}-M_x\ ,\quad Q_L=N_{L1}+N_{L2}-M_x\ ,
\ee
with
\bel{Q2}
Q=Q_R+Q_L\ .
\ee
The charges $Q_R$ and $Q_L$ are conserved separately. The assignment of particle numbers and charges to the different half-filled ground states is indicated in Table~\ref{table:1}.
Similarly, all half-filled configurations $\tau_{\ i}^{(E)}$ have $Q_{R}=Q_{L}=0$.
\begin{table}[ht]
\begin{center}
\begin{tabular}{ |c|c c c c|c c| } 
\hline
Vacuum & $N_{R1}$ & $N_{R2}$ & $N_{L1}$ & $N_{L2}$ & $Q_{R}$ & $Q_{L}$ \\
\hline
$A$ & $M_x$ & $0$ & $0$ & $M_x$ & $0$ & $0$ \\ 
$B$ & $0$ & $M_x$ & $M_x$ & $0$ & $0$ & $0$ \\ 
$C$ & $M_x$ & $0$ & $M_x$ & $0$ & $0$ & $0$ \\ 
$D$ & $0$ & $M_x$ & $0$ & $M_x$ & $0$ & $0$ \\ 
\hline
\end{tabular}
\end{center}
\caption{Particle numbers and charges for various half-filled ground states.}
\label{table:1}
\end{table}

%Absatz
The different configurations with a single generalized particle in the half-filled equipartition vacuum have
\bel{Q3}
Q_R=\pm1\ ,\quad Q_L=0 \ ,
\ee
or
\bel{Q4}
Q_L=\pm1\ ,\quad Q_R=0\ .
\ee
The individual one-particle configurations can be labeled by $\gl Q_R, Q_L\gr$ and the coordinate $x$ at which the charged particle is found.

\subsection{Complex structure for vacua and charged \\ \hspace*{10pt} particles}
\label{subsec: Complex structure for vacua and charged particles}

For the four half-filled vacua $A$, $B$, $C$, $D$ the particle-hole transformation acts as
\bel{Q4a}
K:\quad A\leftrightarrow B\ ,\quad C\leftrightarrow D\ .
\ee
We may associate $A$ and $C$ with $\{\tau'\}$, while $B$ and $D$ span $\{\tau^c\}$ for this sector. In the complex language we have two complex components of the wave function in this sector
\begin{align}
\label{Q5}
\varphi_1^{(0)}=&\frac{1+i}{\sqrt{2}}q^{(A)}+\frac{1-i}{\sqrt{2}}q^{(B)}\ ,\nn\\
\varphi_2^{(0)}=&\frac{1+i}{\sqrt{2}}q^{(C)}+\frac{1-i}{\sqrt{2}}q^{(D)}\ ,
\end{align}
where we assume again $\qtil=\qbar=q$. For a restriction to these four possible vacuum states the real four-component wave functions $\gl q^{(A)}, q^{(B)}, q^{(C)}, q^{(D)}\gr$ is mapped to a two-component complex wave function $\gl\varphi_1^{(0)}, \varphi_2^{(0)}\gr$. The discrete transformation corresponding to the multiplication with $i$ maps
\begin{align}
\label{Q6}
I:\quad &q^{(A)}\to q^{(B)}\ ,\quad q^{(B)}\to -q^{(A)}\ ,\nn\\
&q^{(C)}\to q^{(D)}\ ,\quad q^{(D)}\to -q^{(C)}\ .
\end{align}
The totally empty and totally filled vacuum states are mapped into each other by $K$. We can employ the same construction replacing $q^{(A)}$ and $q^{(B)}$ by the wave functions for the totally empty and filled states.

%Absatz
The particle-hole symmetric vacua correspond to a real wave function, e.g.
\bel{331A}
\frac{1}{\sqrt{2}}\bigl{(}q^{(A)}+q^{(B)}\bigr{)}=\frac{1}{\sqrt{2}}\bigl{(}\varphi_{1}^{(0)}+(\varphi_{1}^{(0)})^{*}\bigr{)}\ .
\ee
The half-filled equipartition vacuum is invariant under the particle-hole transformation. In the complex language $q^{(E)}$ remains a real wave function. We observe that the individual components $q_{\ i}^{(E)}$ can again be grouped into complex wave functions. This defines the action of multiplication with $i$ as a suitable map in the space of real $q_{\ i}^{(E)}$, and therefore also defines formally  $iq^{(E)}$ in the complex formulation.

%Absatz
For the one-particle excitations of the half-filled vacua we associate the states with positive charges $Q_R=1$ or $Q_L=1$ to $\{\tau'\}$, and the states with negative charges $Q_R=-1$ or $Q_L=-1$ to the complement $\{\tau^c\}$. Particle-hole conjugation and the associated complex conjugation $K$ reverses the charges
\bel{Q7}
\big\{ K, Q_R\big\}=0\ ,\quad \big\{ K, Q_L\big\}=0\ .
\ee
The (generalized) one-particle wave functions can be grouped as real fields depending on $x$,
\bel{Q8}
q_\gamma(x)=\pvec{q_\gamma'(x)}{q_\gamma^c(x)}\ ,
\ee
for which the charge operator $Q$ acts as 
\bel{Q9}
Q\pvec{q_\gamma'(x)}{q_\gamma^c(x)}=\pvec{q_\gamma'(x)}{-q_\gamma^c(x)}\ ,\quad Q=\tau_3\ .
\ee
The complex one-particle wave functions read
\bel{Q10}
\varphi_\gamma(x)=\frac{1+i}{\sqrt{2}}q_\gamma'(x)+\frac{1-i}{\sqrt{2}}q_\gamma^c(x)\ .
\ee
These complex wave functions are not eigenstates of the charge operator which mixes $\varphi_\gamma(x)$ and $\varphi_\gamma^*(x)$.

%Absatz
The appropriate range of the index $\gamma$ depends on the particular vacuum state. It always comprises $R$ and $L$. For the equipartition vacuum only four generalized particles with $Q_{R}=\pm 1$ , $Q_{L}=0$ and $Q_{R}=0$, $Q_{L}=\pm 1$ exist, such that $\gamma = R,\, L$. For the particle-hole symmetric combination of the totally empty and filled configurations $\gamma$ takes the four values ($R1$, $R2$, $L1$, $L2$). For the particle hole symmetric combination of the vacua $(A)$ and $(B)$ the range of $\gamma$ depends on which configurations we count as particles.

\subsection{Chiral transformations of the one-particle wave \\ \hspace*{10pt} functions\label{subsec: chiral transf}}

%Absatz
We can perform separate phase transformations on the complex wave function of the right-moving and left-moving one-particle excitations
\bel{335A}
\varphi_{R}\to e^{i\alpha_{R}} \varphi_{R}\, , \quad \varphi_{L}\to e^{i\alpha_{L}} \varphi_{L}\ .
\ee
The infinitesimal chiral transformation
\begin{equation}
\delta\varphi = i\alpha \varphi 
\end{equation}
can be translated directly to the infinitesimal transformation of the real wave functions $q$, $q^{c}$,
\bel{335B}
\delta q=\alpha q^{c}\, ,\quad  \delta q^{c}=-\alpha q'\ .
\ee
This follows directly from the definition
\bel{335C}
\varphi=\frac{1+i}{\sqrt{2}}q'+\frac{1-i}{\sqrt{2}}q^{c}=e^{\tfrac{i\pi}{4}} q'+e^{-\frac{i\pi}{4}}q^{c} \ .
\ee
Thus chiral rotations correspond for our complex structure to a rotation between particles and holes. Since the step evolution operator does not distinguish between particles and holes the evolution is invariant under chiral rotations.

\section{Momentum and uncertainly relation\label{sec: 09}}

%Absatz
We have discussed in sect.~\ref{sec: 08} the simple properties of vacua and one-particle wave functions in the language of the cellular automaton. All these findings translate directly to the discrete fermionic quantum field theory. The wave functions are identical. In the opposite direction we can ask how simple properties of the fermionic quantum field theory translate to the cellular automaton. For one-particle fermionic excitations the characteristic quantity is the momentum of the particle. We will investigate how this observable appears for the cellular automaton. For a single fermion we also have the position observable. The associated operators for momentum and position do not commute, implying the uncertainty relation characteristic for quantum mechanics. The appearance of non-commuting operators for observables in the classical statistical system of the cellular automaton will also shed light on how ``no go theorems" for a classical statistical implementation of quantum systems, as Bell's inequalities~\cite{BELL,CHSH}, are circumvented.

\subsection{Evolution of one-particle wave function\label{subsec: evolution of one-particle wave function}}

%Absatz
The time evolution of the one-particle wave function is very simple. Since a single particle or hole cannot scatter, the generalized particles move on straight lines, with $ \text{d}x=\text{d} t$ for right-movers and $\text{d}x=-\text{d}t$ for left-movers. We may consider a pair of  complex one-particle wave functions, one right-mover $\varphi_{R}(t,x)$ and one left-mover $\varphi_{L}(t,x)$. This applies to all vacua discussed in the preceding section, with an additional index for several species in some cases.
The evolution is given by
\bel{P1}
\varphi(t+2\varepsilon, x)=\widehat{S}\varphi (t,x)\,,\quad \varphi =
\begin{pmatrix}
\varphi_{R} \\ 
\varphi_{L}
\end{pmatrix}
\,,\quad \widehat{S}=
\begin{pmatrix}
\widehat{S}_{R}& 0 \\
 0 & \widehat{S}_{L}
\end{pmatrix}\ ,
\ee
with block diagonal evolution operator generating a shift in the corresponding direction
\bel{P2}
\varphi_{R}(t+2\varepsilon , x)=\varphi_{R}(t, x-2\varepsilon)\,, \quad \varphi_{L}(t+2\varepsilon , x)=\varphi_{L}(t, x+2\varepsilon)\ .
\ee
With discrete ``lattice derivatives"
\begin{align}\label{P3}
\partial_{t}\varphi(t,x)=\frac{1}{4\varepsilon}\bigl{(}\varphi(t+2\varepsilon , x)-\varphi(t-2\varepsilon , x)\bigr{)}\;,\nn\\
\partial_{x}\varphi(t,x)=\frac{1}{4\varepsilon}\bigl{(}\varphi(t, x+2\varepsilon )-\varphi(t, x-2\varepsilon)\bigr{)}\ ,
\end{align}
one obtains the discrete evolution equation
\bel{P4}
\partial_{t}\varphi_{R}=-\partial_{x}\varphi_{R}\, ,\quad \partial_{t}\varphi_{L}=\partial_{x}\varphi_{L}\ .
\ee
The general solutions are functions of $t-x$ or $t+x$, 
\bel{P5}
\varphi_{R}=f_{R}(t-x)\;,\quad \varphi_{L}=f_{L}(t+x)\ .
\ee

%Absatz
By multiplication with $i$ eq.~\eqref{P4} yields the discrete Schrödinger equation
\bel{P6}
i\partial_{t}\varphi =H\varphi\;,\quad H=\begin{pmatrix}
P & 0 \\
0 & -P
\end{pmatrix}
\;,\quad P=-i\partial_{x}\ .
\ee
The continuum limit is rather simple in this case, replacing lattice derivatives by partial derivatives. We recognize the momentum operator $P$. In the real  formulation with $\chi_{R}=(q_{R}' , q_{R}^{c})$ this operator reads
\bel{P7}
\widehat{P}=-I\partial_{x}=\begin{pmatrix}
0 & -\partial_{x} \\
\partial_{x} & 0
\end{pmatrix}
\ ,
\ee
where for discrete lattice points the operator $\varepsilon\partial_{x}$ is a real antisymmetric matrix (cf. ref.~\cite{Wetterich:2020kqi} for details). 
Thus $\widehat{P}$ is symmetric and $P$ therefore hermitian. The
operator $\widehat{P}$ commutes with $\widehat{S}$, such that $\widehat{P}$ corresponds to a conserved quantity.

\subsection{Momentum eigenstates\label{subsec: Momentum eigenstates}}

%Absatz
The eigenfunctions of the momentum operator $P$ are periodic. The ones with positive energy are given by $(p>0)$
\bel{P8}
\varphi_{R}\sim e^{-ip(t-x)}\;,\quad \varphi_{L}\sim e^{-ip(t+x)} \ .
\ee
In the real formulation this corresponds to a rotation between particles and holes
\begin{align}\label{P9}
q_{R}'=\overline{N}\cos\big{[}p(t-x)+\alpha_{R}\big{]}\;,\quad q_{R}^{c}=\overline{N}\sin \big{[}p(t-x)+\alpha_{R}\big{]}\;,\nn\\
q_{L}'=\overline{N}\cos\big{[}p(t+x)+\alpha_{L}\big{]}\;,\quad q_{L}^{c}=\overline{N}\sin\big{[}p(t+x)+\alpha_{L}\big{]} \ ,
\end{align}
with appropriate normalization factor $\overline{N}$. For a finite number $M_{x}$ of sites of a periodic lattice, and lattice distance $2\varepsilon$, the possible values of $p$ are discrete
\bel{P10}
p=\frac{2\pi k}{2\varepsilon M_{x}}=\frac{2\pi k}{L}\;,\quad k\in\mathbb{Z} \ ,
\ee
with $L=2\varepsilon M_{x}$ the circumference of the torus in $x$.
For discrete lattice points the momentum $p$ is periodic with period $\Delta p=\pi/\varepsilon$, since $p$ and $p+\Delta p$ yield the same values for all lattice points separated by $2 \varepsilon$. 

%Absatz
From $(q_{R}')^{2}+(q_{R}^{c})^{2}=\overline{N}^{2}$ we conclude that the probability to find either a right-moving particle or a right-moving hole does not depend on $x$ or $t$. However, the probability to find a particle at $x$ oscillates $\sim \cos^{2}\big{[} p(t-x)+\alpha_{R}\big{]}$. At the maxima of this oscillating probability distribution the probability to find a hole vanishes. For the example of the half-filled equipartition vacuum the particle at $x$ corresponds to the presence of both red and green right-movers at $x$, while a hole at $x$ describes the absence of right movers at $x$. The momentum eigenstates therefore have a very concrete interpretation in terms of probabilities to find various configurations of occupation numbers, and the associated oscillating expectation values.

%Absatz
The precise components between which the oscillation takes place depends on the choice of the complex structure. For a different complex structure the oscillation may be between the two colors.

%Absatz
In the complex picture we can employ the Fourier transform in order to express arbitrary complex wave functions $\varphi_{R}(t,x)$ or $\varphi_{L}(t,x)$ as linear combinations of momentum eigenstates. The Fourier transform is a standard tool for the description of fermionic quantum field theories. Here it finds a corresponding application to probabilistic cellular automata for which its use may not have been obvious without the correspondence to the quantum systems of fermions. 

\subsection{Momentum observable\label{subsec: Momentum Observable}}

%Absatz
In a fermionic quantum field theory the hermitian momentum operator $P$ of a single particle is usually considered to represent an observable. For an observable in quantum mechanics the possible measurement values for an observable correspond to the eigenvalues of the associated operator. 
For discrete space points these eigenvalues are given by
\bel{P11}
\lambda (p)=\frac{\sin (2p\varepsilon)}{2\varepsilon}\ ,
\ee
with discrete values of $p$ given by eq.~\eqref{P10}. In the continuum limit $\varepsilon \to 0$ one simply finds $\lambda (p) =p$.
The expectation value $\langle P\rangle$ is given by the probabilities to find the different eigenvalues, as encoded in the quantum rule
\bel{P12}
\big{\langle} P(t)\big{\rangle} = \tr\big{\lbrace} P\rho (t)\big{\rbrace} \ .
\ee
In particular, for one of the eigenstates~\eqref{P8} for a particular value $p$ the expectation value should coincide with an eigenvalue 
\bel{P13}
\langle P\rangle_{p} =\lambda (p)\ .
\ee
This may be verified by direct computation~\cite{Wetterich:2020kqi} of eq.~\eqref{P12} for the corresponding density matrix $\rho_{p}(t)$. According to our discussion in sect.~\ref{subsec: Change of basis} the probabilities to find given eigenvalues of $\lambda (p)$ can be associated to the diagonal elements of the density matrix in a basis of eigenstates labeled by $p$. This is simply the Fourier-representation of the density matrix. Computing the expectation values with these probabilities amounts to the quantum rule~\eqref{P12}. All these formal requirements are equally obeyed by the fermionic quantum field theory and the probabilistic cellular automaton. They only involve the density matrix and the operator $P$.

%Absatz
What remains to the specified is a measurement description for an observable associated to the operator $P$. This should yield $\lambda(p)$ for individual measurements and reproduce the expectation value $\langle P \rangle$. The problem of finding such a prescription seems to be the same for the fermionic quantum field theory and the probabilistic cellular automaton. Measuring momentum for free relativistic particle is not easy even as a gedanken-experiment. Properties of trajectories cannot be used except for the sign of $p$ - the generalized particles move always on straight lines with light velocity. One possibility for momentum eigenstates could be the counting of oscillations in $x$ for the expectation values of particles and holes. This determines $|p|$ and therefore $\lambda(p)$ for every momentum eigenstate~\cite{Wetterich:2020kqi}. The generalization to arbitrary states would need some ``apparatus" that projects a given state to the eigenstates of $P$ with the required probabilities.

%Absatz
We do not want to deepen here a discussion of possible measurement prescriptions. Whatever can be found, it is clear that the possible measurement values $\lambda(p)$ do not have a given value for the configurations $\tau$. A momentum observable is a ``statistical observable" in the sense that it measures properties of probability distributions (in our case periodicity). In this respect it has common features with quantities as entropy or other characterizations of probability distributions. While not having fixed values in microstates (configurations in our case), it characterizes the probabilistic information. Nevertheless, it can take sharp values $\lambda(p)$ for particular density matrices, and one may associate this with possible outcomes of measurements.

\subsection{Position observable and uncertainly relation\label{subsec: position observable and uncertainly relation}}

Intuitively it is clear that periodic expectation values of occupation numbers are not compatible with sharp positions of particles. This is reflected in Heisenberg's uncertainty relation between position and momentum. Not surprisingly anymore, the quantitative relation will be precisely the same for the probabilistic cellular automaton and the fermionic quantum system. 

%Absatz
For configurations $\tau$ for which one generalized particle (particle or hole) is present at the position $x$ the position observable takes the value $x$. For the one-particle states $\tau^{(1)}$ the position is a standard observable which takes a fixed value $x_{\tau}$ for every one-particle configuration $\tau^{(1)}$. Correspondingly, the associated operator is diagonal
\bel{P14}
\widehat{X}_{\rho\sigma}=\sum_{\tau} x_{\tau}\delta_{\tau\rho}\delta_{\tau\sigma}\ ,
\ee
with $\rho$ and $\sigma$ restricted here to one-particle configurations. The position observable does not make a difference between particles and holes. The corresponding operator $\widehat{X}$ is therefore compatible with the complex structure and takes in the complex formulation again the form~\eqref{P14} (now with a restricted range for $\tau^{(1)} \in \lbrace\tau'\rbrace$).
The position operator also does not distinguish between right- and left- movers, and is color blind in case of several species of one-particle excitations. 

%Absatz
Having defined both the position and momentum operators as suitable matrices we can compute the commutator. In the continuum limit $\varepsilon\to 0$ one obtains in the real and complex formulation, respectively
\bel{P15}
[\widehat{X},\widehat{P}]=I\;,\quad [X,P]=i\ .
\ee
This is the standard result for quantum mechanics. From these Heisenberg's uncertainty relation follows in the standard way. 

%Absatz
Our discussion of momentum and position observables sheds light on the question how no-go theorems for an embedding of quantum mechanics in classical statistics are circumvented. We already know that this must be the case since our probabilistic cellular automaton is fully equivalent to a quantum system, being itself a classical statistical system. Bell's theorem establishes important inequalities for classical correlation functions in a rather general contest. It only assumes the existence of states for which the observables take fixed values, and a positive probability distribution for these states. Quantum experiments show that measured correlations can violate these inequalities. Obviously no contradiction arises if the measured correlation are not given by classical correlation functions. The classical products of observables may simply not exist, as in our case for position and momentum since there are no states for which both position and momentum have a sharp value. More generally, quantum systems are often subsystems for which the probabilistic information is sufficient to compute the expectation values $\langle A\rangle$ and $\langle B\rangle$ for two observables, but insufficient for the computation of their classical correlation. Measurement correlations for a sequence of ideal measurements in such system are typically given by prescriptions different from the classical correlations, typically based on products of the associated operators~\cite{Wetterich:2020kqi}. In this case Bell's inequalities do not lead to contradictions. Such subsystems are characterized by ``incomplete statistics".

\section{Discussion\label{sec: 10}}

%Absatz 
This paper demonstrates that certain fermionic quantum field theories with interactions are equivalent to probabilistic cellular automata. This particular class of fermionic models is realized whenever the quantum evolution operator for discrete time steps is a unique jump matrix with only one element in each row and column equal to one, and all other elements zero. We have established a general way to compute the step evolution operator from the Grassmann functional integral which defines the fermionic quantum field theory. In particular, we have formulated a family of discretized Thirring-type fermionic quantum field theories in 1+1-dimensions for which the step evolution operator is a unique jump matrix and therefore realizes a cellular automaton.

%Absatz
One may consider as an important outcome of this investigation that we have succeeded to construct a genuine quantum theory with interactions as a classical statistical system. All objects and concepts as wave function, density matrix, non-commuting operators and expectation values for observables, are strictly identical for the fermionic quantum field theory and the associated probabilistic cellular automaton. All predictions and quantum rules emerge from the simple laws for probabilities and expectation values for observables in classical statistics. This demonstrates that no-go theorems for the emergence of quantum mechanics from classical statistics cannot apply.

%Absatz
Probabilistic cellular automata for bit systems are synonymous to a type of probabilistic classical computing for which the computational steps of bit-manipulations are deterministic, while initial conditions are probabilistic. Since already these simple forms of probabilistic computing show quantum features, one may ask if forms of quantum computing could be performed by classical probabilistic systems, as static memory materials, artificial neural networks or neuromorphic computing~\cite{CWQC,SEXCW,PMOW,PW}.

%Absatz
On the other side, one may hope that the equivalence to a probabilistic cellular automaton may help for exact or partial solutions of the corresponding fermionic quantum field theories. The updating rules for the cellular automaton often allow for simple exact combinatorial results, as the absence of scattering for doubly occupied lines, that may be less straightforward to find in the Grassmann functional integral description for the fermions. In the present paper we have discussed Dirac fermions. The same methods can be applied for Majorana fermions, Weyl fermions or Mayorana-Weyl fermions. 

%Absatz
In this paper we have focused on a particular discretized Thirring type model. The implementation of more fermion species and new forms of interactions or mass terms seems rather straightforward by use of the recipes of sect.~\ref{sec: 03}. Lorentz-symmetry in the naive continuum limit is not difficult to realize, given the simple transformation rules for the Grassmann variables.

%Absatz
Generalizations of our formalism to three or four dimensions do not seem to encounter problems of principle. What is not yet achieved, however, are cellular automata that realize simple fermionic quantum field theories with Lorentz symmetry in three or four dimensions. So far it remains an open question if our world could be described by a probabilistic cellular automaton, or if more general probabilistic systems are needed.

\appendix

\begin{appendices}

\section{Conventions for Grassmann elements and annihilation/creation operators}
\label{ap.A}

%Absatz
In this appendix we specify an ordering of states and Grassmann variables, a choice of sign for the Grassmann basis elements, and a convention for the creation and annihilation operators. Our results do not depend on this particular choice of conventions. For practical checks it is sometimes useful, however, to have at least one definite and consistent convention. This appendix should help the reader to find all conventions quickly.

%Absatz
We first order the index $\alpha$ for the occupation numbers~$n_{\alpha}$ or the associated Grassmann variables~$\psi_{\alpha}$, with $\alpha=(x,\gamma)$, and $\gamma=(\eta, a)$, $\eta=(R,L)$, $a=(1, 2)$. We start at $\alpha=1$ with $(x_{\textup{in}}, R, 1)$ or $(x_{\textup{in}}, \, \gamma=1)$, and first increase the internal index, $\alpha=2$ corresponding to $(x_{\textup{in}}, \, \gamma=2)$,  $\alpha=4$ to $(x_{\textup{in}}, \, \gamma=4)$. Next we increase $x$ by $\varepsilon$, $\alpha=5$ corresponding to $(x_{\textup{in}}+\varepsilon, \, \gamma=1)$, where $\gamma=1,\dots 4$, with
\begin{align}
\label{A1}
\gamma&=1\quad \widehat{=}(R,1)\nn\\
\gamma&=2\quad \widehat{=}(R,2)\nn\\
\gamma&=3\quad \widehat{=}(L,1)\nn\\
\gamma&=4\quad \widehat{=}(L,2)\;.
\end{align}
With this system, $\alpha$ is given by the integer
\be\label{A2}
\alpha=4m+\gamma=4(x-x_{\textup{in}})/\varepsilon+\gamma\;,\quad m=0,1,2 \dots M_{x}-1\;.
\ee
The variables $n_{\alpha}$ or $\psi_{\alpha}$ can therefore be labeled equivalently by $(x,\gamma)$, $(x,\eta, a)$ or the integer~$\alpha$. We use the index~$\alpha$ without distinction of the different ways of labeling. For $M_{x}$ space points we have $\alpha =1 \dots M$, $M=4M_{x}$.

%Absatz
We next order the $2^{M}$ states denoted by $\tau=1 \dots N$, $N=2^{M}$. A convenient label for each state or configuration of Ising spins, occupation numbers or bits is the bit notation $\lbrace n_{\alpha}\rbrace=(n_{1},n_{2},n_{3}\dots n_{M})$; with $n_{\alpha}=1$ for an occupied generalized site $\alpha$ and $n_{\alpha}=0$ for an empty site, e.g. $(0,1,0)$ for the configuration where $\alpha=2$ is occupied and $\alpha=1,3$ are empty. We can associate to each bit configuration a standard binary number, e.g. $n_{B}=2$ for $(0, 1,0)$, $n_{B}=7$ for $(1,1,1)$.
We order by beginning with the fully occupied state and define the integer
\be\label{A3}
\tau=2^{M}-n_{B}\;.
\ee

%Absatz
Wave functions can be seen as $N$- components vectors with components $q_{\tau}$. In the corresponding vector space we use as basis vectors $v^{(\tau)}$ unit vectors which have one element equal to one and all other elements equal to zero, e.g.
\be\label{A4}
v^{(\tau)}_{\rho}=\delta^{(\tau)}_{\rho}\;.
\ee
We can represent $v^{(\tau)}$ in a ``direct product form"
\be\label{A5}
v^{(\tau)}=\mathcal{N}_{(1)}^{(\tau)}\otimes\mathcal{N}_{(2)}^{(\tau)}\otimes\mathcal{N}_{(3)}^{(\tau)}\dots\otimes\mathcal{N}_{(M)}^{(\tau)}\;,
\ee
with two component unit vectors
\begin{align}\label{A6}
\mathcal{N}_{(\alpha)}^{(\tau)}=\begin{pmatrix}
1\\0
\end{pmatrix} &\quad \textup{if} \quad n_{\alpha}^{(\tau)}=1\;,\nn\\
\mathcal{N}_{(\alpha)}^{(\tau)}=\begin{pmatrix}
0\\1
\end{pmatrix} &\quad \textup{if} \quad n_{\alpha}^{(\tau)}=0\;.
\end{align}
Here $n_{\alpha}^{(\tau)}$ is the occupation number $n_{\alpha}$ corresponding to the state~$\tau$. For the example $M=3$, $N=8$ the components of $v^{(\tau)}$ read $(\sigma_{\alpha}=1,2)$
\be\label{A7}
v_{\rho}^{(\tau)}=\mathcal{N}_{(1)\sigma_{1}}^{(\tau)}\mathcal{N}_{(2)\sigma_{2}}^{(\tau)}\mathcal{N}_{(3)\sigma_{3}}^{(\tau)}\,,\,\rho=4(\sigma_{1}-1)+2(\sigma_{2}-1)+\sigma_{3}\;.
\ee
In particular, one has $v^{(1)}=\mathcal{N}_{(1)}^{(1)}\otimes\mathcal{N}_{(2)}^{(1)}\dots$, with all $\mathcal{N}_{(\alpha)}^{(\tau)}=\begin{pmatrix}
1\\0
\end{pmatrix}$ for the fully occupied state.
For $v^{(2)}$ one changes  $\mathcal{N}_{(M)}^{(2)}=\begin{pmatrix}
0\\1
\end{pmatrix}$, while all $\alpha\neq M$ correspond to  $\mathcal{N}_{(\alpha)}^{(2)}=\begin{pmatrix}
1\\0
\end{pmatrix}$. This direct product form is useful if we represent operators as matrices in a direct product form, as we will do with creation and annihilation operators. 

We next turn to the Grassmann basis elements by fixing the signs~$\tilde{s}_{\tau}$ in eq.~\eqref{06}. The basis element~$g_{\tau}$ contains a factor~$\psi_{\alpha}$ for every $n_{\alpha}^{(\tau)}=0$. We take a plus sign for $\tilde{s}_{\tau}$ if all factors $\psi_{\alpha}$ are ordered with increasing $\alpha$, the smallest $\alpha$ to the left. With the bit notation for $\tau$ one has, for example
\begin{align}\label{A8}
g_{(10011001)}&=\psi_{2}\psi_{3}\psi_{6}\psi_{7} \\
=&\psi_{R2}(x_{\textup{in}})\psi_{L1}(x_{\textup{in}})\psi_{R2}(x_{\textup{in}}+\varepsilon)\psi_{L1}(x_{\textup{in}}+\varepsilon)\;.\nn
\end{align}
The totally filled state, $n_{\alpha}=1$ for all $\alpha$, obeys $g_{(111\dots)}=1$. The other basis elements can be obtained by consecutively taking away particles, starting with the highest $\alpha$. Multiplication of $g_{\tau}$ by $\psi_{\alpha}$ yields zero if $g_{\tau}$ contains already a factor $\psi_{\alpha}$, i.e. for $n_{\alpha}^{(\tau)}=0$. On the other hand, if no $\psi_{\alpha}$ is present, $n_{\alpha}^{(\tau)}=1$, $\psi_{\alpha}g_{\tau}$ is a new basis element $g_{\tau'}$ up to a sign. 
Here $\tau'$ obtains from $\tau$ by annihilating a particle $n_{\alpha}^{(\tau')}=0$ for $n_{\alpha}^{(\tau)}=1$. The sign depends on the number of $\psi_{\beta}$ with $\beta<\alpha$ in $g_{\tau}$.
We conclude that multiplication with $\psi_{\alpha}$ can be associated to the annihilation of a particle $\alpha$. Similarly, $\partial/\partial\psi_{\alpha}$ acts as creation of a particle $\alpha$ if no particle $\alpha$ is present in $\tau$. One has $(\partial/\partial\psi_{\alpha})g_{\tau}=0$ if $n_{\alpha}^{(\tau)}=1$, and 
$(\partial/\partial\psi_{\alpha})g_{\tau}=\pm g_{\tau''}$, with $n_{\alpha}^{(\tau'')}=0$ for $n_{\alpha}^{(\tau)}=1$.

%Absatz
We define a ``right transport operator $\tilde{t}_{\gamma}^{+}(x)$ by (no sum over $\gamma$)
\be\label{A9}
\tilde{t}_{\gamma}^{+}(x)=\frac{\partial}{\partial\psi_{\gamma}(x+\varepsilon)}\psi_{\gamma}(x)\;.
\ee
It annihilates a particle of type $\gamma$ at $x$ and creates one at $x+\varepsilon$, provided that for $g_{\tau}$ a particle $\gamma$ is present at $x$ and no particle $\gamma$ is present at $x+\varepsilon$. Otherwise $\tilde{t}_{\gamma}^{+}(x)g_{\tau}=0$.
The basis element $g_{\tau}$ corresponding to a one-particle state contains all $\psi_{\alpha}$ except for $\alpha=(x, \gamma)$. (At the position $\alpha=(x,\gamma)$ there is a factor $1$ instead of $\psi_{\alpha}$.) For a one particle state with a particle at $(x,\gamma)$ one has
\begin{align}\label{A10}
&\tilde{t}_{\gamma}^{+} (x) g_{\tau}=g_{\tau_{+}}\nn\\
&\tilde{t}_{\gamma'}^{+} (y) g_{\tau}=0 \quad \textup{for} \; x\neq y\,,\; \gamma\neq\gamma' \;.
\end{align}
The state $\tau_{+}$ obtains from $\tau$ by displacing the particle from $x$ to $x+\varepsilon$. In terms of integers $\alpha$ one has
\be\label{A11}
\tilde{t}_{\alpha}^{+}=\frac{\partial}{\partial\psi_{\alpha + 4}}\psi_{\alpha}\;.
\ee
One can check easily that $\tilde{t}_{\gamma}^{+} (x)$ does not introduce a minus sign in the change from $g_{\tau}$ to $g_{\tau_{+}}$. Similarly, the left transport operator $\tilde{t}_{\gamma}^{-} (x)$ obeys for a single particle $\gamma$ at $x$ 
\be\label{A12}
\tilde{t}_{\gamma}^{-} (x)g_{\tau}=\frac{\partial}{\partial\psi_{\gamma}(x-\varepsilon)}\psi(x)g_{\tau}=g_{\tau_{-}}\;,
\ee
with 
$g_{\tau_{-}}$ obtained from $g_{\tau}$ by displacing the single particle from $x$ to $x-\varepsilon$.

%Absatz
Annihilation operators $a\g (x)$ and creation operators $a\g\p (x)=a\g\supt{T}(x)$ are defined in terms of the basis annihilation and creation operators $a$ and $a\p$ which are real $2\times 2$ matrices
\be\label{A13}
a=\begin{pmatrix}
0&0\\1&0
\end{pmatrix}\;,\quad a\p=\begin{pmatrix}
0&1\\0&0
\end{pmatrix}\;,
\ee
acting  on real two component unit vectors as 
\be\label{A14}
\!\!a\!\begin{pmatrix}
1\\0
\end{pmatrix}\!=\!\begin{pmatrix}
0\\1
\end{pmatrix},\; a\!\begin{pmatrix}
0\\1
\end{pmatrix}\!=\! 0,\; a\p\!\begin{pmatrix}
1\\0
\end{pmatrix}\!=\! 0,\; a\p\!\begin{pmatrix}
0\\1
\end{pmatrix}\!=\!\begin{pmatrix}
1\\0
\end{pmatrix}.\!\!
\ee
For $a_{\gamma}(x)$ we choose a direct product representation
\be\label{A15}
a\g (x)=T_{3}\otimes T_{3}\otimes\, \dots\,\otimes T_{3}\otimes a\g\otimes 1\otimes 1\,\otimes \dots\, \otimes 1\;,
\ee
where the $16\times 16$ matrix $a\g$ is placed at the position $x=~x_{\textup{in}}+m\varepsilon$.
The $16\times 16$ matrix $T_{3}$ anticommutes with $a\g, a\g\p$, 
\be\label{A16}
T_{3}= \tau_{3}\otimes \tau_{3}\otimes \tau_{3}\otimes \tau_{3}\;,\lbrace \tau_{3}, a\g\rbrace =0\;,
\ee
and the factors $1$ in eq.~\eqref{A15} stand for unit $16\times 16$ matrices. For $a\g$ we employ
\begin{align}\label{A17}
&a_{R1}=a\otimes 1\otimes 1\otimes 1\;,&&\;a_{R2}=\tau_{3}\otimes a\otimes 1\otimes 1\;,\nn\\
&a_{L1}=\tau_{3}\otimes \tau_{3}\otimes a\otimes 1\;,&&\;a_{L2}=\tau_{3}\otimes \tau_{3}\otimes \tau_{3}\otimes a\;.
\end{align}
With
\be\label{A18}
a\g\p =a\g\supt{T}\;,\quad a\g\p (x)=a\g\supt{T} (x)
\ee
we can obtain $a\g\p (x)$ from $a\g (x)$ by replacing the factor $a$ by $a\p$. We observe the basic anticommutation relations for annihilation and creation operators of fermions
\begin{align}\label{A19}
&\lbrace a\g\p , a_{\delta}\rbrace=\delta_{\gamma\delta}\;,\quad\big{\lbrace} a\g\p (x), a_{\delta}(y)\big{\rbrace}=\delta_{\gamma\delta}\delta_{xy}\;,\nn\\
&\lbrace a\g , a_{\delta}\rbrace =0\; , \quad \big{\lbrace} a\g (x) , a_{\delta}(y)\big{\rbrace} =0\;,\nn\\
&\lbrace a\g\p ,a_{\delta}\p\rbrace =0\;,\quad \big{\lbrace} a\g\p (x), a_{\delta}\p (y)\big{\rbrace} =0\;.
\end{align}

%Absatz
For this convention the right transport operator (no sum over~$\gamma$),
\be\label{A20}
t\g\p(x)=a\g\p(x+\varepsilon)a\g(x)\;,
\ee
transports the single particle wave function for a particle~$\gamma$ at $x$ to a particle $\gamma$ at $x+\varepsilon$ without a change of sign. Noting the relation
\begin{align}\label{A21}
 \tau_{3}a&=-a\tau_{3}=-a\nn\\
\tau_{3}a\p &=-a\p \tau_{3}=a\p\;,
\end{align}
one has
\begin{align}\label{A22}
T_{3}a_{R1}&=-(a\otimes \tau_{3}\otimes\tau_{3}\otimes\tau_{3})\;,\; T_{3}a_{R1}\p=(a\p\otimes \tau_{3}\otimes\tau_{3}\otimes\tau_{3})\;, \nn\\
T_{3}a_{R2}&=-(1\otimes a\otimes\tau_{3}\otimes\tau_{3})\;,\; \;T_{3}a_{R2}\p=(1\otimes a\p\otimes\tau_{3}\otimes\tau_{3})\;,
 \end{align}
 and similarly for $L1$, $L2$. As compared to eq.~\eqref{A17} we note the different $\tau_{3}$ factors.
 For the right transport operator this implies (no sum over $\gamma$)
 \be\label{A23}
 a\g\p (x+\varepsilon)a\g(x)=(1\otimes 1\,\dots \, \otimes T_{3} a\g\otimes a\g\p \otimes 1\otimes 1\,\dots)\;.
 \ee
 Let us consider $\gamma =R1$ and focus on the parts at $x$ and $x+\varepsilon$
 \be\label{A24}
 T_{3}a_{R1}\otimes a_{R1}\p=-(a\otimes\tau_{3}\otimes\tau_{3}\otimes\tau_{3})\otimes (a\p\otimes 1\otimes 1\otimes 1)\;.
 \ee
 The $\tau_{3}$-factors produce a factor $(-1)$ for every absent particle. Since for a single particle of type $R1$ the particles $R2$, $L1$ and $L2$ are absent both at $x$ and $x+\varepsilon$, the three minus signs from the $\tau_{3}$-factors cancel the overall minus sign. We conclude that a $\delta$-type single particle wave function with a particle $R1$ sitting precisely at $x$ is changed by multiplication with $t_{R1}^{+}$ to a similar wave function, now with the particle sitting precisely at $x+\varepsilon$. The situation for the other particle types $R2$, $L1$ and $L2$ is similar. The left transport operator $t\g^{-}(x)$ has the same properties, now with the particle $\gamma$ transported to $x-\varepsilon$,
 \be\label{A25}
 t\g^{-}(x)=a\g\p (x-\varepsilon)a\g (x)\;.
 \ee
 For the part at $(x-\varepsilon , x)$ one has now the product $a\g\p \otimes T_{3}a\g$, such that the change of positions of the factors as compared to $t\g^{+}(x)$ has no influence on the overall sign.
 
 %Absazt
 Applying the annihilation $a\g(x)$ to a Grassmann wave function 
 \be\label{A26}
 g=q_{\tau}g_{\tau}
 \ee
 we can equivalently multiply the vector $q_{\tau}$ with the operator $a\g (x)$, or the vector $g_{\tau}$ with the transposed $a\g\supt{T}(x)=a\g\p (x)$,
 \be\label{A27}
 \big{[}a\g (x)\big{]}_{\tau\rho}q_{\rho}g_{\tau}=q_{\tau}g_{\rho}\gl a\g(x)\gr_{\rho\tau}=q_{\tau}\widehat{g}_{\tau}\;.
 \ee
 We observe the relation
 \be\label{A28}
g_\rho\gl a\g(x)\gr_{\rho\tau}=\big{(}a\g\p (x)\big{)}_{\tau\rho}g_{\rho}=\psi\g (x) g_{\tau}\;,
\ee
which is equivalent to
\be\label{A29}
\Big{[}\big{(}a\g (x)\big{)}_{\tau\rho} q_{\rho}-\psi\g (x)q_{\tau}\Big{]}g_{\tau}=0\;.
\ee
For a proof of the relation~\eqref{A28} we note that only those $\rho$ with $n\g^{(\rho)}(x)=0$ contribute in the sum on the left hand side. The corresponding basis elements $g_{\rho}$ therefore all have a factor $\psi\g (x)$. This is precisely what happens on the r.h.s. of eq.~\eqref{A28} due to the multiplication with $\psi\g (x)$. If $\tau$ is a state with $n\g^{(\tau)}(x)=0$ one has $\psi\g (x) g_{\tau}=0$, while the l.h.s. of eq.~\eqref{A28}vanishes as well due to the vanishing elements $a_{\tau\rho}\p$. On the r.h.s. of eq.\eqref{A28} we only need to consider those $\tau$ for which $n\g^{(\tau)}(x)=1$, and on the l.h.s. only those $g_{\rho}$ with $n\g^{(\rho)}(x)=0$.
The factors of $\psi_{\alpha}$ with $\alpha\neq (x, \gamma)$ are the same for $g_{\tau}$ on the r.h.s. and all terms~$\sim g_{\rho}$ on the l.h.s. The two expressions in eq.~\eqref{A28} can therefore differ at most by a relative minus sign.

%Absazt
For a discussion of a possible sign we start with $M=1$, where
\be\label{A30}
a\p \begin{pmatrix}g_{1}\\g_{2}
\end{pmatrix}=\begin{pmatrix}
\widehat{g}_{1}\\ \widehat{g}_{2}
\end{pmatrix}\;,\quad\quad \widehat{g}_{1}=g_{2}=\psi\,,\quad \widehat{g}_{2}=0\;.
\ee
This coincides with $\widehat{g}_{\tau}=\psi g_{\tau}$, $\widehat{g}_{1}=\psi g_{1}=\psi$, $\widehat{g}_{2}=\psi g_{2}=0$. For $M=2$ we have $a_{1}\p=a\p\otimes 1$, $a_{2}\p=\tau_{3}\otimes a\p$ or
\be\label{A31}
a_{1}\p\begin{pmatrix}
g_{1}\\g_{2}\\g_{3}\\g_{4}
\end{pmatrix}=\begin{pmatrix}
g_{3}\\g_{4}\\0\\0
\end{pmatrix}\;,\quad a_{2}\p\begin{pmatrix}
g_{1}\\g_{2}\\g_{3}\\g_{4}
\end{pmatrix}=\begin{pmatrix}
\phantom{-}g_{2}\\\phantom{-}0\\-g_{4}\\\phantom{-}0
\end{pmatrix}\;.
\ee
On the other hand, one has with eq.~\eqref{18},
\begin{align}\label{A32}
&\psi_{1}\begin{pmatrix}
g_{1}\\g_{2}\\g_{3}\\g_{4}
\end{pmatrix}=\psi_{1}\begin{pmatrix}
1\\\psi_{2}\\\psi_{1}\\\psi_{1}\psi_{2}
\end{pmatrix}=\begin{pmatrix}
\psi_{1}\\\psi_{1}\psi_{2}\\0\\0
\end{pmatrix}=\begin{pmatrix}
g_{3}\\g_{4}\\0\\0
\end{pmatrix}\nn\\
&\psi_{2}\begin{pmatrix}
g_{1}\\g_{2}\\g_{3}\\g_{4}
\end{pmatrix}=\psi_{2}\begin{pmatrix}
1\\\psi_{2}\\\psi_{1}\\\psi_{1}\psi_{2}
\end{pmatrix}=\begin{pmatrix}
\psi_{2}\\0\\\psi_{1}\psi_{2}\\0
\end{pmatrix}=\begin{pmatrix}
\phantom{-}g_{2}\\\phantom{-}0\\-g_{4}\\\phantom{-}0
\end{pmatrix}\;.
\end{align}
Thus for $M=1,2$ eq.~\eqref{A28} is obeyed without an addictional minus sign. Generalizing to arbitrary $M$ we observe 
\be\label{A33}
\psi_{\alpha} g_{\tau}=\widehat{g}_{\tau}=\widehat{\sigma}_{\tau}g_{\tau'}\;,\quad\widehat{\sigma}_{\tau}=\pm 1\;,
\ee
where a minus sign $\widehat{\sigma}_{\tau}=-1$ occurs whenever the number of empty sites, $n_{\beta}^{(\tau)}=0$, for $\beta<\alpha$ is odd. On the l.h.s. of eq.~\eqref{A28} possible minus signs can only arise from the $\tau_{3}$-factors in $a_{\alpha}\p$. For a given $a_{\alpha}\p$ these $\tau_{3}$ factors occur for all $\beta<\alpha$ according to eqs.~\eqref{A15},~\eqref{A17}. Every $\tau_{3}$ factor produces a sign $(-1)$ for $n_{\beta}^{(\tau)}=0$, and a factor $(+1)$ if $n_{\beta}^{(\tau)}=1$. This coincides with the number of minus signs in $\psi_{\alpha}g_{\tau}$, concluding the proof that no additional sign occurs in eq.~\eqref{A28}.

%Absatz
For the Grassmann derivative the relation analogous to eq.~\eqref{A28} reads
\be\label{A34}
g_\rho\gl a\herm(x)\gr_{\rho\tau}=\big{(}a\g(x)\big{)}_{\tau\rho}g_{\rho}=\frac{\partial}{\partial\psi\g (x)}g_{\tau }\;.
\ee
The argument that both sides contain the same $\psi$-factors is similar to the one below eq.~\eqref{A28}. Again, the number minus signs from the anticommutation of $\partial/\partial\psi\g$ with the factors $\psi_{\beta}$, $\beta<\gamma$, equals the number of negative eigenvalues for the $\tau_{3}$-factors in $a\g (x)$.

%Absatz
For sequences of operations of the type~\eqref{A28},~\eqref{A34} we should interprete the action of annihilation and creation operators as matrix multiplication from the right of the basis elements, e.g.
\be\label{A35}
g_{\rho}\big{(}a\g (x)\big{)}_{\rho\tau}=\psi\g (x) g_{\tau}\;,\quad g_{\rho}\big{(}a\g\p (x)\big{)}_{\rho\tau}=\frac{\partial}{\partial\psi\g (x)}g_{\tau}\;.
\ee
This implies relations of the type
\be\label{A36}
g_{\rho}\big{(}a\g\p (x) a_{\beta}(x)\big{)}_{\rho\tau}=\frac{\partial}{\partial\psi\g (x)}\psi_{\beta}g_{\tau}\;.
\ee
For the operator multiplication we have the isomorphism
\be\label{A37}
a\g (x)\widehat{=}\psi\g (x)\;,\quad a\g\p (x)\widehat{=}\frac{\partial}{\partial\psi\g (x)}\;,
\ee
which holds if for every given $(x,\gamma)$ not both $a\g(x)$ and $a\g\herm(x)$ appear in these expressions. (Otherwise the isomorphism holds only for suitably ordered products of annihilation and creation operators.)
Grassmann operators~$\mathcal{A}$ a sequences of factors $\psi_{\alpha}$ and $\partial/\partial\psi_{\beta}$, e.g.
\be\label{A38}
\mathcal{A}_{1}=\psi_{2}\frac{\partial}{\partial\psi_{4}}\psi_{6}\psi_{7}\frac{\partial}{\partial\psi_{10}}\;.
\ee
The associated operator $A$ is a corresponding product of annihilation operators $a_{\alpha}$ and creation operators $a_{\beta}\p$, with the same order, e.g.
\be\label{A39}
A_{1}=a_{2}\, a_{4}\p\, a_{6}\,a_{7}\,a_{10}\p\;.
\ee
The operator $A$ acts on the basis elements from the right, 
\be\label{A40}
g_{\rho}A_{\rho\tau}=\mathcal{A}g_{\tau}\;.
\ee
This procedure implies for arbitrary Grassmann elements
\be\label{A41}
\mathcal{A}g=\mathcal{A}q_{\tau}g_{\tau}=q_{\tau}g_{\rho}A_{\rho\tau}=A_{\tau\rho}q_{\rho}g_{\tau}=q_{\tau}'g_{\tau}\;,
\ee
such that the wave function $q$ is multiplied by the operator $A$ from the left as usual, 
\be\label{A42}
q_{\tau}\rightarrow q_{\tau}'=A_{\tau\rho}q_{\rho}\;.
\ee
In particular, the occupation number operator for a particle~$\gamma$ at $x$ is given in the occupation number basis and as associated Grassmann operator by
\be\label{A43}
\widehat{n}(x)=a\g\p (x)a\g (x)\;,\quad \mathcal{N}\g (x)=\frac{\partial}{\partial\psi\g (x)}\psi\g (x)\;.
\ee

\section{Coarse graining and conjugate\\ Grassmann variables} 
\label{ap.B}

%Absatz
The product of step evolution operators for propagation and interaction can be viewed as a type of coarse grained step evolution operator. Two evolution steps are combined into a common coarse grained step. This doubles the Grassmann variables appearing in the combined step, involving for every coarse grained time both $\psi$ and $\overline{\psi}$. We briefly discuss in this appendix how $\overline{\psi}$ is related to the conjugate Grassmann variables in ref.~\cite{CWFCS}.

The product \eqref{61} is a new local factor $\overline{\mathcal{K}}(t)$ which depends on the Grassmann variables $\psi_{\gamma}(t+2\tilde{\varepsilon},x)$ and $\psi_{\gamma}(x)$. The associated step evolution operator $\widehat{S}(t)$ is given by eq.~\eqref{62}.
Restricting the observables to even $t$ this defines a "coarse grained" fermionic model.
The relation between $\overline{\mathcal{K}}(t)$ and $\widehat{S}(t)$ is given by
\be\label{64}
\overline{\mathcal{K}}(t)=g_{\tau}(t+2\tilde{\varepsilon})\widehat{S}_{\tau\rho}(t)\overline{g}_{\rho}(t)\;.\ee

%Absatz
The units of the time-distance between two neighboring points $m$ and $m+1$ on the time lattice is arbitrary. For our particular construction a non-interacting particle advances one space unit during two time units. The velocity can be normalized to one by choosing the time difference $\tilde{\varepsilon}$ between neighboring lattice points as $\tilde{\varepsilon}=\varepsilon/2$. Eq.~\eqref{64} becomes in the new units
\be\label{65}
\overline{K}(t)=g_{\tau}(t+\varepsilon)\widehat{S}_{\tau\rho}(t)\overline{g}_{\rho}(t).
\ee

%Absatz
As a disadvantage of integrating out the "intermediate Grassmann variables $\psi_{\gamma}(t+\dfrac{\varepsilon}{2},x)$ both the Grassmann basis functions $g_{\tau}$ and the conjugate Grassmann basis functions $\overline{g}_{\tau}$ are needed for an extraction of $\widehat{S}(t)$ from $\overline{\mathcal{K}}(t)$.
Correspondingly the intermediate integration changes the form of $\overline{\mathcal{K}}(t)$, even if $\widehat{S}(t)$ is the identity operation. This is easily seen by the multiplication of unit step evolution operators
\be\label{66}
\tilde{\mathcal{K}}(t+\varepsilon)=\exp(\psi\varphi)\,,\quad\tilde{\mathcal{K}}(t)=\exp(\varphi\psi')\;,
\ee
where
\be\label{67}
\begin{split}
\overline{\mathcal{K}}(t)&=\int\mathcal{D}\varphi\exp(\psi\varphi+\varphi\psi')\\ &=\int\mathcal{D}\varphi\bigl(1+\varphi(\psi'-\psi)\bigr)\\
&=\psi'-\psi=\delta(\psi'-\psi)\;.
\end{split}
\ee

%Absatz
It is advantageous to keep the exponential form of the local factors for the coarse grained view as well. This can be done by doubling the number of Grassmann variables at every $t$.
At every $t$ one has then two sets of variables $\psi_{\gamma}(t,x)$ and $\overline{\psi}_{\gamma}(t,x)$, and the functional integration is over $\psi$ and $\overline{\psi}$. In our case we can simply associate $\overline{\psi}(t)$ with $\psi(t-\varepsilon/2)$ by defining
\begin{align}\label{68}\overline{\psi}_{R, a}(t+\varepsilon,x)=\psi_{R, a}(t+\dfrac{\varepsilon}{2},x)\,, \nn \\
\overline{\psi}_{L, a}(t+\varepsilon,x)=\psi_{L, a}(t+\dfrac{\varepsilon}{2},x)\;.
\end{align}
The Grassmann variables $\psi(t+\dfrac{\varepsilon}{2})$ play a role very similar to the conjugate Grassmann variables used, for example, in ref.\cite{CWFCS}.

%Absatz
We can define even and odd sublattices. With $t=m_{t}\varepsilon$~, $x=m_{x}\varepsilon$ and integer $m_{t}$, $m_{x}$ the even (odd) sublattice contains the points with $ m_{t}+m_{x}$ even (odd). The action of our model does not connect the even and the odd sublattice. Since for every step $\varepsilon$ in $t$ the kinetic terms moves the variables either one place to te right or to the left, it does not mix the sublattices. A particle on the even sublattice remains on the even sublattice. The interaction term is local and does not change the situation. In the following we simply omit the odd sublattice and define $\sum_{t,x}$ as a sum over the points of the even sublattice. The factor $\tilde{D}(t+\varepsilon , x)$ in eq.~\eqref{69} is evaluated on the even sublattice at $t+\varepsilon$.

%Absatz
There exist equivalent alternative formulations that put the right movers on the even sublattice and the left movers on the odd sublattice (see next section), or the green particles on the even sublattice and the red particles on the odd sublattice. This may seem at first sight more economical since one has a full square lattice and only two species at each lattice site. The discussion becomes more complex, however, since often different cases have to be addressed specifically. For this reason we will focus on the formulation on the even sublattice with four species of Grassmann variables and four species of conjugate Grassmann variables at each lattice site.
 
\section{Weyl and Majorana fermions}
\label{ap.C}

%Absatz
In this appendix we discuss the notion of two-dimensional Weyl, Majorana and Majorana-Weyl spinors. This clarifies the particle content for different types of the continuum limit. 

%Absatz
Without a complex structure the ``real" Grassmann variables $\psi_{R}$ and $\psi_{L}$ form right-moving and left-moving Majorana-Weyl spinors. A Majorana spinor  is composed of two Majorana-Weyl spinors $\psi_{+}\,,\, \psi_{-}$  that transform independently
\be\label{94}
\psi_{+}=\dfrac{1+\overline{\gamma}}{2}\psi=\begin{pmatrix}
\psi_{R}\\0
\end{pmatrix}
\,,\quad
\psi_{-}=\dfrac{1-\overline{\gamma}}{2}\psi=\begin{pmatrix}
0\\\psi_{L}
\end{pmatrix}\;,
\ee
where $\overline{\gamma}$ corresponds to $\gamma^{5}$ in four dimensions,
\be\label{95}
\overline{\gamma}=-\gamma^{0}\gamma^{1}=\tau_{3}\,,\quad \lbrace\overline{\gamma},\gamma^{\mu}\rbrace=0\;.
\ee

\subsection{Lorentz invariance of Thirring automaton.}
\label{subsec: Lorentz invariance of Thirring automaton.}

%Absatz
Lorentz transformations scale the Majorana-Weyl spinors in opposite directions
\be\label{96}
\delta\psi_{+}=\dfrac{\eta}{2}\psi_{+}\,,\quad \delta\psi_{-}=-\dfrac{\eta}{2}\psi_{-}\;.
\ee
With $\overline{\psi}_{R}$ and $\overline{\psi}_{L}$ transforming in the same way as $\psi_{R}$ and $\psi_{L}$ any interaction that involves an equal number of right-movers and left-movers is Lorentz invariant.
We observe the relation
\be\label{LD}
\overline{\psi}=(\overline{\psi}_{R},\overline{\psi}_{L})\gamma^{0}\;.
\ee
The kinetic term (first term in the action~\eqref{LB}) involves in addition a derivative that adds to the Lorentz transformation property and makes it invariant.

%Absatz
In order to see the Thirring form and Lorentz-invariance of the interaction term $\mathcal{L}_{\textup{int}}$ we observe the relations
\begin{align}\label{LE}
\overline{\psi}\gamma^{0}\psi&=-(\overline{\psi}_{R}\psi_{R}+\overline{\psi}_{L}\psi_{L})\;,\nn\\
\overline{\psi}\gamma^{1}\psi&=-\overline{\psi}_{R}\psi_{R}+\overline{\psi}_{L}\psi_{L}\;,
\end{align}
and
\be\label{LF}
\overline{\psi}_{d}\gamma^{\mu}\psi_{c}\overline{\psi}_{b}\gamma_{\mu}\psi_{a}=-2(\overline{\psi}_{Rd}\psi_{Rc}\overline{\psi}_{Lb}\psi_{La}+\overline{\psi}_{Rb}\psi_{Ra}\overline{\psi}_{Ld}\psi_{Lc})\;.
\ee
With the naming~\eqref{70} this yields
\begin{align}\label{LG}
2\overline{D}=&-\overline{\psi}_{1}\gamma^{\mu}\psi_{1}\overline{\psi}_{2}\gamma_{\mu}\psi_{2}+\overline{\psi}_{1}\gamma^{\mu}\psi_{2}\overline{\psi}_{2}\gamma_{\mu}\psi_{1}\;,\nn\\
2\overline{C}=&-\frac{1}{2}(\overline{\psi}_{1}\gamma^{\mu}\psi_{1}\overline{\psi}_{1}\gamma_{\mu}\psi_{1}+\overline{\psi}_{2}\gamma^{\mu}\psi_{2}\overline{\psi}_{2}\gamma_{\mu}\psi_{2})\nn\\
&+\frac{\sigma}{2}(\overline{\psi}_{1}\gamma^{\mu}\psi_{2}\overline{\psi}_{1}\gamma_{\mu}\psi_{2}+\overline{\psi}_{2}\gamma^{\mu}\psi_{1}\overline{\psi}_{2}\gamma_{\mu}\psi_{1}\;.
\end{align}
We observe that $\overline{D}$ and $\overline{C}$ are separately Lorentz-invariant, only the color structure depends on the particular combination. We have also used the sign $\sigma$ in the convention~\eqref{82} in order to demonstrate that Lorentz symmetry does not depend on the sign convention. 
The particular combination $\overline{C}+\overline{D}$ can be written in the form
\begin{align}
\label{99A}
&\mathcal{L}_i=2(\overline{C}+\overline{D})=-\frac{1}{2}\overline{\psi}_{a}\gamma^{\mu}\psi_a\overline{\psi}_b\gamma_\mu\psi_b\nn\\
&\ \ +\frac{\sigma}{2}\gl\psibar_1\gamma^\mu\psi_2+\sigma\psibar_2\gamma^\mu\psi_1\gr\gl\psibar_1\gamma_\mu\psi_2+\sigma\psibar_2\gamma_\mu\psi_1\gr\ .
\end{align}
For $\sigma=-1$ this yields eq.~\eqref{LJ}.

%Absatz
Equivalently, we can write $\mathcal{L}_{i}$ in terms of the Lorentz scalars $\psibar_{a}\psi_{b}$ and $\psibar_{a}\overline{\gamma}\psi_{b}$ using
\begin{align}
\label{CDA}
2\overline{D}=&(\psibar_{1}\psi_{2})(\psibar_{2}\psi_{1})- (\psibar_{1}\overline{\gamma}\psi_{2})(\psibar_{2}\overline{\gamma}\psi_{1})\nn\\
&-(\psibar_{1}\psi_{1})(\psibar_{2}\psi_{2})- (\psibar_{1}\overline{\gamma}\psi_{1})(\psibar_{2}\overline{\gamma}\psi_{2})\ ,
\end{align}
and
\begin{align}
\label{CDB}
&2\overline{C}=\frac{1}{2}\Big{\lbrace}(\psibar_{1}\psi_{1})\2-(\psibar_{1}\overline{\gamma}\psi_{1})\2+(\psibar_{2}\psi_{2})\2-(\psibar_{2}\overline{\gamma}\psi_{2})\2\nn\\
&-\sigma\big{[}(\psibar_{1}\psi_{2})\2-(\psibar_{1}\overline{\gamma}\psi_{2})\2+(\psibar_{2}\psi_{1})\2-(\psibar_{2}\overline{\gamma}\psi_{1})\2\big{]}\!\Big{\rbrace}\,\! .\!\!
\end{align}
This is a type of colored Gross-Neveu model.

\subsection{Complex Grassmann variables\label{subsect: Complex Grassmann variables}}

%Absatz
For $\sigma=-1$ we can combine the two Majorana spinors into a complex Dirac spinor. The complex structure for the Grassmann variables is realized by grouping the two colors into a complex Grassmann variable, 
\be\label{LH}
\zeta=\psi_{1}+i\psi_{2}\;,\quad \overline{\zeta}=\psibar_{1}-i\psibar_{2}\;.
\ee
In this case the interaction takes the particularly simple form
\be\label{LI}
\mathcal{L}_i=-\frac{1}{2}(\overline{\zeta}\gamma_{\mu}\zeta)^{*}(\overline{\zeta}\gamma_{\mu}\zeta)\;,
\ee
and the action reads
\be\label{89}
S=-\int_ {t,x}\bigl{\lbrace} Re(\overline{\zeta}\gamma^{\mu}\partial_{\mu}\zeta)+\dfrac{1}{2}(\overline{\zeta}\gamma^{\mu}\zeta)^{*}(\overline{\zeta}\gamma_{\mu}\zeta)\bigr{\rbrace}\;.
\ee

\subsection{Continuity constraint\label{subsect: Continuity constraint}}

%Absatz
The Grassmann variables $\psi(t,x)$ and $\overline{\psi}(t,x)$ are independent. This originates from the association of $\overline{\psi}(t,x)$ to Grassmann variables at $t+\varepsilon$.
In the discrete formulation of the Grassmann functional integral they are independent integration variables. For certain purposes in the continuum limit one may assume by continuity $\psi(t +\varepsilon)=\psi(t)$ and therefore impose the continuity constraint
\bel{C8C}
\psibar_{\gamma}(t,x)=\psi_{\gamma}(t,x)\ .
\ee
This implies
\bel{99E}
\psibar_a\gamma^\mu\psi_b=\psi_a\supt{T}\gamma^0\gamma^\mu\psi_b\ ,
\ee
and therefore
\begin{align}
\label{99F}
\psibar_a\gamma^0\psi_b=&-\gl\psi_{Ra}\psi_{Rb}+\psi_{La}\psi_{Lb}\gr\ ,\nn\\
\psibar_a\gamma^1\psi_b=&-\psi_{Ra}\psi_{Rb}+\psi_{La}\psi_{Lb}\ .
\end{align}
Nonzero contributions require $a\neq b$, such that only the term $\sim\sigma$ in eq.~\eqref{99A} contributes for $\sigma=-1$ to the interaction term
\bel{99G}
\mathcal{L}_i=8\psi_{R1}\psi_{R2}\psi_{L1}\psi_{L2}\ .
\ee
For $\sigma=1$ one has $\mathcal{L}_i=0$.
In the complex formulation the continuity constraint results in
\bel{99C}
\zetabar_\gamma(x)=\zeta_\gamma^*(x)\ .
\ee
With the relation
\bel{99D}
\zeta\herm=\gl\zeta_R^*,\zeta_L^*\gr\ ,\quad \zetabar=\zeta\herm\gamma^0=\gl\zeta_L^*,-\zeta_R^*\gr\ ,
\ee
the identification~\eqref{99C},~\eqref{99D} is compatible with the Lorentz transformations and the time evolution equations or field equations.

%Absatz
In eq.~\eqref{89} the continuity constraint~\eqref{99D} relates $\zetabar$ to~$\zeta^{*}$.
 With
\begin{align}
\label{J8A}
\zeta_+=&\psi\subt{R1}+i\psi\subt{R2}\ ,\quad \zeta_-=\psi\subt{L1}+i\psi\subt{L2}\ ,\quad \zetabar=\gl\zetabar_+,\zetabar_-\gr\ ,\nn\\
\zetabar_+=&\psibar\subt{L1}-i\psi\subt{L2}=\zeta_-^*\ ,\quad \zetabar_-=-\gl\psibar\subt{R1}-i\psibar\subt{R2}\gr=-\zeta_+^*\ ,
\end{align}
one obtains by partial integration $Re(\overline{\zeta}\gamma^{\mu}\partial_{\mu}\zeta)\rightarrow\overline{\zeta}\gamma^{\mu}\partial_{\mu}\zeta$ and $(\overline{\zeta}\gamma^{\mu}\zeta)^{*}\rightarrow -\overline{\zeta}\gamma^{\mu}\zeta$. We end with a fermion model with action
\be\label{98}
S=\int_{t,x}\lbrace -\overline{\zeta}\gamma^{\mu}\partial_{\mu}\zeta+\dfrac{1}{2}(\overline{\zeta}\gamma^{\mu}\zeta)(\overline{\zeta}\gamma_{\mu}\zeta)\rbrace\;.
\ee
This is a particular case of the Thirring model~\cite{THI,KLA,AAR,FAIV} with a particular value of the coupling. The Thirring model is exactly solvable. With the identification~\eqref{99C},~\eqref{99D} the interaction term simplifies
\begin{align}
\label{99}
\overline{C}+\overline{D}=&-\dfrac{1}{4}(\overline{\zeta}\gamma^{\mu}\zeta)^{*}(\overline{\zeta}\gamma_{\mu}\zeta)=
\dfrac{1}{4}(\overline{\zeta}\gamma^{\mu}\zeta)(\overline{\zeta}\gamma_{\mu}\zeta)\nn \\=&\overline{\zeta}_{+}\overline{\zeta}_{-}\zeta_{+}\zeta_{-}=4\psi\subt{R1}\psi\subt{R2}\psi\subt{L1}\psi\subt{L2},
\end{align}
in accordance with eq.~\eqref{99G}. This yields eq.~\eqref{105D}.

\section{Symmetries of the Thirring-type model}
\label{ap.D}

%Absatz
In this appendix we list symmetries of the discrete Thirring type fermionic quantum field theory. They are in direct correspondence to properties of the cellular automaton. The numbers refer to the numbers in sect.~\ref{sec: 04}.
\begin{enumerate}
\item The separately conserved number of left-moving and right moving particles corresponds to global chiral symmetry. In the complex formulation the action is invariant under the chiral transformations
\begin{align}\label{T1}
\psi_{+}&\rightarrow e^{i\alpha_{+}}\psi_{+}\,\,\,\,\,,\quad \psi_{-}\rightarrow e^{i\alpha_{-}}\psi_{-}\,,\nn\\
\overline{\psi}_{+}&\rightarrow e^{-i\alpha_{+}}\overline{\psi}_{+}\,,\quad\overline{\psi}_{-}=e^{-i\alpha_{-}}\overline{\psi}_{-}\;,
\end{align}
as seen directly from
\begin{align}
\label{T2}
&\psi=\begin{pmatrix}
\psi_{+}\\\psi_{-}
\end{pmatrix}
\,,\quad   \overline{\psi}=(\overline{\psi}_{-}\,,\,\overline{\psi}_{+})\;,\nn\\
&\overline{\psi}\gamma^{\mu}\psi=\overline{\psi}_{+}\gamma^{\mu}\psi_{+}+\overline{\psi}_{-}\gamma^{\mu}\psi_{-}\:.
\end{align}
In the formulation~\eqref{LJ} the chiral symmetry corresponds to separate $SO(2) $- color rotations for the right movers and left movers.
\label{i02: 01}
\item
The motion of all particles with velocity $|v|=c=1$ reflects Lorentz symmetry. A particle mass is forbidden by chiral symmetry. For arbitrarily occupied configurations a simple additional particle still moves on a straight line. \label{i02: 02}
\item The fermionic model has bosonic "composite particles" moving freely with speed of light.\label{i02: 03}
\item The interaction in the fermionic model is local, involving precisely the Grassmann variables for four different particles
\be\label{T4}
S_\textup{int}=2\int_{t,x}\overline{\psi}_{-}(x)\overline{\psi}_{+}(x)\psi_{+}(x)\psi_{-}(x)\;.
\ee
This describes $2$ to $2$ scattering. A scattering of bosonic two-particle composite states would involve eight Grassmann variables for boson-boson scattering and six  Grassmann variables for boson-fermion scattering. It is not present.
\label{i02: 04}
\item The action \eqref{98} is invariant under a euclidean rotation in the $t-x$ plane, 
\be\label{T5}
t\rightarrow x\,,\quad x\rightarrow -t\;,
\ee
if the Grassmann variables transform as 
\be\label{T6}
\psi\rightarrow\psi'=-\gamma^{0}\psi\,,\quad\overline{\psi}\rightarrow\overline{\psi}'=\overline{\psi}\gamma^{1}\;.
\ee
\label{i02: 05}
\item The fermionic model is invariant under parity transformations
\begin{align}\label{T7}
&P:\, x\rightarrow -x\,, \\
&\psi(t,x)\rightarrow\gamma^{1}\psi(t,-x)\,,\, \overline{\psi}(t,x)\rightarrow -\overline{\psi}(t, -x)\gamma^{1}\;.\nn
\end{align}
The parity transformation changes $\psi_{L}\leftrightarrow\psi_{R}$. The model is also invariant under time reflections.
\label{i02: 06}
\item The action~\eqref{98} is invariant under the discrete transformations
\be\label{T9}
\psi \rightarrow -\psi\,,\quad\overline{\psi}\rightarrow -\overline{\psi}
\ee
and
\be\label{T10}
R_{2}:\quad\psi_{2}\rightarrow -\psi_{2}\;,\quad\overline{\psi}_{2}\rightarrow -\overline{\psi}_{2}\;.
\ee
The combination of these symmetries implies that the action can only admit terms with an even number of factors for each color. Thus any change of the number of particles with a given color always involves two units.
\label{i02: 07}
\item The ``color exchange symmetry" 
\be\label{T11}
\psi_{1}\leftrightarrow\psi_{2}\,,\quad\overline{\psi}_{1}\leftrightarrow\overline{\psi}_{2}
\ee
is a discrete symmetry beyond the continuous color rotations. It is manifest in eq.~\eqref{LJ}.
\label{i02: 08}
\item
Particle-hole symmetry maps
\be\label{LK}
C:\quad \psi_{\eta\alpha}(t,x)\leftrightarrow\overline{\psi}_{\eta\alpha}(t,x)\;,
\ee
and reads in the doublet notation for Dirac spinors
\be\label{LL}
C:\quad \psi \leftrightarrow\gamma^{0}\overline{\psi}^{T}\;,\quad\overline{\psi}\leftrightarrow\psi^{T}\gamma^{0}\;.
\ee
With $\overline{\psi}\gamma^{\mu}\psi\rightarrow -\overline{\psi}\gamma^{\mu}\psi$ the invariance of the action~\eqref{LJ} is easily verified.
\label{i02: 09}
\end{enumerate}

%Absatz
The one to one correspondence between properties of the cellular automaton and the symmetries of the fermionic model can help to find the fermionic model for a given automaton, or vice versa, if a direct calculation as performed in sects~\ref{sec: 02},~\ref{sec: 03}, is cumbersome.

\section{Conjugate Grassmann wave function}
\label{ap.E}

%Absazt
In this appendix we discuss the conjugate Grassmann wave function. Together with the Grassmann wave function it can be used to evaluate expectation values of observables that are expressed by Grassmann operators. The Grassmann wave function can be extracted from the Grassmann functional integral by integration over Grassmann variables $\psi(t')$ for $t'>t$. For suitable boundary conditions it can be expressed by the wave function $q(t)$. This establishes how observables in the Grassmann formulation and their expectation values can be mapped to observables for the associated cellular automaton.

%Absatz
We can associate~\cite{CWFCS} to the Grassmann wave function~\eqref{G1} a conjugate Grassmann wave function $\widehat{g}(t)$ which depends on $\psibar_{\alpha}(t)$,
\bel{G7}
\widehat{g}(t)=q_{\tau}(t)g_{\tau}'[\,\psibar(t)]\ .
\ee
By use of the identities~\eqref{14},~\eqref{12},~\eqref{16} one has
\bel{G8}
\int\cD\psi(t)\cD\psibar(t)\exp\big{\{}\psibar_{\alpha}(t)\psi_{\alpha}(t)\big{\}}g_{\tau}'[\,\psibar(t)]g_{\rho}[\psi(t)]=\delta_{\tau\rho}\ .
\ee
This implies
\bel{G9}
\int\cD\psi(t)\cD\psibar(t)\exp\big{\{}\psibar_{\alpha}(t)\psi_{\alpha}(t)\big{\}}\widehat{g}(t)g(t)=q_{\tau}(t)q_{\tau}(t)=1\ .
\ee
The conjugate wave function is convenient for the implementation of observables in the fermionic formulation - see ref.~\cite{CWFCS} for details.

%Absatz
For example, the operator for the occupation number of a particle of type $\gamma$ at position $x$ at time $t$ is given by (no sum over $\gamma$ here)
\bel{G10}
\mathcal{N}_{\gamma}=\psibar_{\gamma}(t,x)\psi_{\gamma}(t,x)\ .
\ee
We associate (for $t=t\inn+2m\epstil$) a basis element $g_{\tau}$, for which $\tilde{a}_{\gamma}(x)=\psi_{\gamma}(x)$ in eq.~\eqref{06}, to a microscopic state with no particle of type $\gamma$ present at $x$. For all those basis elements one has $\mathcal{N}_{\gamma}(x)g_{\tau}=0$. On the other hand, the set of basis elements with $\tilde{a}_{\gamma}(x)=1$ corresponds to states for which a particle of type $\gamma$ is present at $x$. Only for those basis elements one has $\mathcal{N}_{\gamma}(x)g_{\tau}=\psibar_{\gamma}(x)\psi_{\gamma}(x)g_{\tau}\neq0$. One infers
\bel{G11}
\int\text{D}\psi g_{\tau}'[\psibar]\mathcal{N}_{\gamma}(x)g_{\rho}[\psi]=\begin{cases} 0\ &\mbox{for no particle $(\gamma,x)$} \\ \delta_{\tau\rho}\ &\mbox{for particle $(\gamma,x)$}\end{cases}\ ,
\ee
where
\begin{align}\label{G12}
\int\text{D}\psi=&\int\cD\psi\cD\psibar\exp\big{\{}\psibar_{\alpha}\psi_{\alpha}\big{\}}\nn\\
=&\prod_{\alpha}\int\text{d}\psi_{\alpha}\text{d}\psibar_{\alpha}(1+\psibar_{\alpha}\psi_{\alpha})\ .
\end{align}
This implies a simple expression for the mean particle number $n_{\gamma}(x)$
\begin{align}\label{G13}
\exval{n_{\gamma}(x)}=&\int\text{D}\psi\widehat{g}[\,\psibar\,]\mathcal{N}_{\gamma}(x)g[\psi]\nn\\
=&\sum_{\tau}q_{\tau}[n_{\gamma}(x)]_{\tau}q_{\tau}=\sum_{\tau}[n_{\gamma}(x)]_{\tau}p_{\tau}\ ,
\end{align}
where $[n_{\gamma}(x)]_{\tau}=(1,0)$ is the particle number in the microscopic state $\tau$. The expression~\eqref{G13} is precisely the mean particle number for the probabilistic cellular automaton. The relations~\eqref{G11}-\eqref{G13} hold for every time $t$. Since the evolution of the wave function is the same for the cellular automaton and the fermionic model, the expectation values are the same for all times. This extends to products of particle numbers and therefore to correlation functions. One can actually construct further interesting observables beyond such products, including a momentum observable~\cite{Wetterich:2020kqi}.

%Absatz
Arbitrary wave functions $q(t)$ are allowed quantum states for the fermionic quantum field theory. These real unit vectors form a Hilbert space once a suitable complex structure is introduced. On the other side, arbitrary probability distributions for the probabilistic cellular automaton can be described by suitable wave functions. We conclude that the probabilistic cellular automaton and the fermionic quantum field theory are equivalent descriptions for the same physical reality or model.

%Absatz
The conjugate Grassmann wave function can be related to the Grassmann functional integral~\cite{CWFCS}, now to the part for $t'>t$,
\begin{align}\label{G14}
\widehat{g}[\,\psibar(t)]=\int&\text{d}\psi'(t)\int\cD\psi(t'\geq t+\epstil)\widehat{g}(t\ff)\nn\\
&\times\Big{(}\prod_{t'\geq t+\epstil}\Ktil(t')\Big{)}\Ktil'(t)\exp\big{\{}-\psibar_{\alpha}(t)\psi_{\alpha}'(t)\big{\}}\ .
\end{align}
Here $\Ktil'(t)$ obtains from $\Ktil(t)$ by replacing $\psi(t)\to\psi'(t)$.
Combining with the expression~\eqref{G9} for the Grassmann wave function this amounts to
\bel{G15}
Z=\int\text{D}\psi(t)\widehat{g}[\,\psibar(t)]g[\psi(t)]\ .
\ee
Indeed, the integration over $\psi'(t)$, combined with the exponential factor in $\text{D}\psi$, yields
\bel{G16}
\int\text{d}\psibar(t)\text{d}\psi'(t)\exp\Big{\{}\psibar_{\alpha}(t)\gl\psi_{\alpha}(t)-\psi_{\alpha}'(t)\gr\Big{\}}\Ktil'(t)=\Ktil(t)\ .
\ee
The product $\widehat{g}[\,\psibar(t)]g[\psi(t)]$ involves the product of local factors $\Ktil(t')$ over all $t'$. Also the functional integral is over the Grassmann variables at all $t'$. One recovers eq.~\eqref{01} with appropriate boundary factors.

%Absatz
For $\widehat{g}(t)$, as defined by eq.~\eqref{G14}, to coincide with the definition~\eqref{G7} we take the boundary factor $\widehat{g}(t\ff)$ to coincide with the expression~\eqref{G7} at $t\ff$. One can show~\cite{CWFCS} that the time evolution of $\widehat{g}(t)$ is the same for the definitions~\eqref{G7} and~\eqref{G14}. The identity therefore holds for all $t$. We therefore find a functional integral expression for the mean particle number
\bel{G17}
\exval{n_{\gamma}(t)}=\int\text{D}\psi\widehat{g}\ff\,\mathcal{N}_{\gamma}(t,x)e^{-S}g\inn\ .
\ee
This extends to observables constructed from occupation numbers. Eq.~\eqref{G17} is a familiar expression for expectation values in fermionic quantum field theories formulated as a Grassmann functional integral. We could have started from this expression and derive the wave function $q(t)$ according to eq.~\eqref{G6}. This demonstrates that wave functions are the appropriate concept for the description of time-local subsystems of an overall probabilistic system formulated for all times~\cite{Wetterich:2020kqi}. This applies directly to the probabilistic cellular automaton that can be seen as a classical statistical system.

\section{Density matrix}
\label{ap.F}

The density matrix is a central tool in quantum mechanics. It permits to describe more general probabilistic states, namely mixed states, in addition to the special case of pure states. In our context it allows more general boundary conditions. Furthermore, the elements of the density matrix can often directly be related to expectation values and correlations to suitable observables. We will see that the density matrix is a suitable object for the introduction of the complex structure of quantum mechanics.
The density matrix can be formulated both for the fermionic quantum field theory and the probabilistic cellular automaton. It is identical for both pictures.

\subsection{Density matrix for fermions}
\label{subsec: Density matrix for fermions}

%Absatz
For general Grassmann wave functions $g(t)$ and conjugate Grassmann wave functions $\ghat(t)$,
\bel{DM1}
g(t)=\tilde{q}_{\tau}(t)g_{\tau}[\psi(t)]\ ,\quad \ghat(t)=\bar{q}_{\tau}(t)g_{\tau}'[\,\psibar(t)]\ ,
\ee
we define the Grassmann density matrix as a bilinear
\begin{align}
\label{DM2}
\tilde{\rho}[\,\psibar(t),\psi(t)]=&g[\psi(t)]\ghat[\,\psibar(t)])\nn\\
=&\tilde{q}_{\tau}(t)\bar{q}_{\rho}(t)g_{\tau}[\psi(t)]g_{\rho}'[\,\psibar(t)]\ .
\end{align}
It depends on the Grassmann variables $\psi(t)$ and $\psibar(t)$. In the original formulation it involves variables at two time layers $t$ and $t-\epstil$.
The coefficients of a double expansion in $\psi(t)$ and $\psibar(t)$ are the elements of the quantum density matrix
\bel{DM3}
\tilde{\rho}[\,\psibar(t),\psi(t)]=\widehat{\rho}_{\tau\rho}(t)g_{\tau}[\psi(t)]g_{\rho}'[\,\psibar(t)]\ ,
\ee
with
\bel{DM4}
\widehat{\rho}_{\tau\rho}(t)=\tilde{q}_{\tau}(t)\bar{q}_{\rho}(t)\ .
\ee

%Absatz
We can define $\widehat{g}[\,\psibar(t)]$ by the functional identity~\eqref{G14} for general final boundary factors $\ghat(t\ff)$. Eq.~\eqref{G15} becomes
\bel{DM5}
Z=\tilde{q}(t)_{\tau}\bar{q}_{\tau}(t)\ ,
\ee
such that $\ghat(t\ff)$ is restricted by the normalization $Z=1$. One would like to identify the diagonal elements of the density matrix with the local probabilities (no sum over $\tau$ here)
\bel{DM6}
p_{\tau}=\widehat{\rho}_{\tau\tau}=\tilde{q}_{\tau}(t)\bar{q}_{\tau}(t)\geq0\ .
\ee
The positivity of $p_{\tau}(t)$ further restricts the possible choices for $\ghat(t\ff)$. We will present below a simple criterion for eq.~\eqref{DM6} based on the positivity of the density matrix. If eq.~\eqref{DM6} holds, the generalization of eq.~\eqref{G17} leads for the corresponding generalization of eq.~\eqref{G13} a probabilistic interpretation
\begin{align}\label{DM7}
&\exval{n_{\gamma}(t,x)}=\int\text{D}\psi\ghat[\,\psibar(t)]\mathcal{N}_{\gamma}(t,x)g[\psi(t)]\nn\\
&=\sum_{\tau}[n_{\gamma}(t,x)]_{\tau}\tilde{q}_{\tau}(t)\bar{q}_{\tau}(t)=\sum_{\tau}[n_{\gamma}(t,x)]p_{\tau}(t)\ .
\end{align}

%Absatz
We can further generalize the boundary conditions by considering a set of probabilistic states with different boundary factors $\gl g\inn^{(\alpha)}(t\inn)$, $\ghat\ff^{(\alpha)}(t\ff)\gr$, labeled by $\alpha$. For each $\alpha$ one has
\bel{DM8}
\widehat{\rho}_{\tau\rho}^{(\alpha)}(t)=\tilde{q}_{\tau}^{(\alpha)}(t)\bar{q}_{\rho}^{(\alpha)}(t)\ .
\ee
Taking a weighed sum over different sets $\alpha$, with positive probabilities $w_{\alpha}\geq0$, $\Sigma_{\alpha}w_{\alpha}=1$, one arrives at the general density matrix
\bel{DM9}
\widehat{\rho}_{\tau\rho}(t)=\sum_{\alpha}w_{\alpha}\tilde{q}_{\tau}^{(\alpha)}(t)\bar{q}_{\rho}^{(\alpha)}(t)\ .
\ee
The normalization of the partition function is kept,
\begin{align}\label{DM10}
Z=&\sum_{\tau}\widehat{\rho}_{\tau\tau}=\sum_{\alpha}w_{\alpha}\tilde{q}_{\tau}^{(\alpha)}(t)\bar{q}_{\tau}^{(\alpha)}(t)\nn\\
=&\sum_{\alpha}w_{\alpha}\sum_{\tau}\widehat{\rho}_{\tau\tau}^{(\alpha)}=\sum_{\alpha}w_{\alpha}=1\ .
\end{align}
The local probabilities are the diagonal elements of the generalized density matrix,
\bel{DM11}
p_{\tau}(t)=\widehat{\rho}_{\tau\tau}(t)\ ,
\ee
and the expectation value of the occupation number takes the form
\bel{DM12}
\exval{n_{\gamma}(t,x)}=\sum_{\tau}[n_{\gamma}(t,x)]_{\tau}\widehat{\rho}_{\tau\tau}(t)=\tr\big{\{}\widehat{n}_{\gamma}(t,x)\widehat{\rho}(t)\big{\}}\ .
\ee
With the definition of the diagonal operator
\bel{DM13}
[\widehat{n}_{\gamma}(t,x)]_{\tau\rho}=[n_{\gamma}(t,x)]_{\tau}\delta_{\tau\rho}
\ee
eq.~\eqref{DM12} is the familiar formula for quantum mechanics in a real language.

%Absatz
Let us consider the transformation
\bel{DM14}
\psi_{\alpha}(t)\leftrightarrow\psibar_{\alpha}(t)\ ,
\ee
combined with a total reordering $R$ of all Grassmann variables. This reordering obeys for the product of two Grassmann elements
\be
\label{DM15A}
R(g_{1}g_{2})=R(g_{2})R(g_{1})\ ,
\ee
and
\be
\label{DM16A}
R(g_{\tau})=g_{\tau}'
\ee
Correspondingly, the Grassmann density matrix transforms as
\begin{align}\label{DM15}
\tilde{\rho}[\psi(t),&\psibar(t)]\to\widehat{\rho}_{\tau\rho}(t)R\gl g_{\tau}[\,\psibar(t)]g_{\rho}'[\psi(t)]\gr\nn\\
=&\widehat{\rho}_{\tau\rho}(t)g_{\rho}[\psi(t)]g_{\tau}'[\,\psibar(t)]=\widehat{\rho}'_{\rho\tau}(t)g_{\rho}[\psi(t)]g_{\tau}'[\,\psibar(t)]\ .
\end{align}
This transformation results in a transposition of the density matrix
\bel{DM16}
\widehat{\rho}(t)\to\widehat{\rho}'(t)=\widehat{\rho}\supt{T}(t)\ ,\quad \widehat{\rho}'_{\tau\rho}(t)=\widehat{\rho}_{\rho\tau}(t)\ .
\ee

\subsection{Time evolution of density matrix}
\label{subsec: Time evolution of density matrix}

%Absatz
For the time evolution of the density matrix $\widehat{\rho}(t)$  we need the time evolution of the conjugate wave function $\bar{q}(t)$. From the definition of the conjugate wave function~\eqref{G14} one infers its relation to the hole wave function $\gbar$
\begin{align}\label{DM17}
\gbar[\psi(t)]=&\int\cD\psibar(t)\exp\{\psibar_{\alpha}(t)\psi_{\alpha}(t)\}\ghat[\,\psibar(t)]\nn\\
=&\int\cD\psi(t'\geq t+\epstil)\ghat(t\ff)\prod_{t'\geq t}\Ktil(t)\nn\\
=&\bar{q}_{\tau}(t)\int\cD\psibar(t)\exp\{\psibar_{\alpha}(t)\psi_{\alpha}(t)\}g_{\tau}'[\,\psibar(t)]\nn\\
=&\bar{q}_{\tau}(t)\gbar_{\tau}[\psi(t)]\ .
\end{align}
With
\begin{align}\label{DM18}
\gbar[&\psi(t-\varepsilon)]=\bar{q}_{\rho}(t-\varepsilon)\bar{g}_{\rho}[\psi(t-\varepsilon)]\nn\\
&=\int\cD\psi(t)\cD\psi(t-\epstil)\bar{g}[\psi(t)]\Ktil(t-\epstil)\Ktil(t-2\epstil)\nn\\
&=\bar{q}_{\tau}(t)\widehat{S}_{\tau\sigma}(t-\epstil)\widehat{S}_{\sigma\rho}((t-2\epstil)\bar{g}_{\rho}[\psi((t-2\epstil)]
\end{align}
one finds
\begin{align}\label{DM19}
\bar{q}_{\rho}(t-\varepsilon)=&\bar{q}_{\tau}(t)\widehat{S}_{\tau\sigma}(t-\epstil)\widehat{S}_{\sigma\rho}(t-2\epstil)\nn\\
=&\bar{q}_{\tau}(t)\gl\widehat{S}\subt{int}\widehat{S}\subt{free}\gr_{\tau\rho}=\bar{q}_{\tau}(t)\widehat{S}_{\tau\rho}(t-\varepsilon)\ ,
\end{align}
where the last expression uses the coarse grained step evolution operator~\eqref{62}. Inversion yields
\bel{DM20}
\bar{q}_{\tau}(\te)=\bar{q}_{\rho}(t)\widehat{S}_{\rho\tau}^{-1}(t)\ ,
\ee
and we conclude the evolution law for the density matrix
\bel{DM21}
\widehat{\rho}(\te)=\widehat{S}(t)\widehat{\rho}(t)\widehat{S}^{-1}(t)=\widehat{S}(t)\widehat{\rho}(t)\widehat{S}\supt{T}(t)\ ,
\ee
where the last expression uses $\widehat{S}\widehat{S}\supt{T}=1$.
This corresponds to the unitary evolution law of quantum mechanics for the case of a real density matrix and real step evolution operator.

\subsection{Density matrix for cellular automaton}
\label{subsec: Density matrix for cellular automaton}

In eq.~\eqref{G14} we have defined the conjugate wave function $\qbar(t)$ in terms of the boundary conditions at final time $t\ff$. Similarly, we can consider the probabilistic cellular automaton as a classical statistical system for which we fix boundary conditions both at $t\inn$ and $t\ff$ in the form of initial and final wave functions $\qtil(t\inn)$ and $\qbar(t\ff)$. For an invertible cellular automaton we can follow $\qbar(t)$ backwards from $t\ff$ to an arbitrary time $t$, and $\qtil(t)$ forwards from $t\inn$ to $t$. The pair of wave functions $\gl\qtil(t),\qbar(t)\gr$ therefore involve boundary information from both boundary conditions at $t\inn$ and $t\ff$. The same holds for the density matrix which we define again by eq.~\eqref{DM4}. The density matrix $\widehat\rho(t)$ contains information beyond the local probabilities $p_\tau(t)$, which correspond to the diagonal elements of $\widehat\rho(t)$ according to eq.~\eqref{DM11}.
We will see below how the information in the off-diagonal elements for $\widehat\rho$ can be used for the computation of the expectation values of additional local observables that are represented by off-diagonal operators. The density matrix~\eqref{DM4} is a pure state density matrix. More general density matrices for mixed states are again defined by eq.~\eqref{DM9}.

%Absatz
The possible differences between $\qbar(t\ff)$ and $\qtil(t\ff)$ will not play an important role for this paper. We will mainly focus on $\qbar(t\ff)=\qtil(t\ff)$, which implies for all $t$
\bel{DMA}
\qtil(t)=\qbar(t)=q(t)\ .
\ee
We will distinguish between $\qtil(t)$ and $\qbar(t)$ only if we want to indicate the different formal status. With eq.~\eqref{DMA} the density matrix is symmetric for all $t$
\bel{DMB}
\widehat\rho^{T}(t)=\widehat\rho(t)\ .
\ee
For the boundary condition~\eqref{DMA} the evolution of the density matrix is directly formulated as an initial value problem. All necessary boundary information is encoded in the initial wave function $q(t\inn)$.

\section{Operators for observables}
\label{ap.G}

The description of observables by non-commuting operators is a characteristic feature of quantum systems. Since our automaton is equivalent to a quantum system for fermions we will find the same non-commuting operator structures for the classical statistical system of the cellular automaton. This constitutes a simple example how non-commuting operator structures emerge in classical statistics~\cite{CWPT,Wetterich:2020kqi}. We will see that the density matrix $\widehat{\rho}(t)$ contains sufficient probabilistic information for the computation of expectation values of observables that are not expressed as functions of occupation numbers at $t$. This includes energy (sect.~\ref{sec: 07}) and momentum (sect.~\ref{sec: 09}).
The operators associated to these observables are typically not diagonal and do not commute with each other. Their expectation values involve the off-diagonal elements of the density matrix.

\subsection{Quantum rule for expectation values}
\label{subsec: Quantum rule for expectation values}

For an observable that takes for the configuration $\tau$ the value $A_\tau$ we define the diagonal operator $\widehat{A}$ with elements
\bel{QA1}
\widehat{A}_{\tau\rho}=A_\tau\delta_{\tau\rho}\ .
\ee
Its expectation value follows the quantum rule
\bel{QA2}
\langle A(t)\rangle=\tr\big{\{}\widehat{A}\widehat{\rho}(t)\big{\}}\ .
\ee
This quantum rule is a consequence of the association of the diagonal elements $\widehat{\rho}_{\tau\tau}(t)$ with the probabilities $p_\tau(t)$ that at time $t$ the configuration $\tau$ is realized.
It follows directly from the basic definition of expectation values in classical statistics
\bel{QA3}
\langle A(t)\rangle=\sum_\tau p_\tau(t) A_\tau=\sum_\tau\rho_{\tau\tau}(t)A_\tau=\tr\big{\{}\widehat{A}\widehat{\rho}(t)\big{\}}\ .
\ee
It is therefore not a separate postulate or axiom.

%Absatz
The observables described by the diagonal operators~\eqref{QA1} can be occupation numbers $n_\gamma(x)$, arbitrary products of the occupation numbers, or more generally arbitrary real functions of occupation numbers. Here occupation numbers all refer to the same time $t$ for which the density matrix $\widehat{\rho}(t)$ is taken in eq.~\eqref{QA2}. The \qq{diagonal observables}~\eqref{QA1} therefore include all equal-time correlation functions of occupation numbers at $t$. The time evolution of expectation values is encoded in the time evolution of the density matrix . This corresponds to the Schrödinger-picture of quantum mechanics For the special case of a pure state density matrix $\widehat\rho_{\tau\rho}(t)=q_\tau(t)q_\rho(t)$ one obtains the quantum rule for expectation values in terms of the wave function
\bel{QA4}
\langle A(t)\rangle=\langle q(t)\widehat{A}q(t)\rangle=q_\tau(t)\widehat{A}_{\tau\rho}q_\rho(t)\ .
\ee
This follows by inserting the pure state density matrix into the quantum rule~\eqref{QA2}. So far we discuss here \qq{real quantum mechanics} with real $\widehat{\rho}$ and $q$. We will introduce a complex structure in the next section. For operators that are compatible with the complex structure the quantum rule for observables will then take the familiar form in terms of a hermitian density matrix and hermitian operators associated to observables.

\subsection{Observables at different times}
\label{subsec: Observables at different times}

The density matrix $\widehat\rho(t)$ contains sufficient probabilistic information for a computation of expectation values of observables at different times. As an example, we consider observables $A(t')$ that are functions of occupation numbers at the time $t'=t-\eps$. We can associate to these observables the operators
\bel{QA5}
\widehat{A}(t'; t)=\Shat(t-\eps)\widehat{A}\Shat^T(t-\eps)\ ,
\ee
or
\bel{QA6}
\widehat{A}(t'; t)_{\tau\rho}=\sum_\sigma A_\sigma\Shat(t-\eps)_{\tau\sigma}\Shat(t-\eps)_{\rho\sigma}\ .
\ee
The expectation value of $A(t')$ can be evaluated by use of the quantum rule~\eqref{QA2} from the density matrix $\widehat\rho(t)$
\bel{QA7}
\langle A(t')\rangle=\tr\big{\{}\widehat{A}(t'; t)\widehat\rho(t)\big{\}}\ .
\ee
Here the first time-argument $t'$ in $\widehat{A}(t'; t)$ refers to the time $t'$ for the occupation numbers $n_\gamma(t',x)$ for which $A(t')$ is defined as a function, while the second time argument $t$ indicates the reference time $t$ for which the density matrix $\widehat\rho(t)$ is taken in eq. ~\eqref{QA2}. Eq.~\eqref{QA7} follows from eq.~\eqref{QA2} at $t-\eps$ and the evolution law~\eqref{DM21} for the density matrix,
\begin{align}
\label{QA8}
\langle A(t'\rangle =&\tr\big\{\widehat{A}\widehat\rho(t')\big\}=\tr\widehat{A}\Shat^T(t-\eps)\widehat\rho(t)\Shat(t-\eps)\big\}\nn\\
=&\tr\big\{\Shat(t-\eps)\widehat{A}\Shat^T(t-\eps)\widehat\rho(t)\big\}\ .
\end{align}
The use of time dependent operators $\widehat{A}(t',t)$ corresponds to the Heisenberg picture in quantum mechanics. In general both the time evolution of $\widehat\rho(t)$ and the $t'$-dependence of $\widehat{A}(t';t)$ contribute to the time dependence of the expectation value $\langle A(t')\rangle$.

For a general $\widehat{S}(t)$ the operator $\widehat{A}(t';t)$  for $t'\neq t$ is no longer diagonal and does not commute with $\widehat{A}$. This is related to the fact that the step evolution operator is not a diagonal matrix. The operators $\widehat{A}$ and $\Shat$ do, in general, not commute. For the particular case where $\widehat{A}$ and $\Shat$ commute one has $\widehat{A}(t';t)=\widehat{A}$ and therefore $\langle A(t')\rangle=\langle A(t)\rangle$.
If $\Shat(t)$ is independent of $t$ and commutes with $\widehat{A}$, the observable $A$ is a conserved quantity with the same expectation value for arbitrary $t'$ and $t$,
\bel{QA9}
\big[\widehat{A},\Shat\big]=0\implies \langle A(t')\rangle=\langle A(t)\rangle\ .
\ee
Non-conserved observables are represented by operators that do not commute with $\widehat{S}$. Nevertheless, for the particular case of  cellular automata for which  $\widehat{S}$ is a unique jump operator the operator $\widehat{A}(t';t)$ is found to be diagonal. It therefore commutes with $\widehat{A}=\widehat{A}(t;t)$. Examples for non-commuting operators will be given later.

%Absatz
The construction of Heisenberg operators $\widehat{A}(t';t)$ is not limited to $t'=t-\eps$. It can be done for the whole range of $t'$ for which the step evolution operator is known in the range between $t'$ and $t$. In particular, one has for constant $\Shat$ and $t'=t+n\eps$
\bel{QA10}
\widehat{A}(t+n\eps, t)=\Shat^{-n}\widehat{A}\Shat^n\ ,
\ee
where we recall $\Shat^{-1}=\Shat^T$. This permits the definition of the time-derivative of observables as
\bel{QA11}
\dot{A}(t)=\frac{1}{4\eps}\Big(A(t+2\eps)-A(t-2\eps)\Big)\ ,
\ee
which is represented by the operator
\bel{QA12}
\dot{\widehat{A}}(t)=\frac{1}{4\eps}\Big[\Shat^{-2}\widehat{A}\Shat^2-\Shat^2\widehat{A}\Shat^{-2}\Big]\ .
\ee
The expectation value of the time-derivative of the observable $A$ follows from the general rule~\eqref{QA2}
\bel{QA13}
\langle \dot{A}(t)\rangle=\tr\big\{\dot{\widehat{A}}(t)\widehat\rho(t)\big\}\ .
\ee

%Absatz
We can further represent products of observables at different times by time-ordered operator products.
For time $t'=t-n\eps$, $n>0$, one has
\bel{QA14}
A(t)B(t')=\widehat{A}\Shat^n\widehat{B}\Shat^{-n}\ ,
\ee
such that the unequal time correlation function reads
\bel{QA15}
\langle A(t)B(t')\rangle=\tr\big\{\widehat{A}\Shat^n\widehat{B}\Shat^{-n}\widehat\rho(t)\big\}\ .
\ee
This is precisely the value that one obtains if we interprete the probabilistic cellular automaton as an overall statistical system for all times~\cite{Wetterich:2020kqi}. It can then be seen as a particular generalized two-dimensional Ising model with boundary conditions. Again, the rule~\eqref{QA15} follows from the standard rule for expectation values in classical statistics.

\subsection{Change of basis}
\label{subsec: Change of basis}

The quantum rule~\eqref{QA2} is invariant under a change of basis
\bel{QA16}
\widehat\rho\to\widehat\rho'=D\widehat\rho D^{-1}\ ,\quad \widehat{A}\to\widehat{A}'=D\widehat{A}D^{-1}\ .
\ee
We focus here on orthogonal $D$, $D^TD=1$, such that a symmetric density matrix remains symmetric in every basis. We can view the density matrix at $t-\eps$ as a basis change of $\widehat\rho(t)$, with $D=\Shat^{-1}$. In the new basis the operator $\widehat{A}(t';t)$ in eq.~\eqref{QA5} becomes
\begin{align}
\label{QA17}
\widehat{A}'(t';t)=&D\Shat\widehat{A}\Shat^{-1}D^{-1}=\widehat{A}\ ,\nn\\
\widehat\rho'(t)=&D\widehat\rho(t)D^{-1}=\rho(t-\eps)\ .
\end{align}
In this basis the expectation value $\langle A(t')\rangle$ follows directly by evaluating eq.~\eqref{QA2} at $t-\eps$.

%Absatz
This simple observation leads to a probabilistic interpretation of the diagonal elements of the transformed density matrix. The diagonal elements $\widehat\rho_{\tau\tau}$ correspond to the probabilities to find the value $A_\tau$ for an observable that is represented in this basis by a diagonal operator $\widehat{A}'$ of the form
\bel{QA18}
\widehat{A}'_{\rho\sigma}=\sum_\tau A_\tau\delta_{\tau\rho}\delta_{\tau\sigma}\ .
\ee
This statement is equivalent to the generalization of the quantum rule to observables that are represented by arbitrary symmetric operators
\bel{QA19}
\langle A\rangle=\sum_\tau\widehat\rho'_{\tau\tau}A_\tau=\tr\big\{\widehat\rho' \widehat{A}'\big\}=\tr\big\{\widehat\rho\widehat{A}\big\}\ .
\ee
Indeed, an arbitrary symmetric matrix $\widehat{A}$ can be diagonalized by an orthogonal transformation~\eqref{QA16} with suitable $D$.

%Absatz
For a positive symmetric density matrix $\widehat\rho$ (all eigenvalues positive semidefinite) also $\widehat\rho'$ is a positive symmetric matrix. This implies $\widehat\rho'_{\tau\tau}\geq0$. The normalization of the probabilities follows from $\tr\widehat\rho'=1$.
In every basis the diagonal elements $\widehat\rho'_{\tau\tau}$ have therefore the properties of a probability distribution. For the choices of basis which diagonalize the operators discussed above the probabilistic interpretation of $\widehat\rho'_{\tau\tau}$ follows from the classical statistical rule for expectation values, employing a complete set of observables.
We will not enter here a discussion of the classical statistical origin of the probabilistic interpretation of $\widehat\rho'_{\tau\tau}$ for an arbitrary choice of basis.
 
\section{Complex structure for operators}
\label{ap.H}

%Absatz
In this appendix we extend the complex structure to operators. This completes the quantum formalism for cellular automata in the usual complex setting.

\subsection{Complex operators\label{subsect: complex operators}}

%Absatz
For real symmetric operators $\widehat{A}$ that are compatible with the complex structure~\eqref{CS8A} the quantum rule~\eqref{QA2} for expectation values translates in the complex formulation to the familiar form
\bel{CS6}
\exval{A(t)}=\tr\{A\rho(t)\}\ .
\ee
With eqs.~\eqref{CS14A},~\eqref{CS15A} we define
\bel{220}
\rho=\rho_{R}-i\rho_{I}\;,\quad \rho_{R}={\rho}'+\rho^{c}\;,\quad \rho_{I}=\tilde{\rho}^{T}-\tilde{\rho} \,
\ee
where
\bel{220A}
\rho_{R}^{T}=\rho_{R}\;,\quad \rho_{I}^{T}=-\rho_{I}\;,\quad A_{R}^{T}=A_{R}\;,\quad A_{I}^{T}=-A_{I}\ .
\ee
Insertion into eq.~\eqref{QA2} yields the standard relation for complex quantum mechanics, 
\bel{220B}
\langle A \rangle  = \tr \lbrace \widehat{A}\widehat{\rho}\rbrace =\tr \lbrace A_{R}\rho_{R}+A_{I}\rho_{I}\rbrace =\tr \lbrace A\rho\rbrace \ .
\ee
The expression~\eqref{CS6} is independent of the choice of basis. It defines the expectation value for a large class of observables beyond the diagonal observables formed from functions of occupation numbers at a fixed time.  General time-local observables are represented by hermitian operators $A(t)=A^{\dagger}(t)$~\cite{Wetterich:2020kqi}. 

%Absatz
Among the diagonal operators only those are compatible with the complex structure associated to the particle-hole transformation that act in the same way on $q'$ and $q^{c}$.
Diagonal operators are described by diagonal $A_{R}$ with $A_{I}=0$, such that the structure~\eqref{CS8A} is block diagonal with the same operator acting on $q'$ and $q^{c}$. This generalizes to all operators that do not mix $q'$ and $q^{c}$. Some simple observables as particle numbers for separate species cannot be represented by operators that are compatible with the complex structure. The particle numbers for $q'$ and $q^{c}$ differ. For such observables the translation to the complex pictures results in operators that mix $\varphi$ and $\varphi^{*}$. There remain, however, many interesting observables that are compatible with the complex structure. A simple example is the position operator for single particle states that we will discuss in sect.~\ref{sec: 10}.

%Absatz
As expected, the Grassmann functional integral for fermions entails all rules of quantum mechanics.
What is perhaps more surprising at first sight is that the same holds for the associated cellular automaton. It follows, however, from the fact that the implementation of boundary conditions in terms of wave functions and, more generally, the density matrix, can be implemented in complete correspondence to the fermionic model. This becomes even more apparent if we formulate the probabilistic cellular automaton as a generalized Ising model~\cite{CWFGI}.

\subsection{Involution for Grassmann variables}
\label{subsec: Involution for Grassmann variables}

%Absatz
The complex structure for the wave function, density matrix and operators can be associated to an involution in the Grassmann algebra. This involution maps $\psi_{\alpha}(t)\rightarrow\psibar_{\alpha}(t)$. In order to recall that this involution is related to complex conjugation we write somewhat formally
\bel{CS1}
\psi_{\alpha}^{*}(t)=\psibar_{\alpha}(t)\ .
\ee
This relation resembles the relation~\eqref{99C},~\eqref{99D} with $\alpha=(x,\gamma)=(x,\eta,a)$, e.g.
\bel{CS1A}
\psi_{a}^{\dagger}=\gl\psi_{Ra}^{*},\psi_{La}^{*}\gr=\gl\psibar_{Ra},\psibar_{La}\gr\ .
\ee
It is compatible with the Lorentz transformations. In contrast to the complex Grassmann variables $\zeta$, $\zetabar$ in eqs.~\eqref{99C},~\eqref{99D} the Grassmann variables $\psi$, $\psibar$ remain ``real" objects.

%Absatz
On the level of the Grassmann wave function the involution $g\rightarrow g^{*}$ maps $g[\psi(t)]$ to $\ghat[\,\psibar(t)]$ and vice versa
\bel{CS1B}
g^{*}[\psi(t)]=\ghat[\,\psibar(t)]=\tilde{q}_{\tau}(t)g_{\tau}'[\,\psibar(t)]\ .
\ee
This operation includes the total reordering $R$ such that
\bel{CS1C}
g_{\tau}^*[\psi(t)]=R\gl g_{\tau}[\,\psibar(t)]\gr=g_{\tau}'[\,\psibar(t)]\ .
\ee
We recover for the Grassmann density matrix the transformation~\eqref{DM14} including the reordering $R$,
\begin{align}\label{CS1D}
\tilde{\rho}^{*}[\,\psibar(t),\psi(t)]&=\widehat{\rho}_{\tau\rho}(t)g_{\rho}[\psi(t)]g_{\tau}'[\,\psibar(t)]\nn\\
&=\widehat{\rho}_{\tau\rho}^{\, \prime}(t)g_{\tau}[\psi(t)]g_{\rho}'[\psi(t)]\ .\phantom{\Big{|}}
\end{align}
Comparison with eq.~\eqref{DM3} establishes again eq.~\eqref{DM16}
\bel{CS1F}
\widehat{\rho}'(t)=\widehat{\rho}\supt{T}(t)\ .
\ee

%Absatz
The complex conjugate Grassmann element $g^{*}[\psi]=\widehat{g}[\overline{\psi}]$ is related to the hole wave function $\overline{g}[\psi]$ by eq.~\eqref{DM17}. Observing that $\overline{g}_{\tau}[\psi]$ and $g_{\tau}^{c}[\psi]$ only differ by possible minus signs we infer the close relation of the involution for Grassmann variables~\eqref{CS1} with the particle-hole transformation. By a suitable choice of signs one can have the two involutions coincide. We could choose a definition of $g_{\tau}^{c}=\overline{g}_{\tau}(\varepsilon_{\tau}^{c}=1)$.

\section{Transport operators in terms of fermionic annihilation and creation operators}
\label{ap.I}

%Absatz
In this appendix we derive the expression of the right- and left-transport operators in terms of annihilation and creation operators. It contains a specific ordering of operators which is necessary in order to maintain the unique jump property for the transport operators. The proof is somewhat involved and takes advantage of the relation between annihilation and creation operators on one side and multiplication of Grassmann elements with Grassmann variables on the other side.

\subsection{Discrete transport operators\label{subsec: Discrete transport operators}}

%Absatz
We concentrate first on the right movers and omit the index $(Ra)$. The evolution of the Grassmann wave function from $g(t)=\tilde{q}_{\tau}(t)g_{\tau}[\psi(t)]$ to $g(t+\epstil)=\Shat_{\tau\rho}q_{\rho}(t)g_{\tau}[\psi(t+\epstil)]$ can be described either by the evolution of the wave function $\tilde{q}_{\tau}(t+\epstil)=\Shat_{\tau\rho}\tilde{q}_{\rho}(t)$ at fixed basis functions $g_{\tau}$, or equivalently by a transformation of basis functions,
\bel{CE38}
g_{\tau}[\psi(t+\epstil)]\to g_{\tau}^{(t)}[\psi(t+\epstil)]=g_{\rho}[\psi(t+\epstil)]\Shat_{\rho\tau}\ ,
\ee
at fixed $\tilde{q}_{\tau}$. Note that for this transformation the operators multiply the basis functions from the right, see the appendix for details. For right movers the new basis functions $g_{\tau}^{(t)}$ obtain by a replacement of all factors $\psi(x)$ in $g_{\tau}$ by $\psi(x+\epsilon)$. We obtain the expression
\bel{CE39}
g_{\rho}[\psi]\Shat_{\rho\tau}=\int\cD\psi'\prod_{x}\delta\gl\psi'(x)-\psi(x+\eps)\gr g_{\tau}[\psi']\ .
\ee
The same expression holds if we replace $g_{\tau}$ by $\gbar'_{\tau}$. Insertion of the expression for $\gbar'_{\tau}\Shat_{\tau\rho}$ into eq.~\eqref{11} for $\Ktil(t)$ reproduces indeed $\Ktil_R$ according to eq.~\eqref{32a}.

%Absatz
The $\delta$-function for Grassmann variables can be expressed as
\begin{align}
\label{CE51}
\delta\gl\psi'(x&)-\psi(x+\eps)\gr=\psi'(x)-\psi(x+\eps)\nn\\
=&\int\text{d}\widehat{\psi}(x+\eps)\exp\Big{\{}\widehat{\psi}(x+\eps)\big{[}\psi'(x)-\psi(x+\eps)\big{]}\Big{\}}\ ,
\end{align}
or
\begin{align}
\label{CE52}
\prod_{x}\delta&\gl\psi'(x)-\psi(x+\eps)\gr\nn\\
=&\int\cD\widehat{\psi}\exp\Big{\{}\sum_{x}\widehat{\psi}(x+\eps)\big{[}\psi'(x)-\psi(x+\eps)\big{]}\Big{\}}\ ,
\end{align}
such that
\begin{align}
\label{CE53}
&(\Shat g)_{\tau}[\psi]=g_{\rho}[\psi]\Shat_{\rho\tau}\nn\\
&=\int\cD\widehat{\psi}\cD\psi'\exp\Big{\{}\sum_{x}\widehat{\psi}(x+\eps)\big{[}\psi'(x)-\psi(x+\eps)\big{]}\Big{\}}g_{\tau}[\psi']\ .
\end{align}
The integral over the product with an arbitrary Grassmann element $\gtil[\psi]$ obeys
\begin{align}
\label{CE54}
&\int\cD\psi\;\gtil[\psi](\Shat g)_{\tau}[\psi]=\int\cD\psi\cD\widehat{\psi}\cD\psi'\;\gtil[\psi]\nn\\
&\times\exp\Big{\{}\sum_{x}\widehat{\psi}(x+\eps)\big{[}\psi'(x+\eps)-\psi(x+\eps)\big{]}\Big{\}}\nn\\
&\times\exp\Big{\{}\sum_{x}\widehat{\psi}(x+\eps)\big{[}\psi'(x)-\psi'(x+\eps)\big{]}\Big{\}}g_{\tau}[\psi']\ .
\end{align}

%Absatz
We next employ the general identity~\cite{CWEM},
\begin{align}
\label{CE55}
&\int\cD\psi\gtil[\psi]\mathcal{F}_{N}\Big{[}\frac{\partial}{\partial_{\psi}},\psi\Big{]}g_{\tau}[\psi]=\int\cD\psi\cD\widehat{\psi}\cD\psi'\;\gtil[\psi]\nn\\
&\times\exp\Big{\{}\sum_{x}\widehat{\psi}(x+\eps)\big{[}\psi'(x+\eps)-\psi(x+\eps)\big{]}\Big{\}}\nn\\
&\times F_{N}\big{[}\widehat{\psi},\psi'\;\big{]}g_{\tau}[\psi'\;]\ .
\end{align}
Here $F_{N}[\widehat{\psi},\psi']$ orders a function $F[\widehat{\psi},\psi']$ such that all factors $\widehat{\psi}_{\alpha}$ are on the left of factors $\psi'_{\alpha}$. The function $\mathcal{F}_{N}\Big{[}\frac{\partial}{\partial_{\psi}},\psi\Big{]}$ obtains from $F_{N}[\widehat{\psi},\psi'\;]$ by the replacements $\widehat{\psi}_{\alpha}\to\partial/\partial\psi_{\alpha}$, $\psi'_{\alpha}\to\psi_{\alpha}$, keeping the ordering of $\partial/\partial\psi_{\alpha}$ to the left of $\psi_{\alpha}$.
In particular, each factor of $\widehat{\psi}_{\alpha}$ for a given $\alpha$ can appear in $F_{N}$ at most once, since $\widehat{\psi}_{\alpha}^{2}=0$. Thus $\partial/\partial\psi_{\alpha}$ can occur in $\mathcal{F}_{N}$ at most once. The same holds for $\psi'_{\alpha}$ in $F_{N}$ and $\psi_{\alpha}$ in $\mathcal{F}_{N}$. We recognize in eq.~\eqref{CE54} the structure of eq.~\eqref{CE55}, with $F_{N}$ given by the second exponential after reordering. In consequence, we find the relation
\begin{align}
\label{CE56}
&\int\cD\psi\;\gtil[\psi](\Shat g)_{\tau}[\psi]=\int\cD\psi\;\gtil[\psi]\nn\\
&\times N\Big{[}\exp\Big{\{}\sum_{x}\frac{\partial}{\partial\psi(x+\eps)}\big{[}\psi(x)-\psi(x+\eps)\big{]}\Big{\}}\Big{]}g_{\tau}[\psi]\ .
\end{align}
Here the ordering operation $N$ applied to the exponential puts for a Taylor expansion $\exp(x)=\sum_{n=0}^{\infty}x^{n}/n!$ all factors $\partial/\partial\psi_{\alpha}$ to the left, without changing their relative order. Furthermore, each term in this Taylor expansion is multiplied by a sign $(-1)^{n-1}$ for $n\geq2$. This minus sign results from the anti-commutation with factors $\psi'_{\alpha}$ when bringing the factors $\widehat{\psi}_{\alpha}$ to the left for the function $F_{N}$ in eq.~\eqref{CE55}. Since the relation~\eqref{CE56} holds for arbitrary $\gtil[\psi]$ we conclude
\bel{CE57}
(\Shat g)_{\tau}[\psi]=N\Big{[}\exp\Big{\{}\sum_{x}\frac{\partial}{\partial\psi(x+\eps)}\big{[}\psi(x)-\psi(x+\eps)\big{]}\Big{\}}\Big{]}g_{\tau}[\psi]\ .
\ee

%Absatz
We finally employ the relation~\eqref{CE14A} between Grassmann variables and annihilation/creation operators,
\bel{CE58}
g_{\rho}[\psi]\Shat_{\rho\tau}=g_{\rho}[\psi]N\Big{[}\exp\Big{\{}\sum_{x}a\herm(x+\eps)\big{[}a(x)-a(x+\eps)\big{]}\Big{\}}\Big{]}_{\rho\tau}\ ,
\ee
where the ordering operation orders now the creation operators $a\herm$ to the left. Multiplication of eq.~\eqref{CE58} with $\gbar_{\tau}[\psi]$ and integrating over $\psi$ yields a $\delta$-function according to eq.~\eqref{12}.
This concludes the proof of eq.~\eqref{CE37} for right movers. For left movers we replace in eq.~\eqref{CE39} $\eps\to-\eps$. This results in $\eps\to-\eps$ for eq.~\eqref{CE37}.

\subsection{Ordering and unique jump property\label{subsec: Ordering and unique}}

%Absatz
The ordering operation $N$ is important for the orthogonality of $\Shat$, which guarantees the unitary evolution. This can be seen if we expand the exponential to second order (using periodicity in $x$)
\begin{align}
\label{CE59}
\Shat_R=&N\Big{[}\exp\Big{\{}\sum_{x}a\herm(x+\eps)\big{[}a(x)-a(x+\eps)\big{]}\Big{\}}\Big{]}\nn\\
=&1+\sum_{x}a\herm(x+\eps)a(x)-\sum_{x}a\herm(x)a(x)+T_{2}+\dots\ ,
\end{align}
where
\begin{align}
\label{CE60}
T_{2}=-\frac{1}{2}&\sum_{x}\sum_{y}a\herm(x+\eps)a\herm(y+\eps)\nn\\
&\times\big{[}a(x)-a(x+\eps)\big{]}\big{[}a(y)-a(y+\eps)\big{]}\ .
\end{align}
We can apply these operators on states with different particle numbers.
For the vacuum state only the term $1$ contributes, and the vacuum is time-translation invariant.

%Absatz
For a single particle located at $z$, with a sharp wave function $\tilde{q}_{1}(x)\sim\delta_{x,z}$, only terms with a single factor $a(z)$ contribute in the sums in eq.~\eqref{CE59}. This eliminates the contribution $T_2$ in eq.~\eqref{CE59} which contains two annihilation operators on the right, yielding zero when applied to a single particle state. The same holds for higher order terms in the expansion. From the first terms only the ones with $x=z$ contribute in the sums,
\bel{CE61}
\Shat_R\tilde{q}_{1}=\big{[}1+a\herm(z+\eps)a(z)-a\herm(z)a(z)\big{]}\tilde{q}_{1}\ .
\ee
For the one particle state one has $a\herm(z)a(z)=n(z)=1$, which cancels the term $1$. The particle at $z$ is therefore transported to $z+\eps$ by the term $a\herm(z+\eps)a(z)$, and $\Shat_R\tilde{q}_{1}$ is indeed again a one particle state, with the particle located now precisely at $z+\eps$.
This simple property would not hold without the ordering operation $N$. We can write
\begin{align}
\label{CE62}
T_2=&\frac{1}{2}\sum_{x,y}a\herm(x+\eps)\big{[}a(x)-a(x+\eps)\big{]}\nn\\
&\quad\times a\herm(y+\eps)\big{[}a(y)-a(y+\eps)\big{]}+T_{2,\text{c}}\ ,\nn\\
T_{2,\text{c}}=&\frac{1}{2}\sum_xa\herm(x+\eps)\big[2a(x)-a(x+\eps)-a(x-\eps)\big]\ .
\end{align}
The first term corresponds to the expansion of the exponential without the ordering operation. Without the ordering operation one would therefore have to replace in eq.~\eqref{CE60} $T_2\to T_2-T_{2,\text{c}}$. The contribution $-T_{2,\text{c}}$ would replace the jump from $z$ to $z+\eps$ by half the wave function staying at $z$, and the other half jumping to $z+2\eps$. For an exponential without the ordering there would be further terms modifying the evolution and broadening the wave function. The action of $\Shat_R$ would no longer be a unique jump operation.

%Absatz
For the ordered exponential the term $T_2$ accounts for the propagation of two particles at two distinct sites $z_1$ and $z_2$. They both jump by one site, to $z_1+\eps$ and $z_2+\eps$, respectively. This term also contains contributions that cancel the effect of the first terms in eq.~\eqref{CE59} on the two particle state.

\section{Continuum evolution and smoothness of probabilistic information}
\label{ap.J}

%Absatz
In this appendix we derive the continuum time evolution in a real formulation, starting from the discrete step evolution operator. This highlights more formally the conditions for a valid continuum limit. We also map to the complex formulation of the time evolution in quantum mechanics.

%Absatz
The evolution law~~\eqref{PC04} is a discrete Schrödinger equation in a real representation. We can write it in a more familiar form by introducing the discrete "time-derivative"
\be\label{PC09}
\partial_{t}q(t)=\dfrac{1}{4\varepsilon}\bigl(q(t+2\varepsilon)-q(t-2\varepsilon)\bigr)\;.
\ee
The advantage of this choice as compared to eq.~\eqref{71} is that $q(t)$, $q(t+2\varepsilon)$ and $q(t-2\varepsilon)$ all involve lattice points with the same $x$, in accordance to our restriction to the even sublattice. (In the second formulation with blocks of four lattice sites the blocks at $t$ and $t+\varepsilon$ are shifted by one unit in $x$, such that it is convenient to combine two time steps as well.) We introduce the operator $\widehat{W}(t)$ by
\be\label{PC10}
e^{2\varepsilon\widehat{W}(t)}=\widehat{S}(t+\varepsilon)\widehat{S}(t)\;.
\ee
By definition, $\widehat{W}(t)$ is antisymmetric for orthogonal $\widehat{S}(t)$,
\be\label{PC11}
\widehat{W}^{T}(t)=-\widehat{W}(t)\;,
\ee
since
\be\label{PC12}
\begin{split}
e^{-2\varepsilon\widehat{W}(t)}=\bigl[\widehat{S}(t+\varepsilon)\widehat{S}(t)\bigr{]}^{-1}=\bigl[\widehat{S}(t+\varepsilon)\widehat{S}(t)\bigr{]}^{T} \\=\bigl[e^{2\varepsilon\widehat{W}(t)}\bigr]^{T}=e^{2\varepsilon\widehat{W}^{T}(t)}\;.
\end{split}
\ee
In terms of $\widehat{W}(t)$ and the discrete derivative~\eqref{PC09} the evolution equation~\eqref{PC04} takes the form
\be\label{PC13}
\partial_{t}q(t)=\dfrac{1}{4\varepsilon}\bigl[e^{2\varepsilon\widehat{W}(t)}-e^{-2\varepsilon\widehat{W}(t-2\varepsilon)}\bigr]\;.
\ee

%Absatz
A continuum limit can be realized whenever $q(t)$ is a sufficiently smooth function of $t$ such that for $\varepsilon\rightarrow 0$ the discrete derivative~\eqref{PC09} can be replaced by a partial derivative $\partial_{t}$ acting on a differentiable real function $q(t)$.
Expanding eq.~\eqref{PC13} for small $\varepsilon$ yields
\be\label{PC14}
\partial_{t}q(t)=\dfrac{1}{2}\bigl[\:\widehat{W}(t)+\widehat{W}(t-2\varepsilon)\bigr]q(t)+0(\varepsilon^{2})\;,
\ee
and the continuum limit becomes
\be\label{PC15}
\partial_{t}q(t)=W(t)q(t)\;,
\ee
where
\be\label{PC16}
W(t)=\dfrac{1}{2}\bigl(\:\widehat{W}(t)+\widehat{W}(t-2\varepsilon)\bigr)\;.
\ee
We emphasize that the formal expansion~\eqref{PC14} has to be handled with care since $\widehat{S}$ is, in general, not close to the unit operator. One has to specify in which sense $\varepsilon \widehat{W}$ is a quantity of the order $\varepsilon$. 

%Absatz
The antisymmetry of $W(t)$ guarantees that the continuous evolution equation~\eqref{PC15} preserves the norm of $q(t)$. Eq.~\eqref{PC15} is the continuous Schrödinger equation in a real representation. In a complex representation an antisymmetric $W(t)$ corresponds to a hermitian Hamilton operator. We observe that the evolution equations~\eqref{PC04},~\eqref{PC13} or ~\eqref{PC16} are all linear in $q$ such that the superposition principale for amplitudes in quantum mechanics follows automatically.

%Absatz
The continuum limit leads to an important simplification. In general $\widehat{S}(t)$ and $\widehat{S}(t+\varepsilon)$ do not commute. Defining
\be\label{PC17}
\widehat{S}(t)=e^{\varepsilon\tilde{W}(t)}
\ee
one has
\begin{align}\label{PC18}
\widehat{S}(t+\varepsilon)\widehat{S}(t)&=e^{\varepsilon\tilde{W}(t+\varepsilon)}e^{\varepsilon\tilde{W}(t)}
\\
&=e^{\varepsilon(\tilde{W}(t+\varepsilon)+\tilde{W}(t))}+0\bigl(\varepsilon^{2}[\tilde{W}(t+\varepsilon),\tilde{W}(t)]\bigr)\;.\nn
\end{align}
In the continuum limit the commutator term $\sim \varepsilon^{2}$ can be neglected, such that
\begin{align}\label{PC19}
&\widehat{W}(t)=\dfrac{1}{2}\bigl(\tilde{W}(t+\varepsilon)+\tilde{W}(t)\bigr)\,, \\
W(t)=&\dfrac{1}{4}\bigl(\tilde{W}(t+\varepsilon)+\tilde{W}(t)+\tilde{W}(t-\varepsilon)+\tilde{W}(t-2\varepsilon)\bigr)\;.\nn
\end{align}

%Absatz
In the presence of a complex structure the antisymmetric matrix $W(t)$ is mapped to $-iH(t)$ with hermitian Hamilton operator $H(t)$. The evolution equation~\eqref{PC15} becomes the Schrödinger equation~\eqref{CE83}. For our generalized Thirring model the Hamilton operator is given by eq.~\eqref{CE79}. Indeed, the relations between the step evolution operator and the creation and annihilation operators hold as well if we replace $i$ by a matrix multiplication with $I$ in the real formulation. In the real formulation one has $W=-IH$. For the kinetic part $-I(H^{(R)}+H^{(L)})$ is a real antisymmetric matrix, as obtained by multiplying eq.~\eqref{CE67} with $-i$. The interaction part $-IH_{\textup{int}}$ involves~$I$, with $\exp (-\varepsilon I H_{\textup{int}})$ the interaction part of the step evolution operator. In the complex formulation one replaces $I\to i$, leading to eq.~\eqref{CE79}. The coefficients $\sim \varepsilon^{-1}$ or $\varepsilon^{-2}$ suggest at first sight that the Hamilton operator diverges in the continuum limit $\varepsilon\to 0$. This divergence is absorbed, however, by a field redefinition or the associated continuum normalization~\eqref{CE80} of the annihilation and creation operators. For this continuum normalization the limit $\varepsilon\to 0$ is straightforward since $H$ does no longer involve $\varepsilon$.

\end{appendices}

%\bibliography{probabilistic_cellular_automata}
%\nocite{*} %Show alle Reference
% Create the reference section using BibTeX: \usepackage{natbib}
\bibliography{refs}

\begin{thebibliography}{66}%
\makeatletter
\providecommand \@ifxundefined [1]{%
 \@ifx{#1\undefined}
}%
\providecommand \@ifnum [1]{%
 \ifnum #1\expandafter \@firstoftwo
 \else \expandafter \@secondoftwo
 \fi
}%
\providecommand \@ifx [1]{%
 \ifx #1\expandafter \@firstoftwo
 \else \expandafter \@secondoftwo
 \fi
}%
\providecommand \natexlab [1]{#1}%
\providecommand \enquote  [1]{``#1''}%
\providecommand \bibnamefont  [1]{#1}%
\providecommand \bibfnamefont [1]{#1}%
\providecommand \citenamefont [1]{#1}%
\providecommand \href@noop [0]{\@secondoftwo}%
\providecommand \href [0]{\begingroup \@sanitize@url \@href}%
\providecommand \@href[1]{\@@startlink{#1}\@@href}%
\providecommand \@@href[1]{\endgroup#1\@@endlink}%
\providecommand \@sanitize@url [0]{\catcode `\\12\catcode `\$12\catcode
  `\&12\catcode `\#12\catcode `\^12\catcode `\_12\catcode `\%12\relax}%
\providecommand \@@startlink[1]{}%
\providecommand \@@endlink[0]{}%
\providecommand \url  [0]{\begingroup\@sanitize@url \@url }%
\providecommand \@url [1]{\endgroup\@href {#1}{\urlprefix }}%
\providecommand \urlprefix  [0]{URL }%
\providecommand \Eprint [0]{\href }%
\providecommand \doibase [0]{http://dx.doi.org/}%
\providecommand \selectlanguage [0]{\@gobble}%
\providecommand \bibinfo  [0]{\@secondoftwo}%
\providecommand \bibfield  [0]{\@secondoftwo}%
\providecommand \translation [1]{[#1]}%
\providecommand \BibitemOpen [0]{}%
\providecommand \bibitemStop [0]{}%
\providecommand \bibitemNoStop [0]{.\EOS\space}%
\providecommand \EOS [0]{\spacefactor3000\relax}%
\providecommand \BibitemShut  [1]{\csname bibitem#1\endcsname}%
\let\auto@bib@innerbib\@empty
%</preamble>
\bibitem [{\citenamefont {Wetterich}(2020)}]{Wetterich:2020kqi}%
  \BibitemOpen
  \bibfield  {author} {\bibinfo {author} {\bibfnamefont {C.}~\bibnamefont
  {Wetterich}},\ }\bibfield  {title} {\enquote {\bibinfo {title} {{The
  probabilistic world}},}\ }\href@noop {} {\  (\bibinfo {year} {2020})},\
  \Eprint {http://arxiv.org/abs/2011.02867} {arXiv:2011.02867 [quant-ph]}
  \BibitemShut {NoStop}%
\bibitem [{\citenamefont {Wetterich}(2010{\natexlab{a}})}]{CWQMCS}%
  \BibitemOpen
  \bibfield  {author} {\bibinfo {author} {\bibfnamefont {C.}~\bibnamefont
  {Wetterich}},\ }\bibfield  {title} {\enquote {\bibinfo {title} {{Quantum
  mechanics from classical statistics}},}\ }\href {\doibase
  10.1016/j.aop.2009.12.006} {\bibfield  {journal} {\bibinfo  {journal} {Annals
  Phys.}\ }\textbf {\bibinfo {volume} {325}},\ \bibinfo {pages} {852} (\bibinfo
  {year} {2010}{\natexlab{a}})},\ \Eprint {http://arxiv.org/abs/0906.4919}
  {arXiv:0906.4919 [quant-ph]} \BibitemShut {NoStop}%
\bibitem [{\citenamefont {Bell}(1964)}]{BELL}%
  \BibitemOpen
  \bibfield  {author} {\bibinfo {author} {\bibfnamefont {J.~S.}\ \bibnamefont
  {Bell}},\ }\bibfield  {title} {\enquote {\bibinfo {title} {{On the
  Einstein-Podolsky-Rosen paradox}},}\ }\href {\doibase
  10.1103/PhysicsPhysiqueFizika.1.195} {\bibfield  {journal} {\bibinfo
  {journal} {Physics Physique Fizika}\ }\textbf {\bibinfo {volume} {1}},\
  \bibinfo {pages} {195--200} (\bibinfo {year} {1964})}\BibitemShut {NoStop}%
\bibitem [{\citenamefont {Clauser}\ \emph {et~al.}(1969)\citenamefont
  {Clauser}, \citenamefont {Horne}, \citenamefont {Shimony},\ and\
  \citenamefont {Holt}}]{CHSH}%
  \BibitemOpen
  \bibfield  {author} {\bibinfo {author} {\bibfnamefont {J.~F.}\ \bibnamefont
  {Clauser}}, \bibinfo {author} {\bibfnamefont {M.~A.}\ \bibnamefont {Horne}},
  \bibinfo {author} {\bibfnamefont {A.}~\bibnamefont {Shimony}}, \ and\
  \bibinfo {author} {\bibfnamefont {R.~A.}\ \bibnamefont {Holt}},\ }\bibfield
  {title} {\enquote {\bibinfo {title} {{Proposed experiment to test local
  hidden variable theories}},}\ }\href {\doibase 10.1103/PhysRevLett.23.880}
  {\bibfield  {journal} {\bibinfo  {journal} {Phys. Rev. Lett.}\ }\textbf
  {\bibinfo {volume} {23}},\ \bibinfo {pages} {880--884} (\bibinfo {year}
  {1969})}\BibitemShut {NoStop}%
\bibitem [{\citenamefont {'t~Hooft}(2010)}]{HOOFT2}%
  \BibitemOpen
  \bibfield  {author} {\bibinfo {author} {\bibfnamefont {G.}~\bibnamefont
  {'t~Hooft}},\ }\bibfield  {title} {\enquote {\bibinfo {title} {{Classical
  cellular automata and quantum field theory}},}\ }\href {\doibase
  10.1142/S0217751X10050469} {\bibfield  {journal} {\bibinfo  {journal} {Int.
  J. Mod. Phys. A}\ }\textbf {\bibinfo {volume} {25}},\ \bibinfo {pages}
  {4385--4396} (\bibinfo {year} {2010})}\BibitemShut {NoStop}%
\bibitem [{\citenamefont {'t~Hooft}(2014)}]{GTH}%
  \BibitemOpen
  \bibfield  {author} {\bibinfo {author} {\bibfnamefont {G.}~\bibnamefont
  {'t~Hooft}},\ }\bibfield  {title} {\enquote {\bibinfo {title} {{The Cellular
  Automaton Interpretation of Quantum Mechanics. A View on the Quantum Nature
  of our Universe, Compulsory or Impossible?}}}\ }\href@noop {} {\  (\bibinfo
  {year} {2014})},\ \Eprint {http://arxiv.org/abs/1405.1548} {arXiv:1405.1548
  [quant-ph]} \BibitemShut {NoStop}%
\bibitem [{\citenamefont {Elze}(2014)}]{ELZE}%
  \BibitemOpen
  \bibfield  {author} {\bibinfo {author} {\bibfnamefont {H.-T.}\ \bibnamefont
  {Elze}},\ }\bibfield  {title} {\enquote {\bibinfo {title} {{Quantumness of
  discrete Hamiltonian cellular automata}},}\ }\href {\doibase
  10.1051/epjconf/20147802005} {\bibfield  {journal} {\bibinfo  {journal} {EPJ
  Web Conf.}\ }\textbf {\bibinfo {volume} {78}},\ \bibinfo {pages} {02005}
  (\bibinfo {year} {2014})},\ \Eprint {http://arxiv.org/abs/1407.2160}
  {arXiv:1407.2160 [quant-ph]} \BibitemShut {NoStop}%
\bibitem [{\citenamefont {'t~Hooft}(2021{\natexlab{a}})}]{HOOFT3}%
  \BibitemOpen
  \bibfield  {author} {\bibinfo {author} {\bibfnamefont {G.}~\bibnamefont
  {'t~Hooft}},\ }\bibfield  {title} {\enquote {\bibinfo {title} {{Fast Vacuum
  Fluctuations and the Emergence of Quantum Mechanics}},}\ }\href {\doibase
  10.1007/s10701-021-00464-7} {\bibfield  {journal} {\bibinfo  {journal}
  {Found. Phys.}\ }\textbf {\bibinfo {volume} {51}},\ \bibinfo {pages} {63}
  (\bibinfo {year} {2021}{\natexlab{a}})},\ \Eprint
  {http://arxiv.org/abs/2010.02019} {arXiv:2010.02019 [quant-ph]} \BibitemShut
  {NoStop}%
\bibitem [{\citenamefont {'t~Hooft}(2021{\natexlab{b}})}]{HOOFT4}%
  \BibitemOpen
  \bibfield  {author} {\bibinfo {author} {\bibfnamefont {G.}~\bibnamefont
  {'t~Hooft}},\ }\bibfield  {title} {\enquote {\bibinfo {title} {{Explicit
  construction of Local Hidden Variables for any quantum theory up to any
  desired accuracy}},}\ }\href@noop {} {\  (\bibinfo {year}
  {2021}{\natexlab{b}})},\ \Eprint {http://arxiv.org/abs/2103.04335}
  {arXiv:2103.04335 [quant-ph]} \BibitemShut {NoStop}%
\bibitem [{\citenamefont {Wetterich}(2017)}]{CWFGI}%
  \BibitemOpen
  \bibfield  {author} {\bibinfo {author} {\bibfnamefont {C.}~\bibnamefont
  {Wetterich}},\ }\bibfield  {title} {\enquote {\bibinfo {title} {{Fermions as
  generalized Ising models}},}\ }\href {\doibase
  10.1016/j.nuclphysb.2017.02.012} {\bibfield  {journal} {\bibinfo  {journal}
  {Nucl. Phys. B}\ }\textbf {\bibinfo {volume} {917}},\ \bibinfo {pages}
  {241--271} (\bibinfo {year} {2017})},\ \Eprint
  {http://arxiv.org/abs/1612.06695} {arXiv:1612.06695 [cond-mat.stat-mech]}
  \BibitemShut {NoStop}%
\bibitem [{\citenamefont {von Neumann}(1951)}]{JVN}%
  \BibitemOpen
  \bibfield  {author} {\bibinfo {author} {\bibfnamefont {J.}~\bibnamefont {von
  Neumann}},\ }\enquote {\bibinfo {title} {The general and logical theory of
  automata.}}\ \ (\bibinfo  {publisher} {Wiley},\ \bibinfo {address} {Oxford,
  England},\ \bibinfo {year} {1951})\ pp.\ \bibinfo {pages} {1--41}\BibitemShut
  {NoStop}%
\bibitem [{\citenamefont {Ulam}(1950)}]{ULA}%
  \BibitemOpen
  \bibfield  {author} {\bibinfo {author} {\bibfnamefont {S.}~\bibnamefont
  {Ulam}},\ }\bibfield  {title} {\enquote {\bibinfo {title} {Random processes
  and transformations},}\ }in\ \href@noop {} {\emph {\bibinfo {booktitle}
  {Proceedings of the International Congress on Mathematics}}},\ Vol.~\bibinfo
  {volume} {2}\ (\bibinfo {year} {1950})\ pp.\ \bibinfo {pages}
  {264--275}\BibitemShut {NoStop}%
\bibitem [{\citenamefont {Zuse}(1969)}]{ZUS}%
  \BibitemOpen
  \bibfield  {author} {\bibinfo {author} {\bibfnamefont {Konrad}\ \bibnamefont
  {Zuse}},\ }\href@noop {} {\emph {\bibinfo {title} {Rechnender Raum}}}\
  (\bibinfo  {publisher} {Vieweg, Teubner Verlag},\ \bibinfo {year} {1969})\
  pp.\ \bibinfo {pages} {1--3}\BibitemShut {NoStop}%
\bibitem [{\citenamefont {Gardner}(1970)}]{GAR}%
  \BibitemOpen
  \bibfield  {author} {\bibinfo {author} {\bibfnamefont {Martin}\ \bibnamefont
  {Gardner}},\ }\bibfield  {title} {\enquote {\bibinfo {title} {Mathematical
  games},}\ }\href@noop {} {\bibfield  {journal} {\bibinfo  {journal}
  {Scientific American}\ }\textbf {\bibinfo {volume} {223}},\ \bibinfo {pages}
  {120--123} (\bibinfo {year} {1970})}\BibitemShut {NoStop}%
\bibitem [{\citenamefont {Lindenmayer}\ and\ \citenamefont
  {Rozenberg}(1976)}]{LIRO}%
  \BibitemOpen
  \bibfield  {author} {\bibinfo {author} {\bibfnamefont {Aristid}\ \bibnamefont
  {Lindenmayer}}\ and\ \bibinfo {author} {\bibfnamefont {Grzegorz}\
  \bibnamefont {Rozenberg}},\ }\bibfield  {title} {\enquote {\bibinfo {title}
  {Automata, languages, development},}\ \ }(\bibinfo  {publisher} {North
  Holland},\ \bibinfo {year} {1976})\BibitemShut {NoStop}%
\bibitem [{\citenamefont {Toom}(1978)}]{TOOM}%
  \BibitemOpen
  \bibfield  {author} {\bibinfo {author} {\bibfnamefont {A.~L.}\ \bibnamefont
  {Toom}},\ }\href@noop {} {\emph {\bibinfo {title} {Locally Interacting
  Systems and their application in Biology}}}\ (\bibinfo  {publisher} {Springer
  Berlin Heidelberg},\ \bibinfo {year} {1978})\BibitemShut {NoStop}%
\bibitem [{\citenamefont {R.~L.~Dobrushin}(1978)}]{DKT}%
  \BibitemOpen
  \bibfield  {author} {\bibinfo {author} {\bibfnamefont {A.~L.~Toom}\
  \bibnamefont {R.~L.~Dobrushin}, \bibfnamefont {V.I.~Kryukov}},\ }\href@noop
  {} {\emph {\bibinfo {title} {Stochastic cellular systems: Ergodicity, Memory,
  Morphogenesis}}}\ (\bibinfo  {publisher} {Manchester University Press},\
  \bibinfo {year} {1978})\BibitemShut {NoStop}%
\bibitem [{\citenamefont {Wolfram}(1983)}]{WOLF}%
  \BibitemOpen
  \bibfield  {author} {\bibinfo {author} {\bibfnamefont {S.}~\bibnamefont
  {Wolfram}},\ }\bibfield  {title} {\enquote {\bibinfo {title} {Statistical
  mechanics of cellular automata},}\ }\href {\doibase
  10.1103/RevModPhys.55.601} {\bibfield  {journal} {\bibinfo  {journal} {Rev.
  Mod. Phys.}\ }\textbf {\bibinfo {volume} {55}},\ \bibinfo {pages} {601--644}
  (\bibinfo {year} {1983})}\BibitemShut {NoStop}%
\bibitem [{\citenamefont {Vichniac}(1984)}]{VICH}%
  \BibitemOpen
  \bibfield  {author} {\bibinfo {author} {\bibfnamefont {G{\'e}rard~Y.}\
  \bibnamefont {Vichniac}},\ }\bibfield  {title} {\enquote {\bibinfo {title}
  {Simulating physics with cellular automata},}\ }\href@noop {} {\bibfield
  {journal} {\bibinfo  {journal} {Physica D: Nonlinear Phenomena}\ }\textbf
  {\bibinfo {volume} {10}},\ \bibinfo {pages} {96--116} (\bibinfo {year}
  {1984})}\BibitemShut {NoStop}%
\bibitem [{\citenamefont {Preston}\ and\ \citenamefont {Duff}(1984)}]{PREDU}%
  \BibitemOpen
  \bibfield  {author} {\bibinfo {author} {\bibfnamefont {Kendall}\ \bibnamefont
  {Preston}}\ and\ \bibinfo {author} {\bibfnamefont {Michael J.~B.}\
  \bibnamefont {Duff}},\ }\href@noop {} {\emph {\bibinfo {title} {Modern
  Cellular Automata}}}\ (\bibinfo  {publisher} {Springer {US}},\ \bibinfo
  {year} {1984})\ pp.\ \bibinfo {pages} {1--15}\BibitemShut {NoStop}%
\bibitem [{\citenamefont {Toffoli}\ and\ \citenamefont
  {Margolus}(1990)}]{TOMA}%
  \BibitemOpen
  \bibfield  {author} {\bibinfo {author} {\bibfnamefont {Tommaso}\ \bibnamefont
  {Toffoli}}\ and\ \bibinfo {author} {\bibfnamefont {Norman~H.}\ \bibnamefont
  {Margolus}},\ }\bibfield  {title} {\enquote {\bibinfo {title} {Invertible
  cellular automata: A review},}\ }\href@noop {} {\bibfield  {journal}
  {\bibinfo  {journal} {Physica D: Nonlinear Phenomena}\ }\textbf {\bibinfo
  {volume} {45}},\ \bibinfo {pages} {229--253} (\bibinfo {year}
  {1990})}\BibitemShut {NoStop}%
\bibitem [{\citenamefont {LOUIS}\ and\ \citenamefont {Nardi}(2018)}]{FLN}%
  \BibitemOpen
  \bibfield  {author} {\bibinfo {author} {\bibfnamefont {Pierre-Yves}\
  \bibnamefont {LOUIS}}\ and\ \bibinfo {author} {\bibfnamefont {Francesca~R.}\
  \bibnamefont {Nardi}},\ }\href@noop {} {\emph {\bibinfo {title}
  {{Probabilistic Cellular Automata}}}}\ (\bibinfo  {publisher} {Springer},\
  \bibinfo {year} {2018})\BibitemShut {NoStop}%
\bibitem [{\citenamefont {Hedlund}(1969)}]{HED}%
  \BibitemOpen
  \bibfield  {author} {\bibinfo {author} {\bibfnamefont {G.~A.}\ \bibnamefont
  {Hedlund}},\ }\bibfield  {title} {\enquote {\bibinfo {title} {Endomorphisms
  and automorphisms of the shift dynamical system},}\ }\href@noop {} {\bibfield
   {journal} {\bibinfo  {journal} {Mathematical systems theory}\ }\textbf
  {\bibinfo {volume} {3}},\ \bibinfo {pages} {320--375} (\bibinfo {year}
  {1969})}\BibitemShut {NoStop}%
\bibitem [{\citenamefont {Richardson}(1972)}]{RICH}%
  \BibitemOpen
  \bibfield  {author} {\bibinfo {author} {\bibfnamefont {D.}~\bibnamefont
  {Richardson}},\ }\bibfield  {title} {\enquote {\bibinfo {title}
  {Tessellations with local transformations},}\ }\href@noop {} {\bibfield
  {journal} {\bibinfo  {journal} {Journal of Computer and System Sciences}\
  }\textbf {\bibinfo {volume} {6}},\ \bibinfo {pages} {373--388} (\bibinfo
  {year} {1972})}\BibitemShut {NoStop}%
\bibitem [{\citenamefont {Amoroso}\ and\ \citenamefont {Patt}(1972)}]{AMPA}%
  \BibitemOpen
  \bibfield  {author} {\bibinfo {author} {\bibfnamefont {S.}~\bibnamefont
  {Amoroso}}\ and\ \bibinfo {author} {\bibfnamefont {Y.N.}\ \bibnamefont
  {Patt}},\ }\bibfield  {title} {\enquote {\bibinfo {title} {Decision
  procedures for surjectivity and injectivity of parallel maps for tessellation
  structures},}\ }\href@noop {} {\bibfield  {journal} {\bibinfo  {journal}
  {Journal of Computer and System Sciences}\ }\textbf {\bibinfo {volume} {6}},\
  \bibinfo {pages} {448--464} (\bibinfo {year} {1972})}\BibitemShut {NoStop}%
\bibitem [{\citenamefont {Hardy}\ \emph {et~al.}(1976)\citenamefont {Hardy},
  \citenamefont {de~Pazzis},\ and\ \citenamefont {Pomeau}}]{HPP}%
  \BibitemOpen
  \bibfield  {author} {\bibinfo {author} {\bibfnamefont {J.}~\bibnamefont
  {Hardy}}, \bibinfo {author} {\bibfnamefont {O.}~\bibnamefont {de~Pazzis}}, \
  and\ \bibinfo {author} {\bibfnamefont {Y.}~\bibnamefont {Pomeau}},\
  }\bibfield  {title} {\enquote {\bibinfo {title} {Molecular dynamics of a
  classical lattice gas: Transport properties and time correlation
  functions},}\ }\href@noop {} {\bibfield  {journal} {\bibinfo  {journal}
  {Phys. Rev. A}\ }\textbf {\bibinfo {volume} {13}},\ \bibinfo {pages}
  {1949--1961} (\bibinfo {year} {1976})}\BibitemShut {NoStop}%
\bibitem [{\citenamefont {Creutz}(1986)}]{CREU}%
  \BibitemOpen
  \bibfield  {author} {\bibinfo {author} {\bibfnamefont {Michael}\ \bibnamefont
  {Creutz}},\ }\bibfield  {title} {\enquote {\bibinfo {title} {Deterministic
  ising dynamics},}\ }\href@noop {} {\bibfield  {journal} {\bibinfo  {journal}
  {Annals of Physics}\ }\textbf {\bibinfo {volume} {167}},\ \bibinfo {pages}
  {62--72} (\bibinfo {year} {1986})}\BibitemShut {NoStop}%
\bibitem [{\citenamefont {Lenz}(1920)}]{LENZ}%
  \BibitemOpen
  \bibfield  {author} {\bibinfo {author} {\bibfnamefont {W.}~\bibnamefont
  {Lenz}},\ }\bibfield  {title} {\enquote {\bibinfo {title} {Beitrag zum
  verst{\"a}ndnis der magnetischen erscheinungen in festen k{\"o}rpern},}\
  }\href@noop {} {\bibfield  {journal} {\bibinfo  {journal} {Z. Phys.}\
  }\textbf {\bibinfo {volume} {21}},\ \bibinfo {pages} {613--615} (\bibinfo
  {year} {1920})}\BibitemShut {NoStop}%
\bibitem [{\citenamefont {Ising}(1925)}]{ISING}%
  \BibitemOpen
  \bibfield  {author} {\bibinfo {author} {\bibfnamefont {E.}~\bibnamefont
  {Ising}},\ }\bibfield  {title} {\enquote {\bibinfo {title} {Beitrag zur
  theorie des ferromagnetismus},}\ }\href@noop {} {\bibfield  {journal}
  {\bibinfo  {journal} {Zeitschrift f{\"u}r Physik}\ }\textbf {\bibinfo
  {volume} {31}},\ \bibinfo {pages} {253--258} (\bibinfo {year}
  {1925})}\BibitemShut {NoStop}%
\bibitem [{\citenamefont {Binder}(2001)}]{BINDER}%
  \BibitemOpen
  \bibfield  {author} {\bibinfo {author} {\bibfnamefont {K.}~\bibnamefont
  {Binder}},\ }\bibfield  {title} {\enquote {\bibinfo {title} {Ising model},}\
  }\href@noop {} {\bibfield  {journal} {\bibinfo  {journal} {Hazewinkel,
  Michiel, Encyclopedia of Mathematics. Springer, Berlin, Heidelberg}\ }
  (\bibinfo {year} {2001})}\BibitemShut {NoStop}%
\bibitem [{\citenamefont {Shannon}(1948)}]{SHA}%
  \BibitemOpen
  \bibfield  {author} {\bibinfo {author} {\bibfnamefont {C.~E.}\ \bibnamefont
  {Shannon}},\ }\bibfield  {title} {\enquote {\bibinfo {title} {{A mathematical
  theory of communication}},}\ }\href@noop {} {\bibfield  {journal} {\bibinfo
  {journal} {Bell Syst. Tech. J.}\ }\textbf {\bibinfo {volume} {27}},\ \bibinfo
  {pages} {379--423} (\bibinfo {year} {1948})}\BibitemShut {NoStop}%
\bibitem [{\citenamefont {Plechko}(2005)}]{PLECH}%
  \BibitemOpen
  \bibfield  {author} {\bibinfo {author} {\bibfnamefont {V.~N.}\ \bibnamefont
  {Plechko}},\ }\bibfield  {title} {\enquote {\bibinfo {title} {{Fermions and
  correlations in the two-dimensional Ising model}},}\ }\href@noop {}
  {\bibfield  {journal} {\bibinfo  {journal} {Phys. Part. Nucl.}\ }\textbf
  {\bibinfo {volume} {36}},\ \bibinfo {pages} {S203--S208} (\bibinfo {year}
  {2005})},\ \Eprint {http://arxiv.org/abs/hep-th/0512263}
  {arXiv:hep-th/0512263} \BibitemShut {NoStop}%
\bibitem [{\citenamefont {Berezin}(1966)}]{BER1}%
  \BibitemOpen
  \bibfield  {author} {\bibinfo {author} {\bibfnamefont {F.~A.}\ \bibnamefont
  {Berezin}},\ }\bibfield  {title} {\enquote {\bibinfo {title} {The method of
  second quantization. nauka{\i}, moscow (1965)},}\ }\href@noop {} {\bibfield
  {journal} {\bibinfo  {journal} {English translation Academic Press, New
  York}\ } (\bibinfo {year} {1966})}\BibitemShut {NoStop}%
\bibitem [{\citenamefont {Berezin}(1969)}]{BER2}%
  \BibitemOpen
  \bibfield  {author} {\bibinfo {author} {\bibfnamefont {F.~A.}\ \bibnamefont
  {Berezin}},\ }\bibfield  {title} {\enquote {\bibinfo {title} {The plane ising
  model},}\ }\href {\doibase 10.1070/rm1969v024n03abeh001346} {\bibfield
  {journal} {\bibinfo  {journal} {Russian Mathematical Surveys}\ }\textbf
  {\bibinfo {volume} {24}},\ \bibinfo {pages} {1--22} (\bibinfo {year}
  {1969})}\BibitemShut {NoStop}%
\bibitem [{\citenamefont {Samuel}(1980)}]{SAM}%
  \BibitemOpen
  \bibfield  {author} {\bibinfo {author} {\bibfnamefont {S.}~\bibnamefont
  {Samuel}},\ }\bibfield  {title} {\enquote {\bibinfo {title} {{The Use of
  Anticommuting Integrals in Statistical Mechanics. 1.}}}\ }\href {\doibase
  10.1063/1.524404} {\bibfield  {journal} {\bibinfo  {journal} {J. Math.
  Phys.}\ }\textbf {\bibinfo {volume} {21}},\ \bibinfo {pages} {2806--2814}
  (\bibinfo {year} {1980})}\BibitemShut {NoStop}%
\bibitem [{\citenamefont {Itzykson}(1982)}]{ITS}%
  \BibitemOpen
  \bibfield  {author} {\bibinfo {author} {\bibfnamefont {C.}~\bibnamefont
  {Itzykson}},\ }\bibfield  {title} {\enquote {\bibinfo {title} {Ising fermions
  (i). two dimensions},}\ }\href {\doibase
  https://doi.org/10.1016/0550-3213(82)90173-0} {\bibfield  {journal} {\bibinfo
   {journal} {Nuclear Physics B}\ }\textbf {\bibinfo {volume} {210}},\ \bibinfo
  {pages} {448--476} (\bibinfo {year} {1982})}\BibitemShut {NoStop}%
\bibitem [{\citenamefont {Plechko}(1985)}]{PLE1}%
  \BibitemOpen
  \bibfield  {author} {\bibinfo {author} {\bibfnamefont {V.~N.}\ \bibnamefont
  {Plechko}},\ }\bibfield  {title} {\enquote {\bibinfo {title} {Simple solution
  of two-dimensional ising model on a torus in terms of grassmann integrals},}\
  }\href@noop {} {\bibfield  {journal} {\bibinfo  {journal} {Theoretical and
  Mathematical Physics}\ }\textbf {\bibinfo {volume} {64}},\ \bibinfo {pages}
  {748--756} (\bibinfo {year} {1985})}\BibitemShut {NoStop}%
\bibitem [{\citenamefont {Furuya}\ \emph {et~al.}(1982)\citenamefont {Furuya},
  \citenamefont {Gamboa~Saraví},\ and\ \citenamefont {Schaposnik}}]{FUR}%
  \BibitemOpen
  \bibfield  {author} {\bibinfo {author} {\bibfnamefont {K.}~\bibnamefont
  {Furuya}}, \bibinfo {author} {\bibfnamefont {R.E.}\ \bibnamefont
  {Gamboa~Saraví}}, \ and\ \bibinfo {author} {\bibfnamefont {F.A.}\
  \bibnamefont {Schaposnik}},\ }\bibfield  {title} {\enquote {\bibinfo {title}
  {Path-integral formulation of chiral invariant fermion models in two
  dimensions},}\ }\href {\doibase https://doi.org/10.1016/0550-3213(82)90191-2}
  {\bibfield  {journal} {\bibinfo  {journal} {Nuclear Physics B}\ }\textbf
  {\bibinfo {volume} {208}},\ \bibinfo {pages} {159 -- 181} (\bibinfo {year}
  {1982})}\BibitemShut {NoStop}%
\bibitem [{\citenamefont {Na\'on}(1985)}]{NAO}%
  \BibitemOpen
  \bibfield  {author} {\bibinfo {author} {\bibfnamefont {C.~M.}\ \bibnamefont
  {Na\'on}},\ }\bibfield  {title} {\enquote {\bibinfo {title} {Abelian and
  non-abelian bosonization in the path-integral framework},}\ }\href {\doibase
  10.1103/PhysRevD.31.2035} {\bibfield  {journal} {\bibinfo  {journal} {Phys.
  Rev. D}\ }\textbf {\bibinfo {volume} {31}},\ \bibinfo {pages} {2035--2044}
  (\bibinfo {year} {1985})}\BibitemShut {NoStop}%
\bibitem [{\citenamefont {Coleman}(1975)}]{COL}%
  \BibitemOpen
  \bibfield  {author} {\bibinfo {author} {\bibfnamefont {S.}~\bibnamefont
  {Coleman}},\ }\bibfield  {title} {\enquote {\bibinfo {title} {Quantum
  sine-gordon equation as the massive thirring model},}\ }\href {\doibase
  10.1103/PhysRevD.11.2088} {\bibfield  {journal} {\bibinfo  {journal} {Phys.
  Rev. D}\ }\textbf {\bibinfo {volume} {11}},\ \bibinfo {pages} {2088--2097}
  (\bibinfo {year} {1975})}\BibitemShut {NoStop}%
\bibitem [{\citenamefont {Damgaard}\ \emph {et~al.}(1992)\citenamefont
  {Damgaard}, \citenamefont {Nielsen},\ and\ \citenamefont {Sollacher}}]{DNS}%
  \BibitemOpen
  \bibfield  {author} {\bibinfo {author} {\bibfnamefont {P.H.}\ \bibnamefont
  {Damgaard}}, \bibinfo {author} {\bibfnamefont {H.B.}\ \bibnamefont
  {Nielsen}}, \ and\ \bibinfo {author} {\bibfnamefont {R.}~\bibnamefont
  {Sollacher}},\ }\bibfield  {title} {\enquote {\bibinfo {title} {Smooth
  bosonization: The cheshire cat revisited},}\ }\href {\doibase
  https://doi.org/10.1016/0550-3213(92)90100-P} {\bibfield  {journal} {\bibinfo
   {journal} {Nuclear Physics B}\ }\textbf {\bibinfo {volume} {385}},\ \bibinfo
  {pages} {227 -- 250} (\bibinfo {year} {1992})}\BibitemShut {NoStop}%
\bibitem [{\citenamefont {Wetterich}(2021{\natexlab{a}})}]{CWPCA}%
  \BibitemOpen
  \bibfield  {author} {\bibinfo {author} {\bibfnamefont {C.}~\bibnamefont
  {Wetterich}},\ }\bibfield  {title} {\enquote {\bibinfo {title} {Probabilistic
  cellular automata for interacting fermionic quantum field theories},}\
  }\href@noop {} {\bibfield  {journal} {\bibinfo  {journal} {Nuclear Physics
  B}\ }\textbf {\bibinfo {volume} {963}},\ \bibinfo {pages} {115296} (\bibinfo
  {year} {2021}{\natexlab{a}})},\ \Eprint {http://arxiv.org/abs/2007.06366}
  {arXiv:2007.06366 [quant-ph]} \BibitemShut {NoStop}%
\bibitem [{\citenamefont {Wetterich}(2010{\natexlab{b}})}]{CWFCS}%
  \BibitemOpen
  \bibfield  {author} {\bibinfo {author} {\bibfnamefont {C.}~\bibnamefont
  {Wetterich}},\ }\bibfield  {title} {\enquote {\bibinfo {title} {{Fermions
  from classical statistics}},}\ }\href {\doibase 10.1016/j.aop.2010.07.003}
  {\bibfield  {journal} {\bibinfo  {journal} {Annals Phys.}\ }\textbf {\bibinfo
  {volume} {325}},\ \bibinfo {pages} {2750--2786} (\bibinfo {year}
  {2010}{\natexlab{b}})},\ \Eprint {http://arxiv.org/abs/1006.4254}
  {arXiv:1006.4254 [hep-th]} \BibitemShut {NoStop}%
\bibitem [{\citenamefont {Wetterich}(2021{\natexlab{b}})}]{CWFCB}%
  \BibitemOpen
  \bibfield  {author} {\bibinfo {author} {\bibfnamefont {C.}~\bibnamefont
  {Wetterich}},\ }\bibfield  {title} {\enquote {\bibinfo {title} {{Quantum
  fermions from classical bits}},}\ }\href@noop {} {\  (\bibinfo {year}
  {2021}{\natexlab{b}})},\ \Eprint {http://arxiv.org/abs/2106.15517}
  {arXiv:2106.15517 [quant-ph]} \BibitemShut {NoStop}%
\bibitem [{\citenamefont {Thirring}(1958)}]{THI}%
  \BibitemOpen
  \bibfield  {author} {\bibinfo {author} {\bibfnamefont {Walter~E.}\
  \bibnamefont {Thirring}},\ }\bibfield  {title} {\enquote {\bibinfo {title} {A
  soluble relativistic field theory},}\ }\href {\doibase
  https://doi.org/10.1016/0003-4916(58)90015-0} {\bibfield  {journal} {\bibinfo
   {journal} {Annals of Physics}\ }\textbf {\bibinfo {volume} {3}},\ \bibinfo
  {pages} {91 -- 112} (\bibinfo {year} {1958})}\BibitemShut {NoStop}%
\bibitem [{\citenamefont {Klaiber}(1968)}]{KLA}%
  \BibitemOpen
  \bibfield  {author} {\bibinfo {author} {\bibfnamefont {B.}~\bibnamefont
  {Klaiber}},\ }\bibfield  {title} {\enquote {\bibinfo {title} {{The thirring
  model}},}\ }\href@noop {} {\bibfield  {journal} {\bibinfo  {journal} {Lect.
  Theor. Phys. A}\ }\textbf {\bibinfo {volume} {10}},\ \bibinfo {pages}
  {141--176} (\bibinfo {year} {1968})}\BibitemShut {NoStop}%
\bibitem [{\citenamefont {Abdalla}\ \emph {et~al.}(1991)\citenamefont
  {Abdalla}, \citenamefont {Abdalla},\ and\ \citenamefont {Rothe}}]{AAR}%
  \BibitemOpen
  \bibfield  {author} {\bibinfo {author} {\bibfnamefont {E.}~\bibnamefont
  {Abdalla}}, \bibinfo {author} {\bibfnamefont {M.C.B.}\ \bibnamefont
  {Abdalla}}, \ and\ \bibinfo {author} {\bibfnamefont {K.D.}\ \bibnamefont
  {Rothe}},\ }\href@noop {} {\emph {\bibinfo {title} {{Nonperturbative methods
  in two-dimensional quantum field theory}}}}\ (\bibinfo {year}
  {1991})\BibitemShut {NoStop}%
\bibitem [{\citenamefont {Faber}\ and\ \citenamefont {Ivanov}(2001)}]{FAIV}%
  \BibitemOpen
  \bibfield  {author} {\bibinfo {author} {\bibfnamefont {M.}~\bibnamefont
  {Faber}}\ and\ \bibinfo {author} {\bibfnamefont {A.N.}\ \bibnamefont
  {Ivanov}},\ }\bibfield  {title} {\enquote {\bibinfo {title} {{On the solution
  of the massless Thirring model with fermion fields quantized in the chiral
  symmetric phase}},}\ }\href@noop {} {\  (\bibinfo {year} {2001})},\ \Eprint
  {http://arxiv.org/abs/hep-th/0112183} {arXiv:hep-th/0112183} \BibitemShut
  {NoStop}%
\bibitem [{\citenamefont {Wetterich}(2018{\natexlab{a}})}]{CWIT}%
  \BibitemOpen
  \bibfield  {author} {\bibinfo {author} {\bibfnamefont {C.}~\bibnamefont
  {Wetterich}},\ }\bibfield  {title} {\enquote {\bibinfo {title} {{Information
  transport in classical statistical systems}},}\ }\href {\doibase
  10.1016/j.nuclphysb.2017.12.008} {\bibfield  {journal} {\bibinfo  {journal}
  {Nucl. Phys. B}\ }\textbf {\bibinfo {volume} {927}},\ \bibinfo {pages}
  {35--96} (\bibinfo {year} {2018}{\natexlab{a}})},\ \Eprint
  {http://arxiv.org/abs/1611.04820} {arXiv:1611.04820 [cond-mat.stat-mech]}
  \BibitemShut {NoStop}%
\bibitem [{\citenamefont {Wetterich}(2018{\natexlab{b}})}]{CWQF}%
  \BibitemOpen
  \bibfield  {author} {\bibinfo {author} {\bibfnamefont {C.}~\bibnamefont
  {Wetterich}},\ }\bibfield  {title} {\enquote {\bibinfo {title} {{Quantum
  formalism for classical statistics}},}\ }\href {\doibase
  10.1016/j.aop.2018.03.022} {\bibfield  {journal} {\bibinfo  {journal} {Annals
  Phys.}\ }\textbf {\bibinfo {volume} {393}},\ \bibinfo {pages} {1--70}
  (\bibinfo {year} {2018}{\natexlab{b}})},\ \Eprint
  {http://arxiv.org/abs/1706.01772} {arXiv:1706.01772 [quant-ph]} \BibitemShut
  {NoStop}%
\bibitem [{\citenamefont {Baxter}(1982)}]{BAX}%
  \BibitemOpen
  \bibfield  {author} {\bibinfo {author} {\bibfnamefont {R.J.}\ \bibnamefont
  {Baxter}},\ }\href@noop {} {\emph {\bibinfo {title} {{Exactly solved models
  in statistical mechanics}}}}\ (\bibinfo {year} {1982})\BibitemShut {NoStop}%
\bibitem [{\citenamefont {Fuchs}(1990)}]{FUC}%
  \BibitemOpen
  \bibfield  {author} {\bibinfo {author} {\bibfnamefont {Norman~H.}\
  \bibnamefont {Fuchs}},\ }\bibfield  {title} {\enquote {\bibinfo {title}
  {Transfer-matrix analysis for ising models},}\ }\href {\doibase
  10.1103/PhysRevB.41.2173} {\bibfield  {journal} {\bibinfo  {journal} {Phys.
  Rev. B}\ }\textbf {\bibinfo {volume} {41}},\ \bibinfo {pages} {2173--2183}
  (\bibinfo {year} {1990})}\BibitemShut {NoStop}%
\bibitem [{\citenamefont {Wetterich}(2012)}]{CWPT}%
  \BibitemOpen
  \bibfield  {author} {\bibinfo {author} {\bibfnamefont {C.}~\bibnamefont
  {Wetterich}},\ }\bibfield  {title} {\enquote {\bibinfo {title}
  {{Probabilistic Time}},}\ }\href {\doibase 10.1007/s10701-012-9675-3}
  {\bibfield  {journal} {\bibinfo  {journal} {Found. Phys.}\ }\textbf {\bibinfo
  {volume} {42}},\ \bibinfo {pages} {1384--1443} (\bibinfo {year} {2012})},\
  \Eprint {http://arxiv.org/abs/1002.2593} {arXiv:1002.2593 [hep-th]}
  \BibitemShut {NoStop}%
\bibitem [{\citenamefont {Wetterich}(2011{\natexlab{a}})}]{Wetterich:2010ni}%
  \BibitemOpen
  \bibfield  {author} {\bibinfo {author} {\bibfnamefont {C.}~\bibnamefont
  {Wetterich}},\ }\bibfield  {title} {\enquote {\bibinfo {title} {{Spinors in
  euclidean field theory, complex structures and discrete symmetries}},}\
  }\href {\doibase 10.1016/j.nuclphysb.2011.06.013} {\bibfield  {journal}
  {\bibinfo  {journal} {Nucl. Phys. B}\ }\textbf {\bibinfo {volume} {852}},\
  \bibinfo {pages} {174--234} (\bibinfo {year} {2011}{\natexlab{a}})},\ \Eprint
  {http://arxiv.org/abs/1002.3556} {arXiv:1002.3556 [hep-th]} \BibitemShut
  {NoStop}%
\bibitem [{\citenamefont {Wetterich}(2011{\natexlab{b}})}]{CWQFT}%
  \BibitemOpen
  \bibfield  {author} {\bibinfo {author} {\bibfnamefont {C.}~\bibnamefont
  {Wetterich}},\ }\bibfield  {title} {\enquote {\bibinfo {title} {{Quantum
  field theory from classical statistics}},}\ }\href@noop {} {\  (\bibinfo
  {year} {2011}{\natexlab{b}})},\ \Eprint {http://arxiv.org/abs/1111.4115}
  {arXiv:1111.4115 [hep-th]} \BibitemShut {NoStop}%
\bibitem [{\citenamefont {Wetterich}(2011{\natexlab{c}})}]{Wetterich:2011dt}%
  \BibitemOpen
  \bibfield  {author} {\bibinfo {author} {\bibfnamefont {C.}~\bibnamefont
  {Wetterich}},\ }\bibfield  {title} {\enquote {\bibinfo {title} {{Classical
  probabilities for Majorana and Weyl spinors}},}\ }\href {\doibase
  10.1016/j.aop.2011.04.005} {\bibfield  {journal} {\bibinfo  {journal} {Annals
  Phys.}\ }\textbf {\bibinfo {volume} {326}},\ \bibinfo {pages} {2243--2293}
  (\bibinfo {year} {2011}{\natexlab{c}})},\ \Eprint
  {http://arxiv.org/abs/1102.3586} {arXiv:1102.3586 [hep-th]} \BibitemShut
  {NoStop}%
\bibitem [{\citenamefont {Gross}\ and\ \citenamefont {Neveu}(1974)}]{GN}%
  \BibitemOpen
  \bibfield  {author} {\bibinfo {author} {\bibfnamefont {David~J.}\
  \bibnamefont {Gross}}\ and\ \bibinfo {author} {\bibfnamefont {Andr\'e}\
  \bibnamefont {Neveu}},\ }\bibfield  {title} {\enquote {\bibinfo {title}
  {Dynamical symmetry breaking in asymptotically free field theories},}\ }\href
  {\doibase 10.1103/PhysRevD.10.3235} {\bibfield  {journal} {\bibinfo
  {journal} {Phys. Rev. D}\ }\textbf {\bibinfo {volume} {10}},\ \bibinfo
  {pages} {3235--3253} (\bibinfo {year} {1974})}\BibitemShut {NoStop}%
\bibitem [{\citenamefont {Wetzel}(1985)}]{WWE}%
  \BibitemOpen
  \bibfield  {author} {\bibinfo {author} {\bibfnamefont {Werner}\ \bibnamefont
  {Wetzel}},\ }\bibfield  {title} {\enquote {\bibinfo {title} {Two-loop
  $\beta$-function for the gross-neveu model},}\ }\href {\doibase
  https://doi.org/10.1016/0370-2693(85)90551-9} {\bibfield  {journal} {\bibinfo
   {journal} {Physics Letters B}\ }\textbf {\bibinfo {volume} {153}},\ \bibinfo
  {pages} {297--299} (\bibinfo {year} {1985})}\BibitemShut {NoStop}%
\bibitem [{\citenamefont {Rosenstein}\ \emph {et~al.}(1991)\citenamefont
  {Rosenstein}, \citenamefont {Warr},\ and\ \citenamefont {Park}}]{RWP}%
  \BibitemOpen
  \bibfield  {author} {\bibinfo {author} {\bibfnamefont {Baruch}\ \bibnamefont
  {Rosenstein}}, \bibinfo {author} {\bibfnamefont {Brian~J.}\ \bibnamefont
  {Warr}}, \ and\ \bibinfo {author} {\bibfnamefont {Seon~H.}\ \bibnamefont
  {Park}},\ }\bibfield  {title} {\enquote {\bibinfo {title} {Dynamical symmetry
  breaking in four-fermion interaction models},}\ }\href {\doibase
  https://doi.org/10.1016/0370-1573(91)90129-A} {\bibfield  {journal} {\bibinfo
   {journal} {Physics Reports}\ }\textbf {\bibinfo {volume} {205}},\ \bibinfo
  {pages} {59--108} (\bibinfo {year} {1991})}\BibitemShut {NoStop}%
\bibitem [{\citenamefont {Stoll}\ \emph {et~al.}(2021)\citenamefont {Stoll},
  \citenamefont {Zorbach}, \citenamefont {Koenigstein}, \citenamefont {Steil},\
  and\ \citenamefont {Rechenberger}}]{SZKSR}%
  \BibitemOpen
  \bibfield  {author} {\bibinfo {author} {\bibfnamefont {Jonas}\ \bibnamefont
  {Stoll}}, \bibinfo {author} {\bibfnamefont {Niklas}\ \bibnamefont {Zorbach}},
  \bibinfo {author} {\bibfnamefont {Adrian}\ \bibnamefont {Koenigstein}},
  \bibinfo {author} {\bibfnamefont {Martin~J.}\ \bibnamefont {Steil}}, \ and\
  \bibinfo {author} {\bibfnamefont {Stefan}\ \bibnamefont {Rechenberger}},\
  }\href@noop {} {\enquote {\bibinfo {title} {Bosonic fluctuations in the $( 1
  + 1 )$-dimensional gross-neveu(-yukawa) model at varying $\mu$ and $t$ and
  finite $n$},}\ } (\bibinfo {year} {2021}),\ \Eprint
  {http://arxiv.org/abs/2108.10616} {arXiv:2108.10616 [hep-ph]} \BibitemShut
  {NoStop}%
\bibitem [{\citenamefont {Wetterich}(2010{\natexlab{c}})}]{CWQPCS}%
  \BibitemOpen
  \bibfield  {author} {\bibinfo {author} {\bibfnamefont {C.}~\bibnamefont
  {Wetterich}},\ }\bibfield  {title} {\enquote {\bibinfo {title} {{Quantum
  particles from classical statistics}},}\ }\href {\doibase
  10.1002/andp.201000088} {\bibfield  {journal} {\bibinfo  {journal} {Annalen
  Phys.}\ }\textbf {\bibinfo {volume} {522}},\ \bibinfo {pages} {807} (\bibinfo
  {year} {2010}{\natexlab{c}})},\ \Eprint {http://arxiv.org/abs/0904.3048}
  {arXiv:0904.3048 [quant-ph]} \BibitemShut {NoStop}%
\bibitem [{\citenamefont {Wetterich}(2019)}]{CWQC}%
  \BibitemOpen
  \bibfield  {author} {\bibinfo {author} {\bibfnamefont {C.}~\bibnamefont
  {Wetterich}},\ }\bibfield  {title} {\enquote {\bibinfo {title} {{Quantum
  computing with classical bits}},}\ }\href {\doibase
  10.1016/j.nuclphysb.2019.114776} {\bibfield  {journal} {\bibinfo  {journal}
  {Nucl. Phys. B}\ }\textbf {\bibinfo {volume} {948}},\ \bibinfo {pages}
  {114776} (\bibinfo {year} {2019})},\ \Eprint
  {http://arxiv.org/abs/1806.05960} {arXiv:1806.05960 [quant-ph]} \BibitemShut
  {NoStop}%
\bibitem [{\citenamefont {Sexty}\ and\ \citenamefont
  {Wetterich}(2018)}]{SEXCW}%
  \BibitemOpen
  \bibfield  {author} {\bibinfo {author} {\bibfnamefont {D.}~\bibnamefont
  {Sexty}}\ and\ \bibinfo {author} {\bibfnamefont {C.}~\bibnamefont
  {Wetterich}},\ }\bibfield  {title} {\enquote {\bibinfo {title} {{Static
  memory materials}},}\ }\href@noop {} {\  (\bibinfo {year} {2018})},\ \Eprint
  {http://arxiv.org/abs/1802.08596} {arXiv:1802.08596 [cond-mat.stat-mech]}
  \BibitemShut {NoStop}%
\bibitem [{\citenamefont {Pehle}\ \emph {et~al.}(2018)\citenamefont {Pehle},
  \citenamefont {Meier}, \citenamefont {Oberthaler},\ and\ \citenamefont
  {Wetterich}}]{PMOW}%
  \BibitemOpen
  \bibfield  {author} {\bibinfo {author} {\bibfnamefont {C.}~\bibnamefont
  {Pehle}}, \bibinfo {author} {\bibfnamefont {K.}~\bibnamefont {Meier}},
  \bibinfo {author} {\bibfnamefont {M.}~\bibnamefont {Oberthaler}}, \ and\
  \bibinfo {author} {\bibfnamefont {C.}~\bibnamefont {Wetterich}},\ }\bibfield
  {title} {\enquote {\bibinfo {title} {{Emulating quantum computation with
  artificial neural networks}},}\ }\href@noop {} {\  (\bibinfo {year}
  {2018})},\ \Eprint {http://arxiv.org/abs/1810.10335} {arXiv:1810.10335
  [quant-ph]} \BibitemShut {NoStop}%
\bibitem [{\citenamefont {Pehle}\ and\ \citenamefont {Wetterich}(2021)}]{PW}%
  \BibitemOpen
  \bibfield  {author} {\bibinfo {author} {\bibfnamefont {C.}~\bibnamefont
  {Pehle}}\ and\ \bibinfo {author} {\bibfnamefont {C.}~\bibnamefont
  {Wetterich}},\ }\bibfield  {title} {\enquote {\bibinfo {title} {Neuromorphic
  quantum computing},}\ }\href@noop {} {\  (\bibinfo {year} {2021})},\ \Eprint
  {http://arxiv.org/abs/2005.01533} {arXiv:2005.01533 [cond-mat.dis-nn]}
  \BibitemShut {NoStop}%
\bibitem [{\citenamefont {Wetterich}(2009)}]{CWEM}%
  \BibitemOpen
  \bibfield  {author} {\bibinfo {author} {\bibfnamefont {C.}~\bibnamefont
  {Wetterich}},\ }\bibfield  {title} {\enquote {\bibinfo {title} {{Emergence of
  quantum mechanics from classical statistics}},}\ }\href {\doibase
  10.1088/1742-6596/174/1/012008} {\bibfield  {journal} {\bibinfo  {journal}
  {J. Phys. Conf. Ser.}\ }\textbf {\bibinfo {volume} {174}},\ \bibinfo {pages}
  {012008} (\bibinfo {year} {2009})},\ \Eprint {http://arxiv.org/abs/0811.0927}
  {arXiv:0811.0927 [quant-ph]} \BibitemShut {NoStop}%
\end{thebibliography}%
\end{document}